\documentclass[11pt]{article}

\usepackage{booktabs} 

\pdfoutput=1
\usepackage{latexsym}
\usepackage{epsfig}
\usepackage[mathscr]{eucal}
\usepackage{amsfonts}
\usepackage{amscd}
\usepackage{cite}
\usepackage{array}
\usepackage{bbold}
\usepackage{amssymb}
\usepackage{colordvi}
\usepackage[centertags]{amsmath}
\usepackage{enumerate}
\usepackage{graphicx}
\usepackage{booktabs}
\usepackage{theorem}
\usepackage{caption}
\usepackage{soul}
\usepackage{url}
\usepackage[hidelinks]{hyperref}
\usepackage{mcite}
\usepackage{slashed}
\usepackage[dvipsnames]{xcolor}
\usepackage{ulem}
\usepackage{bbm}
\usepackage[utf8]{inputenc}
\usepackage{scrextend}
\usepackage{listings}
\usepackage{multirow}
\usepackage{textpos}  
\usepackage{slashed}
\usepackage{tikz}
\usepackage{tikz-feynman}
\usepackage{booktabs}
\usepackage{float}
\usepackage{cancel}

\definecolor{codegreen}{rgb}{0,0.6,0}
\definecolor{codegray}{rgb}{0.5,0.5,0.5}
\definecolor{codepurple}{rgb}{0.58,0,0.82}
\definecolor{backcolour}{rgb}{0.95,0.95,0.92}

\lstdefinestyle{mystyle}{
    backgroundcolor=\color{backcolour},   
    commentstyle=\color{codegreen},
    keywordstyle=\color{blue},
    numberstyle=\tiny\color{codegray},
    stringstyle=\color{codepurple},
    basicstyle=\ttfamily\footnotesize,
    breakatwhitespace=false,         
    breaklines=true,                 
    captionpos=b,                    
    keepspaces=true,                 
    numbers=left,                    
    numbersep=5pt,                  
    showspaces=false,                
    showstringspaces=false,
    showtabs=false,                  
    tabsize=2
}

\newcommand{\etab}{\overline{\eta}_2}

\allowdisplaybreaks{}

\numberwithin{equation}{section}

\newcommand{\be}{\begin{equation}}
\newcommand{\ee}{\end{equation}}
\newcommand{\beq}{\begin{eqnarray}}
\newcommand{\eeq}{\end{eqnarray}}

\newcommand{\RNum}[1]{\uppercase\expandafter{\romannumeral #1\relax}}

\newcommand{\si}[1]{\text}
\newcommand{\SI}[1]{\text}

\newcommand{\s}{\newline \vspace*{-3.5mm}}

\usepackage{comment}

\begin{document}

\lstset{style=mystyle}

\title{

  \vspace{-1.3cm} 
    \hfill {\small \hfill KA-TP-11-2026}\\[1cm]
    \centering
    {A Deep Dive into Baryon Asymmetry -- the C2HDM} 
  
}

\date{\today}
\author{
Margarete M\"{u}hlleitner$^{1\,}$\footnote{E-mail:
	\texttt{margarete.muehlleitner@kit.edu}},
Johann Plotnikov$^{1\,}$\footnote{E-mail:  \texttt{johann.plotnikov@online.de}},
Rui Santos$^{2,3\,}$\footnote{E-mail:  \texttt{rasantos@fc.ul.pt}},
João Viana$^{2,3}$\footnote{E-mail: \texttt{jfvvchico@hotmail.com}}
\\[9mm]
{\small\it
$^1$Institute for Theoretical Physics, Karlsruhe Institute of Technology,} \\
{\small\it Wolfgang-Gaede-Str. 1, 76131 Karlsruhe, Germany.}\\[3mm]
{\small\it
$^2$Departamento de F\'{\i}sica, Faculdade de Ci\^{e}ncias,} \\
{\small \it   Universidade de Lisboa, 1749-016 Lisboa, Portugal} \\[3mm]
{\small\it
$^3$Centro de F\'{\i}sica Te\'{o}rica e Computacional,
    Faculdade de Ci\^{e}ncias,} \\
{\small \it    Universidade de Lisboa, Campo Grande, Edif\'{\i}cio C8
  1749-016 Lisboa, Portugal} \\[3mm]
}

\maketitle
\begin{abstract}
In this paper, we present our new implementation of the computation of the baryon asymmetry in the code \texttt{BSMPT}. It is based on the WKB ansatz generalizing the transport equations to an arbitrary number of moments. Two different truncation schemes are implemented, and the profile of the vacuum expectation value (VEV) is  derived from the equations of motion in addition to the modeling with the kink profile. We validate our implementation with a simple benchmark model and perform a detailed analysis within the CP-violating 2-Higgs-Doublet Model (C2HDM). Barring the collision term, however, our implementation can readily be applied to any extended Higgs sector with an arbitrary number of VEV directions. We study in detail the dependencies of the baryon asymmetry on the number of moment equations, the applied truncation scheme, the wall velocity, the wall velocity times wall width, the VEV profile, the strength of the phase transition, and the amount of CP violation in the model and present a detailed uncertainty analysis. We investigate the interplay of the generated baryon asymmetry and the gravitational waves signal at \texttt{LISA}. Our results guide the way for future improvements in the computation of the baryon asymmetry and give directions for model building. The uncertainty analysis is the basis for any investigation aiming at deducing model parameters from cosmological processes.
\end{abstract}
\newpage

\tableofcontents

\newpage
\section{Introduction}
 One of the most prominent open questions in particle physics is why there is is more matter than antimatter in the universe. The slight excess of matter is quantified by the baryon asymmetry divided by the entropy density measured by Planck \cite{Planck:2018vyg} as
\begin{equation}
        \eta_\text{obs}=\frac{n_B-n_{\bar{B}}}{s}\simeq\frac{n_B}{s}\simeq(8.690\pm0.053)\cdot10^{-11}\label{eq: BAU}\;.
\end{equation} 
A dynamical generation of this baryon asymmetry of the Universe (BAU) is given by the mechanism of electroweak baryogenesis \cite{Kuzmin:1985mm,Cohen:1990it,Cohen:1993nk,Quiros:1994dr,Rubakov:1996vz,Funakubo:1996dw,Trodden:1998ym,Bernreuther:2002uj,Morrissey:2012db}, provided the three Sakharov conditions \cite{Sakharov:1967dj} -- baryon number violation, C and CP violation, departure from thermal equilibrium -- are fulfilled. \s

In this paper, we will consider non-local baryogenesis, where the baryon asymmetry is generated away from the bubble wall. Roughly sketched, it proceeds in three steps. During the evolution of the universe, starting from a symmetric vacuum with a zero vacuum expectation value (VEV), bubbles with a non-zero VEV start to form as the universe cools down  and expand into the thermal plasma with a wall velocity $v_w$. In the first step, left- and right-handed particles in the plasma of the symmetric vacuum collide with the expanding bubble wall, with wall width $L_w$. Since the left- and right-handed particles have different reflection coefficients because of the assumed CP violation, we generate a chiral asymmetry outside the bubble.
 In the second step, the abundance is converted through weak sphaleron processes into a baryon asymmetry. 
This transient baryon asymmetry is transferred, in the third step, into the expanding bubble of the broken vacuum, where electroweak sphaleron processes are sufficiently suppressed to prevent the asymmetry from being washed out.
In the presence of a strong first order electroweak phase transition, inside the bubble the suppression of sphaleron processes is sufficiently large to preserve the generated baryon asymmetry, leaving us with the observed matter–antimatter asymmetry. \s 

The goal of this work is to compute the generated amount of baryon asymmetry from first principles. For this we updated the implementation of the calculation of the baryon asymmetry in the CP-violating 2-Higgs Doublet Model (C2HDM)\footnote{For recent studies that include the calculation of the BAU with different levels of approximations in the C2HDM see for instance~\cite{Fromme:2006cm, Cline:2011mm, Goncalves:2023svb, Kanemura:2023juv, Aiko:2025tbk,Biekotter:2025fjx}.} in our code \texttt{BSMPT}v2 \cite{Basler:2020nrq}, which will be released as version \texttt{BSMPT}v4 of \texttt{BSMPT} \cite{Basler:2018cwe} in a follow-up of the present paper. With this code, we are able to go through the whole chain starting from a particle physics model at zero temperature to the generated baryon asymmetry at the electroweak phase transition: We define our model of interest of today and check it with our code \texttt{ScannerS} \cite{Coimbra:2013qq,Ferreira:2014dya,Costa:2015llh,Muhlleitner:2016mzt,Muhlleitner:2020wwk} for the relevant theoretical and experimental constraints. With the last release \texttt{BSMPT}v3 \cite{Basler:2024aaf} of our code \texttt{BSMPT}, we trace the vacuum history of the model at non-zero temperature. The code computes the transition rates for the found transitions from the respective false vacuum to the true vacuum. It calculates the relevant thermal parameters and the related gravitational waves spectra. For the transitions from a symmetric (false) vacuum into an electroweak broken (true) vacuum, the generated amount of baryon asymmetry is computed. To the best of our knowledge, this is the only code that performs the entire chain from a particle physics model to the computation of the baryon asymmetry.\footnote{The code \texttt{BARYONET} \cite{Barni:2025ifb} is a code that solely computes the baryon asymmetry.} \s

The implementation of the whole process in one program allows not only to be self-contained but above all to be consistent. We thus have full control of all used  algorithms and the applied approximations, we ensure that the various quantities can be defined consistently, we can react flexibly and quickly to new developments in the literature and implement them accordingly. The consistent approach allows us to provide meaningful uncertainty estimates\footnote{For a discussion of approximations and uncertainties, cf.~e.g.~\cite{Athron:2023xlk,Kainulainen:2024qpm}.}, allowing us to interpret the obtained results appropriately. This also gives a guide to directions to be taken to further improve the derivations on the one hand and for model building on the other hand. \s

The new features with respect to our previous implementation \cite{Basler:2021kgq} of the calculation of the baryon asymmetry in  \texttt{BSMPT}v2 are:
\begin{itemize}
\item[$\diamond$] Implementation of the transport equations based on the WKB ansatz with the recent development given in Ref.~\cite{Kainulainen:2024qpm}, which generalizes the transport equations systematically to an arbitrary number of moments, of which we implemented $50$ in \texttt{BSMPT}.
\item[$\diamond$] The proper implementation of the boundary conditions for the solution of the transport equations.
\item[$\diamond$] The implementation of two different truncation schemes in the solution of the transport equations.
\item[$\diamond$] The possibility to derive and apply the VEV profile from the solution of the equations of motion.
\item[$\diamond$] Our new computation of the top-Yukawa rate at non-zero temperature.
\end{itemize}
In this paper, we will describe in detail this new implementation and the approximations applied. We will then move on to validate our code by applying it to the benchmark model introduced in \cite{Kainulainen:2024qpm}. We investigated the impact of number $n$ of moment equations, of the truncation constant, the wall velocity, of the size of the parameter $L_wT_n$, where $T_n$ is the nucleation temperature, and the strength of the phase transition and compare with the literature where possible. \s

We will then move on to a dedicated analysis in the framework of the C2HDM, for a generated sample of points that were tested against the relevant theoretical and experimental constraints. Starting from a benchmark sample we successively vary various parameters to perform a dedicated study of the uncertainties related to the used number $n$ of transport equations, the parameter $L_w T_p$, where $T_p$ denotes the percolation temperature, the choice of the transition temperature and the wall velocity. Subsequently, we will move on to the phenomenological study where we analyse the impact of CP violation and investigate the relation between baryon asymmetry and a possible gravitational waves signal. We will conclude the study with the dedicated analysis of the impact of the VEV profiles on the $z$ dependence of the top quark mass and its CP-violating phase as well as the relation between generated baryon asymmetry and the chosen number $n$ of transport equations and the transition temperature, respectively. While we perform studies for the benchmark model and the C2HDM, our implementation is set up such that it can  readily be applied to any extended Higgs sector model with several VEV directions barring the model-dependent collision term.
\s

The organization of the paper is as follows. In the subsequent Sec.~\ref{sec:model}, we introduce the models used in our analysis, the benchmark model and the C2HDM. In Sec.~\ref{sec:transequ} we derive the transport equations needed for the computation of the BAU, using the WKB ansatz. Section~\ref{sec:momexp} introduces the moment expansion applied to solve the transport equations. In Sec.~\ref{sec:bau} we present the formulae for the computation of the amount of generated baryon asymmetry. Section~\ref{sec:bubbleprofile} describes the bubble profiles that we apply. After these theory chapters, we move on to the numerical analysis, starting with the benchmark model in Sec.~\ref{sec:numbench} and moving on to the C2HDM in Sec.~\ref{sec:c2hdmresults}. We finish with our conclusions in Sec.~\ref{sec:concl}. In App.~\ref{app: yuk}, we describe our computation of the top Yukawa rate and in App.~\ref{app: universal_funcs} we give the general formula of the universal functions needed to solve the transport equations.
\s


\section{Investigated Models \label{sec:model}}
In the following numerical analysis, we will consider two different models. In the benchmark model introduced in Ref.~\cite{Kainulainen:2024qpm}, we will study the effects of different transition temperatures, wall velocities, wall thicknesses and truncation schemes and compare with the literature. In the CP-violating 2-Higgs-Doublet Model (C2HDM)~\cite{Branco:1985aq,Ginzburg:2002wt,Gunion:2002zf,Khater:2003wq,Ginzburg:2004vp,ElKaffas:2006gdt,Arhrib:2010ju,Barroso:2012wz,Inoue:2014nva,Fontes:2014xva,Grzadkowski:2014ada}, we will investigate the impact of the various C2HDM potential parameters on the BAU. We start by briefly introducing the models. 

\subsection{Benchmark Model \label{sec:benchmark}}
The benchmark model includes a massless left-chiral bottom quark and a massless Higgs boson as well as a heavy top quark. They are coupled via gauge and Yukawa interactions. CP-violation is sourced by the complex mass of the top quark with the usual Yukawa coupling to the Higgs field $h$, and by a dimension-5 operator,
\begin{align}
i \frac{s}{\Lambda} \bar{Q}_3 H t_R, 
\end{align}
which couples the top to the additionally introduced singlet field $s$. Here, $Q_3$ and $H$ are the left-handed third generation quark doublet and the Higgs doublet, respectively, and $\Lambda$ denotes the new physics scale. When both $h$ and $s$ vary accross the bubble wall, the top quark obtains an effective spatially varying complex mass term, given by 
\begin{align}
m_t (z) = y_t h(z) \left( 1 + i \frac{s(z)}{\Lambda} \right) \,,
\end{align}
with $y_t$ denoting the strength of the top-quark Yukawa coupling. The absolute value and phase of the top-quark mass are given by
\begin{align}
|m_t (z)| = y_t h(z) \sqrt{1 + \frac{s(z)^2}{\Lambda^2}} \;, \quad \theta(z) = \arctan \frac{s(z)}{\Lambda} \;.
\end{align}
The VEV profiles of the scalar fields $h(z)$ and $s(z)$ are modeled by the kink ansatz as
\begin{align}
h(z) = \frac{v_n}{2} \left( 1- \tanh \frac{z}{L_w} \right) \;, \quad s(z) = \frac{\omega_n}{2} \left( 1 + \tanh \frac{z}{L_w} \right) \;.
\end{align}
The input parameters that we will use in the numerical analysis are chosen as
\begin{align}
v_n = \frac{1}{2} \omega_n = T_n \,, \quad \Lambda = 1\mbox{ TeV} \,, \quad L_w = L_s = \frac{5}{T_n} \,, \quad T_n = 100 \mbox{ GeV} \;.
\end{align}

\subsection{The C2HDM}
In the C2HDM, a second Higgs doublet is added to the SM Higgs sector which belongs to the same gauge representation as the SM Higgs doublet. Following the notation introduced in Ref.~\cite{Fontes:2017zfn}, the most general gauge-invariant renormalizable Higgs potential with a softly broken $\mathrm{Z}_2$ symmetry can be written as 
\begin{eqnarray}
    V &=& m_{11}^2 |\Phi_1|^2 + m_{22}^2 |\Phi_2|^2
- \left(m_{12}^2 \, \Phi_1^\dagger \Phi_2 + h.c.\right)
+ \frac{\lambda_1}{2} (\Phi_1^\dagger \Phi_1)^2 +
\frac{\lambda_2}{2} (\Phi_2^\dagger \Phi_2)^2 \nonumber \\
&& + \lambda_3
(\Phi_1^\dagger \Phi_1) (\Phi_2^\dagger \Phi_2) + \lambda_4
(\Phi_1^\dagger \Phi_2) (\Phi_2^\dagger \Phi_1) +
\left[\frac{\lambda_5}{2} (\Phi_1^\dagger \Phi_2)^2 + h.c.\right] \;,
\end{eqnarray}
where because of hermiticity all couplings are real, except for $m_{12}^2$ and $\lambda_5$ which are chosen to be complex.
Expanding the doublets about the real vacuum expectation values (VEVs) of the neutral components, 
$\langle \Phi_i^0 \rangle =v_i / \sqrt{2}\,\,(i=1,2)$,
with $v^2=v_1^2+v_2^2\approx 246~{\rm GeV}$, they can be written as
\begin{equation}
\Phi_1 = \left(
\begin{array}{c}
\phi_1^+ \\
\frac{v_1 + \rho_1 + i \eta_1}{\sqrt{2}}
\end{array}
\right) \qquad \mbox{and} \qquad
\Phi_2 = \left(
\begin{array}{c}
\phi_2^+ \\
\frac{v_2 + \rho_2 + i \eta_2}{\sqrt{2}}
\end{array}
\right) \;, \label{eq:2hdmdoubletexpansion}
\end{equation}
in terms of the real fields $\rho_i$ and $\eta_i$ and the charged fields $\phi_i^+$. 
The minimum conditions read
\begin{eqnarray}
  m_{11}^2 v_1 + \frac{\lambda_1}{2} v_1^3 + \frac{\lambda_{345}}{2} v_1
v_2^2 &=& \textrm{Re} \left(m_{12}^2\right) v_2 \;, \label{eq:mincond1} \\
m_{22}^2 v_2 + \frac{\lambda_2}{2} v_2^3 + \frac{\lambda_{345}}{2} v_1^2
v_2 &=& \textrm{Re} \left(m_{12}^2\right) v_1 \;, \label{eq:mincond2} \\
2\, \mbox{Im} (m_{12}^2) &=& v_1 v_2 \mbox{Im} (\lambda_5)
\;, \label{eq:mincond3}  
\end{eqnarray}
which, for non-zero VEVs $v_1$ and $v_2$,
ensure one independent CP-violating phase if the condition 
$\mbox{Im} \left\{\lambda_5^* \left(m_{12}^2\right)^2\right\} \neq 0$ is fulfilled~\cite{Ginzburg:2002wt,Gunion:2002zf,Barroso:2012wz}. To obtain the mass eigenstates, first a rotation to the Higgs basis is performed~\cite{Georgi:1978ri,Donoghue:1978cj,Lavoura:1994fv,Botella:1994cs},
\begin{equation}
     \left( \begin{array}{c} {\cal H}_1 \\ {\cal H}_2 \end{array} \right) =
 R^T_H \left( \begin{array}{c} \Phi_1 \\
     \Phi_2 \end{array} \right) \equiv
 \left( \begin{array}{cc} c_\beta & s_\beta \\ - s_\beta &
     c_\beta \end{array} \right) \left( \begin{array}{c} \Phi_1 \\
     \Phi_2 \end{array} \right) \;,
\end{equation}
with
\begin{equation}
 \tan \beta \equiv \frac{v_2}{v_1} \;.
\end{equation}
In the Higgs basis, the doublets, with the Goldstone bosons $G^\pm$ and $G^0$ in ${\cal H}_1$, can be written as
\begin{equation}
 {\cal H}_1 = \left( \begin{array}{c} G^\pm \\ \frac{1}{\sqrt{2}} (v + H^0
     + i G^0) \end{array} \right) \quad \mbox{and} \qquad
 {\cal H}_2 = \left( \begin{array}{c} H^\pm \\ \frac{1}{\sqrt{2}} (R_2
     + i I_2) \end{array} \right) \;.
\end{equation}
The mass matrix for the neutral Higgs states is defined for $\rho_1$, $\rho_2$ and $\rho_3=I_2$ as
\begin{equation}
 \;({\cal M}^2)_{ij} = \left\langle \frac{\partial^2 V}{\partial \rho_i
  \partial \rho_j} \right\rangle \;.
\label{eq:c2hdmmassmat}
\end{equation}
In can be diagonalized by a orthogonal matrix $R$ as~\cite{ElKaffas:2007rq}
\begin{equation}
R {\cal M}^2 R^T = \mbox{diag} (m_{1}^2, m_{2}^2, m_{3}^2) \;,
\end{equation}
with
\begin{equation}
R =
\left(
\begin{array}{ccc}
c_1 c_2 & s_1 c_2 & s_2\\
-(c_1 s_2 s_3 + s_1 c_3) & c_1 c_3 - s_1 s_2 s_3  & c_2 s_3\\
- c_1 s_2 c_3 + s_1 s_3 & -(c_1 s_3 + s_1 s_2 c_3) & c_2 c_3
\end{array}
\right)\, ,
\label{matrixR}
\end{equation}
where we used the notation $s_i \equiv \sin{\alpha_i}$,
$c_i \equiv \cos{\alpha_i}$ ($i = 1, 2, 3$), and with the angles chosen in the ranges
\begin{equation}
- \pi/2 < \alpha_1 \leq \pi/2,
\hspace{5ex}
- \pi/2 < \alpha_2 \leq \pi/2,
\hspace{5ex}
- \pi/2 < \alpha_3 \leq \pi/2.
\label{range_alpha}
\end{equation}
The three neutral mass eigenstates $h_{1,2,3}$, with the masses ordered such that $m_{h_1} \le m_{h_2} \le m_{h_3}$, do not carry a definite CP quantum number. We furthermore have two charged Higgs bosons $H^\pm$ in the spectrum. The Higgs potential can be parametrized by nine independent input parameters, which we choose as 
\begin{eqnarray}
v, \, \tan \beta,\, 
\alpha_1,\, \alpha_2,\, \alpha_3,\, m_{h_1},\, m_{h_2}, \, m_{H^\pm},\, \mbox{and } \textrm{Re}(m_{12}^2).
\end{eqnarray}
The mass of the heaviest neutral scalar, $m_{h_3}$, is a derived quantity and given by
\begin{equation}
m_{h_3}^2 = \frac{m_{h_1}^2\, R_{13} (R_{12} \tan{\beta} - R_{11})
+ m_{h_2}^2\ R_{23} (R_{22} \tan{\beta} - R_{21})}{R_{33} (R_{31} - R_{32} \tan{\beta}) }\;.
\label{m3_derived}
\end{equation}
The requirement that $m_{3}^2$ has to be a positive quantity is implemented as a constraint in our scanning procedure for the input parameters~\cite{deSouza:2025uxb,deSouza:2025bpl,Boto:2025ovp}. \s

By extending the imposed $\mathbb{Z}_2$ symmetry to the Yukawa sector such that each of the three families of fermions couples to one and only one scalar field, tree-level flavour-changing neutral couplings are forbidden~\cite{Glashow:1976nt,Paschos:1976ay}. 
For  the doublet $\Phi_i\,(i=1,2)$ that, respectively, couples to up-type, down-type and charged lepton doublets $\Phi_u$, $\Phi_d$, and $\Phi_l$, respectively, there are four possible Yukawa types of the softly-broken $\mathbb{Z}_2$ symmetric 2HDM, which are
\begin{itemize}
\item
Type-I:
$\Phi_u=\Phi_d=\Phi_\ell \equiv \Phi_2$
\item
Type-II:
$\Phi_u \equiv \Phi_2 \neq \Phi_d=\Phi_\ell \equiv \Phi_1$
\item
Lepton-Specific (LS)
$\Phi_u=\Phi_d \equiv \Phi_2 \neq \Phi_\ell \equiv \Phi_1$
\item
Flipped
$\Phi_u=\Phi_\ell \equiv \Phi_2 \neq \Phi_d \equiv \Phi_1$.
\end{itemize}
The Yukawa Lagrangian for the neutral scalars can be written as
\begin{equation}
{\cal L}_Y = - \sum_{i=1}^3 \frac{m_f}{v} \bar{\psi}_f \left[ c^e(h_i
  ff) + i c^o(h_i ff) \gamma_5 \right] \psi_f h_i \;, \label{eq:yuklag}
\end{equation}
where the $\psi_f$ denote the fermion fields with mass $m_f$. The
coefficients $c^e(h_i ff)$ of the CP-even part and $c^o(h_i ff)$ of the CP-odd part of the Yukawa
coupling, $c^e$ and $c^o (h_i ff)$, are given in Tab.~\ref{tab:yukcoup}. The label $f$ denotes the fermion type, with $f=t$ ($f=b$) standing for up-type (down-type) quarks, and $f=l$ for leptons. 
\begin{table}
\begin{center}
\begin{tabular}{rccc} \toprule
& up-type & down-type & leptons \\ \midrule
Type-I & $\frac{R_{i2}}{s_\beta} - i \frac{R_{i3}}{t_\beta} \gamma_5$
& $\frac{R_{i2}}{s_\beta} + i \frac{R_{i3}}{t_\beta} \gamma_5$ &
$\frac{R_{i2}}{s_\beta} + i \frac{R_{i3}}{t_\beta} \gamma_5$ \\
Type-II & $\frac{R_{i2}}{s_\beta} - i \frac{R_{i3}}{t_\beta} \gamma_5$
& $\frac{R_{i1}}{c_\beta} - i t_\beta R_{i3} \gamma_5$ &
$\frac{R_{i1}}{c_\beta} - i t_\beta R_{i3} \gamma_5$ \\
Lepton-Specific & $\frac{R_{i2}}{s_\beta} - i \frac{R_{i3}}{t_\beta} \gamma_5$
& $\frac{R_{i2}}{s_\beta} + i \frac{R_{i3}}{t_\beta} \gamma_5$ &
$\frac{R_{i1}}{c_\beta} - i t_\beta R_{i3} \gamma_5$ \\
Flipped & $\frac{R_{i2}}{s_\beta} - i \frac{R_{i3}}{t_\beta} \gamma_5$
& $\frac{R_{i1}}{c_\beta} - i t_\beta R_{i3} \gamma_5$ &
$\frac{R_{i2}}{s_\beta} + i \frac{R_{i3}}{t_\beta} \gamma_5$ \\ \bottomrule
\end{tabular}
\caption{Yukawa coupling coefficients of the Higgs
  bosons $h_i$ in the C2HDM. The expressions correspond to
  $[c^e(h_i ff) +i c^o (h_i ff) \gamma_5]$ from
  Eq.~\eqref{eq:yuklag}. \label{tab:yukcoup}}
\end{center}
\end{table}
In this work, we are only going to consider the Type-I and Type-II C2HDM. 
\section{Transport Equations \label{sec:transequ}}
In the following, we will derive the transport equations required for the computation of the BAU. We will use the WKB ansatz developed in Refs.~\cite{Cline:1997vk,Cline_2000,Kainulainen_2001,Kainulainen_2002,Fromme_2007,PhysRevD.101.063525}. More specifically, we will adopt the most recent developments given in Ref.~\cite{Kainulainen:2024qpm}, which include a larger moment expansion of the transport equations compared to previous treatments. \s

We want to describe how the distribution functions of the particles and anti-particles in the plasma  evolve as they come into contact with a moving bubble wall. To do this, we will consider the Boltzmann equation depending on the distance $z$ to the bubble wall for a distribution function $f_h^\pm$, 
\begin{equation}
    \mathcal{L}[f_h^\pm] =\mathcal{C}_h^\pm\left[f_h^\pm\right]\;, \label{eq: boltzmann}
\end{equation}
where $h$ represents the helicity of the particle, the "$\pm$" represents a particle or an anti-particle, and the Liouville operator $\mathcal{L}[f_h^\pm]$ is given by
\begin{equation}
\mathcal{L}[f_h^\pm] \equiv\left(v_h^\pm\partial_z+F_h^\pm\partial_{p_z}\right)f_h^\pm \;.
\end{equation}
We will define the collision operator $\mathcal{C}_h^\pm$ in Sec.~\ref{sec: collision_operator}. For a fermion that moves in a CP-violating VEV background with a $z$-dependent mass 
\begin{align}
\hat{m}(z)=m(z)\text{exp}(i\theta(z)) \;,
\end{align}
where $\theta(z)$ is a CP-violating phase.
The $z$-dependent fermion mass $m(z)$ is obtained from the $z$-dependent VEVs generating the respective mass value, times the corresponding Yukawa coupling, where the Yukawa coupling is chosen such that the phenomenological values of the fermion masses are reproduced at $T=0$. For the top mass we choose $m_t = 172.5$~GeV.
The group velocity $v_h^\pm$ and the semiclassical force $F_h^\pm$ are determined to be~\cite{PhysRevD.101.063525}
\begin{align}
    v_h^\pm=\frac{p_z}{E_h^\pm}&\approx\frac{p_z}{E}\pm s_h\frac{|m|^2\theta'}{2EE_z}\;,\\
    F_h^\pm=-\frac{(|m|^2)'}{2E_h^\pm}\pm s_h\frac{(|m|^2\theta')'}{2EE_z}&\approx-\frac{(|m|^2)'}{2E}\pm s_h\left(\frac{(|m|^2\theta')'}{2EE_z}-\frac{|m|^2(|m|^2)'\theta'}{4E^3E_z}\right)\;,\\
    \text{with}\quad E_h^\pm&\approx E\mp s_h\frac{|m|^2\theta'}{2EE_z}\equiv E\pm\Delta E_h\;,
\end{align}
up to second order in the $z$-derivatives denoted by the primes. With $E=\sqrt{\mathbf{p}^2+m^2}$ we have  $E_z=\sqrt{p_z^2+m^2}$, where $\mathbf{p}$ is the physical momentum, and we get the spin factor $s_h$, 
\begin{equation}
    s_h=h\gamma_\parallel\frac{p_z}{|\mathbf{p}|}\;,
\end{equation}
which contains the helicity $h=\pm1$ and the boost factor $\gamma_\parallel=E/E_z$. In order to solve Eq.~(\ref{eq: boltzmann}) we take the following ansatz for our distribution function in the wall reference frame,
\begin{equation}
    f_h^\pm=f_\text{FD/BE}[\gamma_w(E_h^\pm+v_wp_z)-\mu^\pm]+\delta f_h^\pm\;,\label{eq: BEQ_ansatz}
\end{equation}
where we have the Lorentz factor $\gamma_w=1/\sqrt{1-v_w^2}$ of the wall velocity $v_w$, the chemical potential $\mu^\pm$ and the out-of-kinetic-equilibrium perturbations $\delta f_h^\pm$ on which we impose the condition
\begin{equation}
    \int\text{d}^3p \, \delta f_h^\pm=0\;.
\end{equation}
This condition leads to $\mu^\pm(z)$ carrying all the information about deviations of the number density from equilibrium. The Fermi-Dirac $(+)$ and Bose-Einstein $(-)$ distribution functions are given by 
\begin{equation}
    f_\text{FD/BE}(x)=\frac{1}{e^{\beta x}\pm1}\;,\label{eq: fermi/bose}
\end{equation}
where $\beta=1/T$ is the inverse of the temperature $T$. We want to treat $\mu^\pm$ and $\Delta E_h$ as small deviations around the equilibrium distribution $f_{0w}$ given by
\begin{equation}
    f_{0w}=f_\text{FD/BE}[\gamma_w(E+v_wp_z)]\;.\label{eq: equilibrium_distribution}
\end{equation}
Expanding Eq.~(\ref{eq: BEQ_ansatz}) in $\mu$ and $\Delta E_h$ up to second order in the $z$-derivatives gives
\begin{equation}
    f_h^\pm\approx f_{0w}+f_{0w}'(\gamma_w\Delta E_h-\mu^\pm)+\frac{1}{2}f_{0w}''(\gamma_w\Delta E_h-\mu^\pm)^2+\delta f_h^\pm\;,
\end{equation}
where the derivative of the distribution function $f_{0w}$ is with respect to its entire argument given in Eq.~(\ref{eq: equilibrium_distribution}). At this stage, it is convenient to split the perturbations into CP-even ($e$) and CP-odd ($o$) parts such that
\begin{align}
    \mu^\pm=\mu_e\pm\mu_o\quad,\quad\delta f_h^\pm=\delta f_e\pm\delta f_o\;.
\end{align}
With this, we can represent the distribution $f_h^\pm$ via a CP-even and a CP-odd contribution, 
\begin{equation}
    f_h^\pm=f_e\pm f_o\;,
\end{equation}
which are given by
\begin{align}
    f_e&=f_{0w}-f_{0w}'\mu_e+\frac{1}{2}f_{0w}''\gamma_w^2\Delta E_h^2+\delta f_e\;,\label{eq: distribution_even}\\
    f_o&=f_{0w}'(\gamma_w\Delta E_h-\mu_o)-f_{0w}''\gamma_w\Delta E_h\mu_e+\delta f_o\;.\label{eq: distribution_odd}
\end{align}
To compute the BAU, we care about the CP-odd contributions to the Boltzmann equation which we can obtain via the linear combination
\begin{equation}
    \mathcal{L}[f_h^\pm]\Big|_\text{CP-odd}=\frac{1}{2}\left(\mathcal{L}[f_h^+]-\mathcal{L}[f_h^-]\right)=\frac{1}{2}\left(\mathcal{C}[f_h^+]-\mathcal{C}[f_h^-]\right)=\mathcal{C}[f_h^\pm]\Big|_\text{CP-odd}\;.
\end{equation}
This will result in several terms that are proportional to either $\mu_e$ or $\delta f_e$, which are sub-leading contributions as their prefactors are higher order in $\partial_z$, and can be neglected. This leaves us with the following equation
\begin{equation}
    -\frac{p_z}{E}f_{0w}'\partial_z\mu_o+v_w\gamma_w\frac{(|m|^2)'}{2E}f_{0w}''\mu_o+\frac{p_z}{E}\partial_z\delta f_o-\frac{(|m|^2)'}{2E}\partial_{p_z}\delta f_o=S_{oh}+\mathcal{C}[f_h^\pm]\Big|_\text{CP-odd}\;,\label{eq: cp_odd_equation}
\end{equation}
with the CP-violating source term $S_{oh}$ being defined as
\begin{equation}
    S_{oh}=-v_w\gamma_ws_h\left[\frac{(|m|^2\theta')'}{2EE_z}f_{0w}'-\frac{(|m|^2)'|m|^2\theta'}{4E^3E_z}(f_{0w}'-\gamma_wEf_{0w}'')\right]\;.
\end{equation}
Equation~(\ref{eq: cp_odd_equation}) determines the evolution of a single particle species. Although the derivation assumes a CP-violating fermion mass, the equation can also be used for scalar bosons by simply setting $s_h=0$, which removes the source term. We want to consider a set of such equations for different particle species that are relevant for the BAU. These equations become coupled through the collision operator giving us a network of transport equations.

\section{Moment Expansion \label{sec:momexp}}
Finding a unique solution to the transport network constructed by Eq.~(\ref{eq: cp_odd_equation}) is impossible at this point, since we will always have more unknowns ($\mu$, $\delta f$) than equations. To avoid this problem, we will perform a moment expansion of this equation by integrating over the momenta with a weight $(p_z/E)^\ell$ and an average factor $N_1$, for different moments $\ell$. 
Finding the solution of the transport network constructed by Eq.~(\ref{eq: cp_odd_equation}) is an extremely difficult task as the chemical potential $\mu$ and the perturbation $\delta f$ depend both on the 3-momentum and on all spatial components. For this reason, we use an alternative approach dubbed moment expansion which consists of integrating over the momenta with a weight $(p_z/E)^\ell$, where $\ell=0,1,\cdots,n-1$ and $n\ge2$ is the highest moment equation considered. We furthermore normalize the moments with $N_1$, in order to simplify the expressions. More explicitly, we define the average of the quantity $X$ via
\begin{equation}
    \langle X\rangle\equiv\frac{1}{N_1}\int\text{d}^3pX\quad\text{with}\quad N_1=\int\text{d}^3p \, f_{0w,\text{FD}}'\big|_{m=0}=-\gamma_w\frac{2\pi^3}{3}T^2\;.\label{eq: average}
\end{equation}
With this, we can write down the equation for the $\ell$-th moment, given in Eq.~(\ref{eq: cp_odd_equation}), as~\cite{Kainulainen:2024qpm}
\begin{align}
    &-\left\langle\frac{p_z^{\ell+1}}{E^{\ell+1}}f_{0w}'\right\rangle\partial_z\mu_o+v_w\gamma_w(|m|^2)'\left\langle\frac{p_z^\ell}{2E^{\ell+1}}f_{0w}''\right\rangle\mu_o+\partial_z\left\langle\frac{p_z^{\ell+1}}{E^{\ell+1}}\delta f_o\right\rangle+\ell(|m|^2)'\left\langle\frac{p_z^{\ell-1}}{2E^{\ell+1}}\delta f_o\right\rangle\nonumber\\
    &=\left\langle\frac{p_z^\ell}{E^\ell}\left(S_{oh}+\mathcal{C}[f_h^\pm]\big|_\text{CP-odd}\right)\right\rangle\;.\label{eq: lth-moment}
\end{align}
At this stage we introduce the $\ell$-th velocity moment $u_{o,\ell}$ and the dimensionless mass $x$ via
\begin{equation}
    u_{o,\ell}\equiv\left\langle\frac{p_z^\ell}{E^\ell}\delta f_o\right\rangle \quad\text{and}\quad x=\frac{m}{T}\;,
\end{equation}
as well as the dimensionless universal kernels 
\begin{align}
    D_\ell\equiv\left\langle\left(\frac{p_z}{E}\right)^\ell f_{0w}'\right\rangle\quad,\quad Q_\ell&\equiv T^2\left\langle\frac{p_z^{\ell-1}}{2E^{\ell}}f_{0w}''\right\rangle\quad,\quad Q_{h,\ell}^{8o}\equiv T^2\left\langle\frac{s_hp_z^{\ell-1}}{2E^{\ell}E_z}f_{0w}'\right\rangle\nonumber\\
    \text{and}\quad Q_{h,\ell}^{9o}&\equiv T^4\left\langle\frac{s_hp_z^{\ell-1}}{2E^{\ell+2}E_z}(f_{0w}'-\gamma_wEf_{0w}'')\right\rangle\;.\label{eq: universal_kernels}
\end{align}
With these definitions, Eq.~(\ref{eq: lth-moment}) becomes 
\begin{equation}
    -D_{\ell+1}\mu_o'+u_{o,\ell+1}'+v_w\gamma_w(|x|^2)'Q_{\ell+1}\mu_o+\ell(|x|^2)'\bar{R}u_{o,\ell}=\hat{S}_{oh,\ell}+\hat{\mathcal{C}}_{\ell}\;,
\end{equation}
where the $\ell$-th dimensionless source term $\hat{S}_{oh,\ell}$ and collision term $\hat{\mathcal{C}}_\ell$ are given by
\begin{equation}
    \hat{S}_{oh,\ell}=-v_w\gamma_w\left[(|x|^2\theta')'Q_{h,\ell}^{8o}-(|x|^2)'|x|^2\theta'Q_{h,\ell}^{9o}\right]\quad\text{and}\quad
    \hat{C}_\ell=\left\langle\left(\frac{p_z}{E}\right)^\ell\mathcal{C}\right\rangle\;.\label{eq: dimensionless_source}
\end{equation}
Furthermore, we adopted the factorization rule from Ref.~\cite{Kainulainen:2024qpm}, which is defined as
\begin{equation}
    T^2\left\langle\frac{p_z^{\ell-1}}{2E^{\ell+1}}\delta f_o\right\rangle\rightarrow\left[\frac{T^2}{2p_zE}\right]u_{o,\ell}\equiv\bar{R}u_{o,\ell}\quad\text{with}\quad[X]\equiv\frac{1}{N_0}\int\text{d}^3pXf_{0w}\;,\label{eq: factorization_rule}
\end{equation}
where
\begin{equation}
    N_0=\int\text{d}^3pf_{0w}=\gamma_w\int\text{d}^3pf_0\equiv\gamma_w\hat{N}_0\quad\text{and}\quad\bar{R}=\frac{\pi T^2}{\gamma_w^2\hat{N}_0}\int\text{d}E\;\text{ln}\left|\frac{p-v_wE}{p+v_wE}\right|f_0\;.
\end{equation}
The different moment equations are coupled to each other giving us a total amount of $n=\ell+1$ equations per particle species. In Ref.~\cite{Kainulainen:2024qpm} it was shown that $n$ has to increase in steps of four to observe a converging pattern. Therefore, we adopt in this paper the sequence $n=2+4k$ $(k\in\mathbb
N)$ to investigate the moment dependence. While solving this transport network we have to evaluate the universal functions given in Eq.~(\ref{eq: universal_kernels}) many times. However, since these are model independent, we can construct a 2-dimensional grid in $x$ and $v_w$ and store them in separate files. This allows us to load the necessary functions and access their values via a cubic spline during the computation of the BAU.

\subsection{Truncation Schemes}\label{sec: truncation}
At this stage, the equation with the highest moment still has a velocity moment derivative $u_{o,n}'$ that needs to be determined as we do not have enough equations. For this reason, a truncation scheme has to be introduced that fixes $u_{o,n}'$. A common scheme choice is to use the factorization defined in Eq.~(\ref{eq: factorization_rule}) which leads to \cite{Fromme_2007,PhysRevD.101.063525}
\begin{equation}
    u_{o,n}'=Ru_{o,n-1}'\;,\label{eq: simple_truncation}
\end{equation}
with $R=-v_w$. However, we have to remember that this choice for the truncation scheme is arbitrary. One could also set $R=0$, setting the last moment to zero, or set $R=-1$, i.e. opposite to the previous moment. We will refer to the truncation choice, where the last moment is related to the previous one only via a constant as the \textit{constant truncation scheme}. In case this truncation scheme is chosen, the user can set the value for the constant in \texttt{BSMPT}v4.\s

A more sophisticated truncation scheme was introduced in Ref.~\cite{Kainulainen:2024qpm}, which relates the highest moment velocity to all the previous ones by setting the $n$'th variance to zero such that
\begin{equation}
    \left\langle\left(\frac{p_z}{E}-u_{o,1}\right)^n\delta f_o\right\rangle=0\;.
\end{equation}
This leads to a generalization of Eq.~(\ref{eq: simple_truncation}) via
\begin{equation}
    u_{o,n}'=\sum_{i=1}^{n-1}R_iu_{o,i}'\;,\label{eq: variance_truncation}
\end{equation}
with the coefficients $R_i$ given by
\begin{align}
    R_1&=-n^2(-u_{o,1})^{n-1}+\sum_{k=2}^{n-1}
    \begin{pmatrix}
        n\\k
    \end{pmatrix}
    (n-k)u_{o,k}(-u_{o,1})^{n-k-1}\;,\\
    R_i&=-
    \begin{pmatrix}
        n\\i
    \end{pmatrix}
    (-u_{o,1})^{n-i}\quad,\;i\in[2,n-1]\;.
\end{align}
In the following, we will refer to this truncation scheme choice as \textit{variance} scheme. \s

Putting everything together, we can write down the coupled system of moment equations in a clearer form by introducing the vector $w_o=(\mu_o,u_{o,1},\dots,u_{o,n-1})$. Using this vector, the equation network for a single particle species with the general truncation rule from Eq.~(\ref{eq: variance_truncation}) is given by
\begin{equation}
    \hat{\mathcal{A}}w_o'+\hat{\mathcal{B}}w_o=\hat{\mathcal{S}_o}+\hat{\mathcal{C}}\;,\label{eq: single_particle_network_1}
\end{equation}
where $\hat{\mathcal{A}}$ and $\hat{\mathcal{B}}$ are $n\times n$-dimensional matrices defined via
\begin{equation}
    \hat{\mathcal{A}}=
    \begin{pmatrix}
        -D_1 & 1 & \cdots & 0 & 0\\
        -D_2 & 0 & \cdots & 0 & 0\\
        \vdots & \vdots & \ddots & \vdots & \vdots\\
        -D_{n-1} & 0 & \cdots & 0 & 1\\
        -D_n & R_1 & \cdots & R_{n-2} &R_{n-1}
    \end{pmatrix}
    \quad,\quad
    \hat{\mathcal{B}}=
    (|x|^2)'\begin{pmatrix}
        v_w\gamma_wQ_1 & 0 & 0 & \cdots  & 0\\
        v_w\gamma_wQ_2 & \bar{R} & 0 & \cdots & 0\\
        v_w\gamma_wQ_3 & 0 & 2\bar{R} & \cdots & 0\\
        \vdots & \vdots & \vdots & \ddots & \vdots\\
        v_w\gamma_wQ_n & 0 & 0 & \cdots & (n-1)\bar{R}
    \end{pmatrix}\;,\label{eq: A_B_matrices}
\end{equation}
and $\hat{\mathcal{S}}_o$ is the source vector with its components given by Eq.~(\ref{eq: dimensionless_source}). We will define the collision vector $\hat{\mathcal{C}}$ in the next section. By inverting $\hat{\mathcal{A}}$ we can cast Eq.~(\ref{eq: single_particle_network_1}) into a more numerically suitable form as follows, 
\begin{equation}
    w_o'=\hat{\mathcal{A}}^{-1}\left[\hat{\mathcal{S}}_o+\hat{\mathcal{C}}-\hat{\mathcal{B}}w_o\right]\;,
\end{equation}
with 
\begin{align}
    \hat{\mathcal{A}}^{-1}=\frac{1}{\mathcal{D}_n}
    \begin{pmatrix}
        R_1 & R_2 & \cdots & R_{n-1}  & -1\\
        R_1D_1 & R_2D_1 & \cdots & R_{n-1}D_1 & -D_1\\
        \vdots & \vdots & \ddots & \vdots & \vdots\\
        R_1D_{n-1} & R_2D_{n-1} & \cdots & R_{n-1}D_{n-1} & -D_{n-1}
    \end{pmatrix}
    +
    \begin{pmatrix}
        0 & 0 & \cdots & 0\\
        1 & 0 & \cdots & 0\\
        \vdots & \ddots & \ddots & \vdots\\
        0 & \cdots & 1 & 0
    \end{pmatrix}\;,\label{eq: ainv}
\end{align}
and
\begin{equation}
    \quad\mathcal{D}_n\equiv(-1)^n\text{det}(\hat{\mathcal{A}})=D_n-\sum_{i=1}^{n-1}R_iD_i\;.
\end{equation}
To include multiple particles species we can combine the individual particle vectors into a larger $(N\times n)$-dimensional vector $\mathcal{W}^T=((w_o^{1})^T,(w_o^{2})^T,\dots,(w_o^{N})^T)$, where $N$ is the number of involved particles. This gives us the full equation network
\begin{equation}
    \hat{\mathcal{A}}\mathcal{W}'+\hat{\mathcal{B}}\mathcal{W}=\hat{\mathcal{S}}+\hat{\mathcal{C}}[\mathcal{W}]\;,\label{eq: full_network}
\end{equation}
where $\hat{\mathcal{A}}=\text{diag}(\hat{\mathcal{A}}_1,\dots,\hat{\mathcal{A}}_N)$ and $\hat{\mathcal{B}}=\text{diag}(\hat{\mathcal{B}}_1,\dots,\hat{\mathcal{B}}_N)$ are block diagonal matrices whose sub-matrices are defined via Eq.~(\ref{eq: A_B_matrices}). Further, $\hat{\mathcal{S}}$ and $\hat{\mathcal{C}}[\mathcal{W}]$ are correspondingly defined $(N \times n)$-dimensional vectors. Due to the block diagonal structure of Eq.~(\ref{eq: full_network}), it is easy to invert, leading to
\begin{equation}
    \mathcal{W}'=\hat{\mathcal{A}}^{-1}\left[\hat{\mathcal{S}}+\hat{\mathcal{C}}[\mathcal{W}]-\hat{\mathcal{B}}\mathcal{W}\right]\;,\label{eq: full_network_inv}
\end{equation}
with $\hat{\mathcal{A}}^{-1}=\text{diag}(\hat{\mathcal{A}}_1^{-1},\dots,\hat{\mathcal{A}}_N^{-1})$ and $\hat{\mathcal{A}}^{-1}_i$ defined in Eq.~(\ref{eq: ainv}).\par

\subsection{Collision Operator}\label{sec: collision_operator}
In the collision operator all relevant rates need to be included that drive the system back into its equilibrium state, such as decays, annihilation and scattering processes. These rates depend on the model under consideration and their calculation can become rather involved. In any case, the collision operators has to be averaged for different moments. We denote the rates contributing to the evolution of a single particle species by $\bar{C}^i$. With this the $\ell$'th collision moment can be approximated as~\cite{Kainulainen:2024qpm}
\begin{equation}
    \hat{C}_\ell^i=-\kappa^i K_\ell\bar{C}^i-\kappa^i\hat{\Gamma}_\text{tot}^iu_{o,\ell}\;,
\end{equation}
where the universal function $K_\ell$ is given by
\begin{equation}
    K_\ell=\left\langle\left(\frac{p_z}{E}\right)^\ell f_{0w}\right\rangle\;,
\end{equation}
and $\kappa$ is a normalization constant based on the particle species, given by $\kappa_\text{FD}\approx0.912$ and $\kappa_\text{BE}\approx0.685$ for fermions and boson, respectively. Additionally we have the total interaction rate $\hat{\Gamma}_\text{tot}$ between the considered particle species and the plasma. Throughout this paper we will consider the following collision network, which involves the left-/ and right-handed top quark ($t_L/t_R$), the left-handed bottom quark ($b_L$) and the Higgs boson ($h$). Their respective collision operators $\bar{C}^i$ are given by~\cite{Barni:2025ifb}
\begin{align}
    \bar{\mathcal{C}}^{t_L}&=\hat{\Gamma}_y(\mu_{t_L}-\mu_{t_R}+\mu_h)+2\hat{\Gamma}_m(\mu_{t_L}-\mu_{t_R})+\hat{\Gamma}_W(\mu_{t_L}-\mu_{b_L})+\hat{\Gamma}_{ss}[\mu_i]\;,\nonumber\\
    \bar{\mathcal{C}}^{b_L}&=\hat{\Gamma}_y(\mu_{b_L}-\mu_{t_R}+\mu_h)+\hat{\Gamma}_W(\mu_{b_L}-\mu_{t_L})+\hat{\Gamma}_{ss}[\mu_i]\;,\nonumber\\
    \bar{\mathcal{C}}^{t_R}&=-\hat{\Gamma}_y(\mu_{t_L}+\mu_{b_L}-2\mu_{t_R}-2\mu_h)+2\hat{\Gamma}_m(\mu_{t_R}-\mu_{t_L})-\hat{\Gamma}_{ss}[\mu_i]\;,\nonumber\\
    \bar{\mathcal{C}}^h&=\frac{3}{2}\hat{\Gamma}_y(\mu_{t_L}+\mu_{b_L}-2\mu_{t_R}+2\mu_h)+\hat{\Gamma}_h\mu_h\;.\label{eq: collision_network}
\end{align}
For the collision rates we use the values given in Ref.~\cite{Barni:2025ifb}, i.e.
\begin{equation}
    \hat{\Gamma}_m\approx\frac{0.26|m_t(z)|^2}{T}\;,\quad\hat{\Gamma}_h\approx\frac{1.5|m_W(z)|^2}{T}\;,\quad
    \hat{\Gamma}_{ss}\approx2.7\cdot10^{-3}T\;,\quad\hat{\Gamma}_W=\hat{\Gamma}_\text{tot}^h\;,
\end{equation}
with the $W$ boson mas $m_W(z)=gh(z)/2$, where $g=0.65$ denotes the weak interaction coupling of the $SU(2)_L$,  and the total rates for the quarks $q$ and the SM-like Higgs boson $h$  are~\cite{Fromme_2007,PhysRevD.101.063525,Barni:2025ifb}
\begin{equation}
    \hat{\Gamma}^i_\text{tot}\equiv\frac{D_2^i}{D_0^i\hat{D}^i}\quad\text{with}\quad\hat{D}^{q_L}=\frac{7.1}{T}\quad,\quad\hat{D}^{q_R}=\frac{7.6}{T}\quad,\quad\hat{D}^{h}=\frac{14}{T}\;,
\end{equation}
with the $D_{0,2}^i$ defined in Eq.~(\ref{eq: universal_kernels}) for the respective particle species $i$.
For the top Yukawa rate $\hat{\Gamma}_y$ we use an updated value from our computation presented in App.~\ref{app: yuk} which results in
\begin{align}
\hat{\Gamma}_y\approx6\cdot10^{-3}T \;.
\end{align}

\section{Calculation of the BAU}\label{sec:bau}
The transport equations allow us to compute the chemical potential of the relevant particles as a function of the distance $z$ to the bubble wall. In the next step, the generated left-handed chemical potential is converted through the weak sphaleron processes into a baryon asymmetry. Noting that the sphalerons only couple to the left-handed particles, we introduce the baryon chemical potential generated from the left-handed particles,
\begin{equation}
\mu_{B_L} \equiv \frac{1}{2} \sum_q \mu_{q_L} (z) \;.
\end{equation}
For the light quarks, where the strong sphaleron rate is the only relevant one, we can assume that they have the same chemical potential,
\begin{equation}
    \mu_{q_L} = - \mu_{q_R} \equiv \mu_q \;.
\end{equation}
Using this relation, we can eliminate the light quarks and obtain
\begin{equation}
    \mu_{B_L} = \frac{1}{2} (1 + 4 D_0^t) \mu_{t_L} 
    + \frac{1}{2} (1 + 4 D_0^b) \mu_{b_L} + 2 D_0^t \mu_{t_R} \;.
\end{equation}
The baryon asymmetry $n_B$ is then obtained from the diffusion equation~\cite{Cline_2000,Barni:2025ifb}
\begin{equation}
    n_B'(z)=\frac{n_f}{2\gamma_wv_w}\Gamma_{ws}(z)(N_c\mu_{B_L}(z)T^2-An_B(z))\;,\label{eq: baryon_equation}
\end{equation}
where $n_f$ is the number of flavors, $N_c$ the number of colors, $A=15/2$ and $\Gamma_{ws}(z)$ is the weak sphaleron rate given by
\begin{equation}
    \Gamma_{ws}=\Gamma_\text{sph}f_\text{sph}(z)\quad\text{with}\quad f_\text{sph}(z)=\text{min}\left(1,\frac{1.7T}{\Gamma_\text{sph}}e^{-37v(z)/T}\right)\,,\label{eq: weak_sphaleron_rate}
\end{equation}
with $\Gamma_{\text{sph}}= \Gamma_{sph}=8\cdot 10^{-7}T$.
The function $f_\text{sph}(z)$ is defined such that the sphaleron rate is interpolated between the symmetric and the broken phase. Since Eq.~(\ref{eq: baryon_equation}) is a first-order linear equation it can be solved exactly, 
\begin{equation}
    n_B(z)=-\int_z^\infty\text{d}z'\frac{n_fN_c}{2v_w\gamma_w}\Gamma_{ws}(z')\mu_{B_L}(z')T^2\text{exp}\left[-\frac{An_f}{2v_w\gamma_w}\int_z^{z'}\Gamma_{ws}(z'')\text{d}z''\right]\;,
\end{equation}
where we set $n_B(+\infty)=0$. To relate $n_B$ to the BAU given in Eq.~(\ref{eq: BAU}) we need to evaluate $n_B(-\infty)$ and divide it by the entropy density 
\begin{equation}
s=h_\text{eff}\frac{2\pi^2}{45}T^3 \;,
\end{equation}
where $h_\text{eff}$ denotes the number of the effective degrees of freedom for the entropy density, leading to
\begin{equation}
    \eta_B=-\frac{45n_fN_c}{4\pi^2v_w\gamma_wh_\text{eff}T}\int_{-\infty}^\infty\text{d}z \,\Gamma_{ws}(z)\,\mu_{B_L}(z)\,\text{exp}\left[-\frac{An_f}{2v_w\gamma_w}\int_{-\infty}^{z}\Gamma_{ws}(z')\text{d}z'\right]\;.\label{eq: BAU_formula}
\end{equation}
Note, that the integral in the exponent of Eq.~(\ref{eq: BAU_formula}) diverges since the VEV in Eq.~(\ref{eq: weak_sphaleron_rate}) reaches a constant value inside the bubble. This is because in our approach we consider a time-independent solution. If the change in temperature, respectively time, is taken into account, the cooling of the universe would further suppress the sphaleron rate until it nearly vanishes at $T=0$, i.e.~today. Furthermore, the time between the phase transition and today is finite so that the particles only traverse a finite volume and not an infinite one. In practice, we simply truncate the integration boundaries at some large constant relative to the bubble wall thickness. Variations around this large constant value change the final result only negligibly. 

\section{Bubble profile}
\label{sec:bubbleprofile}
In the calculation of the baryon asymmetry the VEV profile across the bubble wall plays an important role. 
In the literature, it is common to consider the so-called kink profile. For general Higgs extended models, which contain $i=1,..., n_{\text{Higgs}}$ Higgs fields that develop an electroweak VEV $\phi_i (z)$, where $z$ is the distance to the bubble wall, the $i$-th component of the profile in the bubble wall reference frame is given by
\begin{align}
\phi_i(z) = \phi_{t,i} + (\phi_{f,i}-\phi_{t,i})\frac{1}{2}\left(1+\tanh\left(\frac{z-\delta_i}{L_{w,i}}\right) \right) \;,
\end{align}
where $\vec\phi_{t}$ and $\vec\phi_{f}$ are the true and false vacuum, respectively, and $\vec{L}_w$ and $\vec{\delta}$ are the vectors that contain the bubble width and displacement, respectively. \s

In the previous \texttt{BSMPT} version v2~\cite{Basler:2020nrq} we assumed $\vec{L}_w$ and $\vec{\delta}$ to be scalar quantities, i.e. the same for all fields.\footnote{This allowed us to set $\vec{\delta}=0$ due to spatial invariance} This simplification implies a major limitation, however, as we get a constant phase $\theta$ for the fermion masses, 
%
%
which would make the source terms vanish and no BAU production would take place.\footnote{This is true if the false vacuum is $\vec\phi_f=\vec{0}$, which is desirable for BAU calculations.} 
%
To circumvent this problem we decided to also model the phase with a kink profile given by
\begin{align}
    \theta(z)=\left(\theta_{\text{brk}}-{\frac{\theta_{\text{brk}}-\theta_{\text{sym}}}{2}}\left(1+\operatorname{tanh}{\frac{z}{L_{w}}}\right)\right)\,,
\end{align}
where the phases in the broken and in the symmetric phase, $\theta_{\text{brk}}$ and $\theta_{\text{sym}}$, respectively, are determined by the path that minimizes the potential on the planes orthogonal to the straight line between the true and the false vacuum. Formally, this path is given by
\begin{align}
\tilde{\vec{\phi}}(t) = \arg\min_{\vec\phi } V(\vec\phi)  \quad \text{such that}\quad (\vec\phi-\vec\gamma(t))\cdot\frac{d\vec\gamma}{dt}=0\,,t\in[0,1] \;,
\label{eq:pathprevious}
\end{align}
where $\gamma(t) = \vec\phi_t + t (\vec\phi_f - \vec\phi_t)$ is the straight line between the true and the false vacuum. This allows us to calculate the symmetric phase $\theta_{\text{sym}}$ by calculating the phase of $\tilde{\vec{\phi}}(t)$ as $t\to1$. More details can be found in~\cite{Basler:2020nrq}. We slightly updated this algorithm such that the phase is still calculated in the same way but the potential barrier is calculated on the straight line that connects the true and false vacuum.  This provided a more stable and faster method to estimate the length scale $L_w$. 
It is obtained from
\begin{align}
L_w = \frac{v_t}{\sqrt{8V_b}} \;,
\label{eq:lwcalc}
\end{align}
where $v_t\equiv \sqrt{\sum_i\omega_{i}(z=-\infty)}$ is the electroweak VEV in the broken phase, and $V_b$ is the potential barrier. The impact on $L_w$ of using  the straight line instead of Eq.~(\ref{eq:pathprevious}) is of at most 8\%, while being much faster and reliable computationally.
\s

In this new \texttt{BSMPT} version we decided to improve the calculation of the bubble profile by solving the equations of motion (EOMs) of the fields. We start with the EOM of a scalar field~\cite{Moore:1995si} in the plasma,   
\begin{align}
    \square \vec\phi+\nabla V(\vec\phi)+\underbrace{\sum_i \nabla_\phi m_i\int\frac{d^{3}p}{(2\pi)^{3}2E_i}\delta f(p,z)}_\text{``friction"}=0 \;,
\end{align}
where $i$ sums over the particles in the plasma. The $\delta f$ are the perturbations of the equilibrium distribution $f$ and act as a ``friction" term that prevents the bubble wall velocity from reaching the speed of light. Solving this equation far beyond the scope of this paper. Instead, we replace the friction term by a velocity dependent friction term $\frac{\mu}{v_w\gamma}\frac{d\vec\phi}{dt}$, where $\mu$ is the friction coefficient that controls the wall velocity~\cite{Liu:1992tn} and $\gamma$ is the Lorentz factor. We expect the bubble to reach a terminal velocity and shape, i.e. we expect the bubble profile to be a solution of our new EOM. It becomes obvious that the field, which in the plasma reference frame depends on $t$ and $z$, must depend on a single spatial component in the wall reference frame.\footnote{We assume a planar bubble profile with a $z$ dependence.} Thus we can assume that the profile only depends on $z \equiv \gamma(\tilde{z}-v_w \tilde{t})$, where the tilde denotes quantities in the plasma reference frame. Substituting $\vec\phi(z)\equiv\vec\phi(\gamma(\tilde{z}-v_w \tilde{t}))$ in the EOM yields
\begin{align}
    \frac{d^2\vec\phi}{dz^2} +  \mu\frac{d\vec\phi}{dz}=\nabla V(\vec\phi)\,.
    \label{eq:vacuumprofile}
\end{align}
It is important to note here that $\mu$ is not a free parameter. In fact, there is a relationship between the profile $\vec\phi(z)$ and $\mu$. To see this we multiply Eq.~{\ref{eq:vacuumprofile}} by $\frac{d\vec\phi}{dz}$ and integrate from $-\infty$ to $\infty$. We obtain 
\begin{align}
    \mu\int_{-\infty}^\infty \left[\frac{d\vec\phi}{dz}\right]^2dz = \Delta V\,,
\end{align}
where $\Delta V = V(\vec\phi(\infty))-V(\vec\phi(-\infty))$ is the difference between the potential values at the false and the true minimum, respectively. This leads to the conclusion that the energy dissipated by the friction term is exactly the energy lost between the two minima. Additionally, for domain walls we have that $\Delta V = 0$ so that there is no need to add a friction term. The field EOM have a soliton solution without it.

\section{Numerical Analysis of the Benchmark Model \label{sec:numbench}}
In this section we will study the influence of the choice of the wall velocity, of the truncation constant $R$, of the expansion parameter $L_w T_n$, i.e.~the wall width times the transition temperature, chosen as the nucleation temperature, of the strength of the phase transition and the stability of the result with respect to the moment expansion. Where possible, we will compare our results with the literature. We perform the analysis within the benchmark model given in Ref.~\cite{Kainulainen:2024qpm} which, to be self-contained, we introduced in Sec.~\ref{sec:benchmark}. \s

\begin{figure}
    \centering
    \includegraphics[width=1\linewidth]{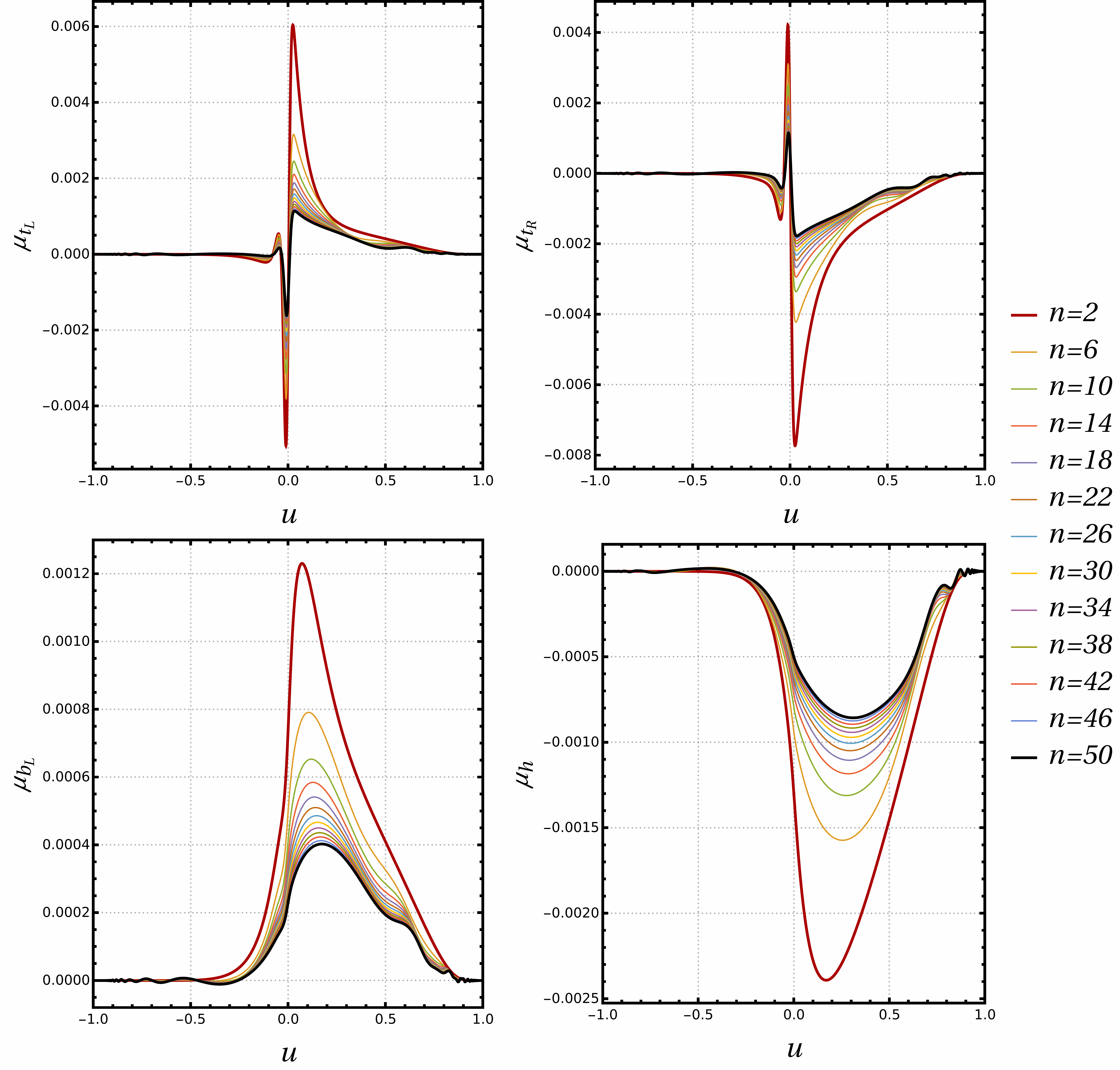}
    \caption{Computed chemical potentials of the particles involved in the fluid network as a function of $u$, defined in Eq.~(\ref{eq: ztou}), for up to $n=50$ moment equations, in the constant truncation scheme with $R=-v_w$.}
    \label{fig: chemical}
\end{figure}

In Fig.~\ref{fig: chemical} we show the chemical potentials of the involved particles, i.e.~the left- and right-handed top-quark $t_{L,R}$, the left-handed bottom quark $b_L$ and the Higgs boson $h$, up to $n=50$ moment equations as a function of
\begin{equation}
    u=\frac{\sqrt{1+z^2}-1}{z}\;,\label{eq: ztou}
\end{equation}
which compactifies the wall distance $z\in(-\infty,\infty)$ to the interval $u\in(-1,1)$. We choose the constant truncation scheme with $R=-v_w$.
We observe the same behavior as in Ref.~\cite{Kainulainen:2024qpm}, where an increase in moments leads to smaller magnitudes of the chemical potentials and to oscillatory modes, which persist far outside and deep inside the bubble wall, i.e.~towards $u=1$ and $-1$, respectively.\s

\begin{figure}
    \centering
    \includegraphics[width=1\linewidth]{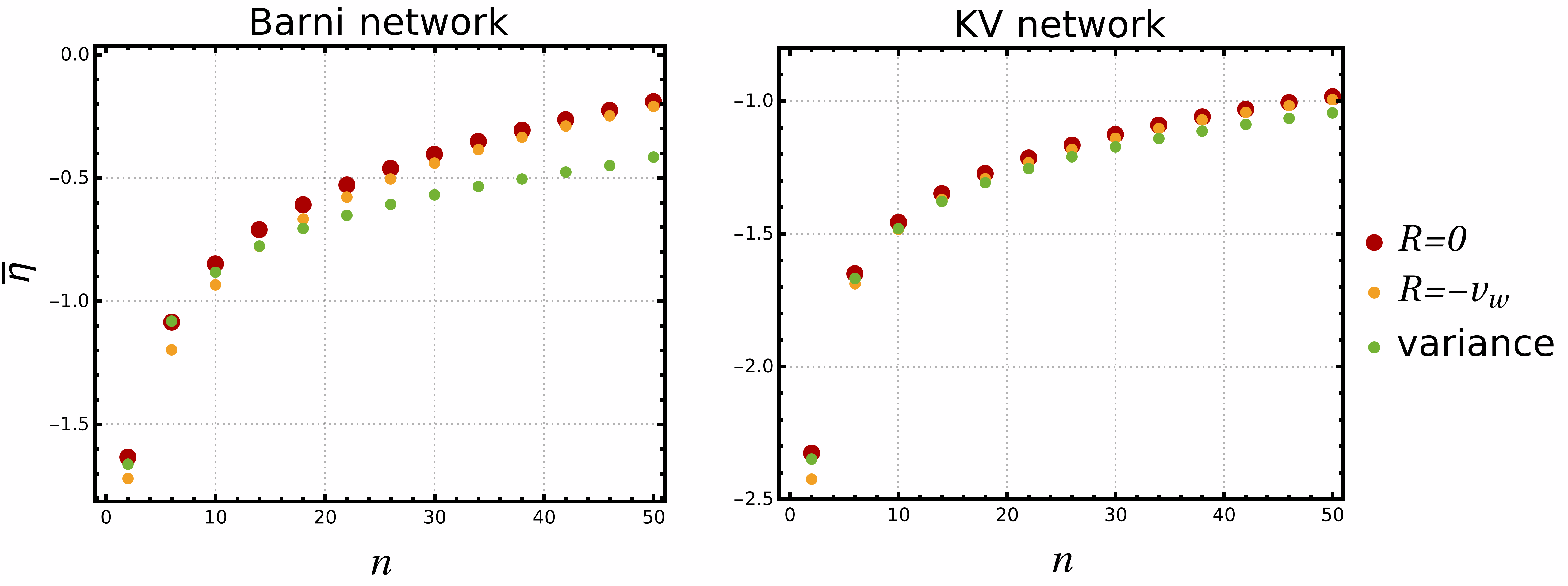}
    \caption{The normalized BAU $\bar{\eta}\equiv\eta/\eta_\text{obs}$ over the number $n$ of moment equations. Left: the results for the collision network defined in Eq.~(\ref{eq: collision_network}) and the rates presented in Sec.~\ref{sec: collision_operator}; right: results for the collision network and rates used in Ref.~\protect\cite{Kainulainen:2024qpm}. The different colors represent different truncation scheme choices, where $R=0,-v_w$ correspond to the constant truncation choice for two different truncation values, and the green points to the variance truncation introduced in Sec.~\ref{sec: truncation}.}
    \label{fig: BAU_moment}
\end{figure}

With these chemical potentials we can compute the BAU via Eq.~(\ref{eq: BAU_formula}). In Fig.~\ref{fig: BAU_moment} we show the BAU normalized to the observed value, i.e. $\overline{\eta}\equiv\eta/\eta_{obs}$, as a function of the number of moments for three different truncation schemes which we introduced in Sec.~\ref{sec: truncation}. On the left, we show the numerical results obtained using the collision network and rates that we defined in Sec.~\ref{sec: collision_operator} and which we call the Barni network, as apart from the top-Yukawa rate, which we calculated ourselves, it is analogous to the implementation in  Ref.~\cite{Barni:2025ifb}.
On the right, we show our results using the collision network and rates from Ref.~\cite{Kainulainen:2024qpm}, which we call the KV network.\footnote{The default scheme in \texttt{BSMPT}v4 is the Barni scheme. It implements the most recent rates and also differs in some prefactors in the collision network.}
We observe that the BAU generated from the Barni network is smaller in magnitude compared to the result obtained with the KV network. This is expected since the collision rates that equilibrate the chemical potentials are larger in the Barni network compared to the ones in~\cite{Kainulainen:2024qpm}. Nonetheless, both networks exhibit the same behavior as we increase $n$. Namely, the absolute value of $\bar{\eta}$ decreases as we increase the number of moments, but does not stabilize to a constant value after taking into account a maximum of $n=50$ moment equations, independent of the truncation scheme choice. We will investigate the scheme dependence in further detail in the next subsection. For the variance truncation, the result we obtained using the KV network agrees very well with the result presented in Ref.~\cite{Kainulainen:2024qpm}. The reason for the minor differences may be attributed to a different equation used to obtain the final BAU (Eq.~(\ref{eq: BAU_formula}) in our case). However, a significant discrepancy is found w.r.t.~Ref.~\cite{Kainulainen:2024qpm} for the constant truncation scheme. We observe a truncation scheme independence, if we apply the KV network, in contrast to Ref.~\cite{Kainulainen:2024qpm}. This may require further investigation in future.

\subsection{Truncation Scheme Dependence}
\begin{figure}[h]
    \centering
    \includegraphics[width=0.9\linewidth]{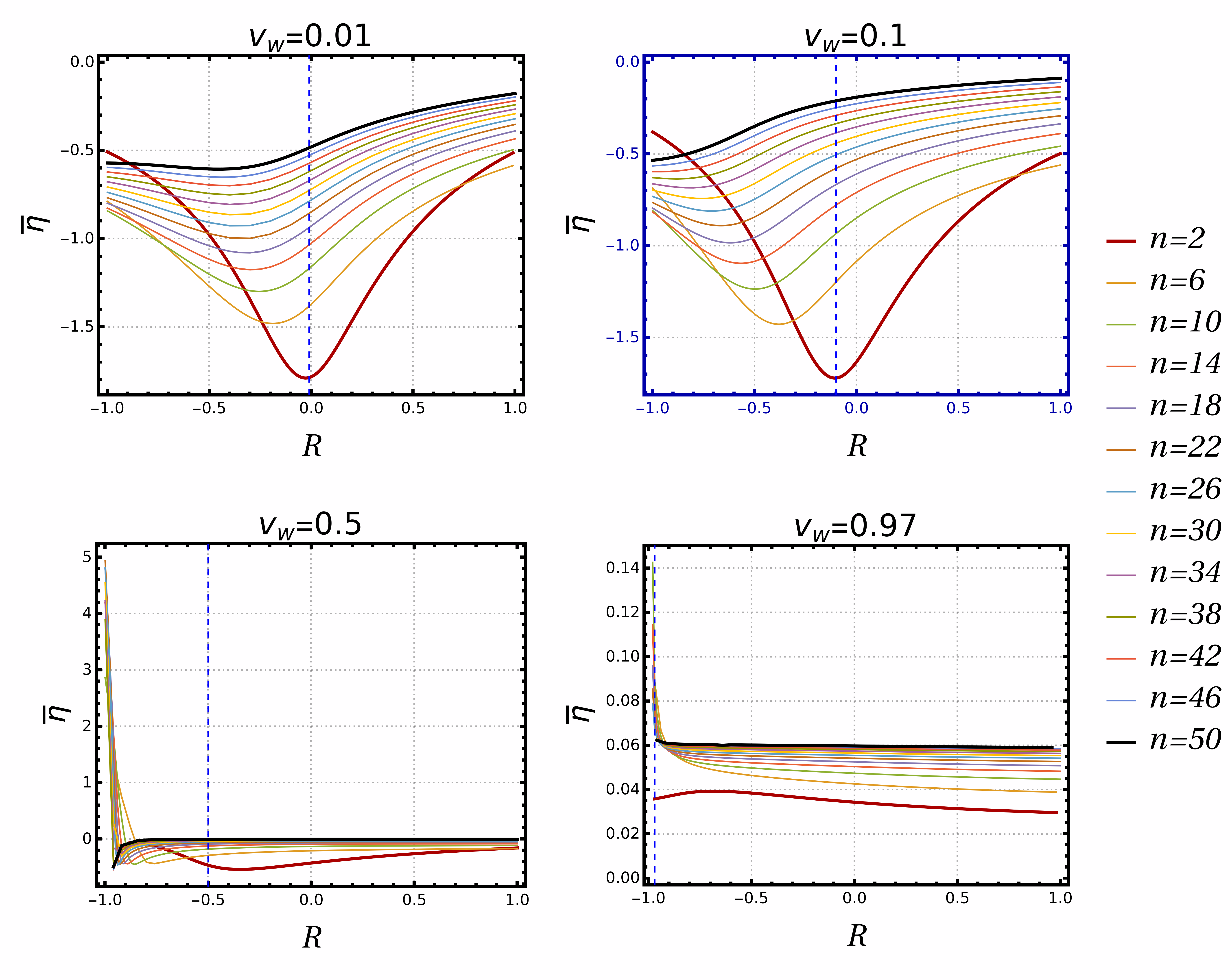}
    \caption{The normalized BAU $\bar{\eta}$ as a function of $R$ in the constant truncation scheme for different wall velocities. The lines represent the largest number $n$ of moment equations taken into account, and the vertical dashed blue line marks the truncation choice $R=-v_w$. The blue box marks the base scenario of the benchmark model.}
    \label{fig: vw_truncation_dependence}
\end{figure}

We saw in Fig.~\ref{fig: BAU_moment} (left), where we apply the Barni scheme, which is the default scheme in \texttt{BSMPT}v4, that depending on the truncation scheme the moment expansion may lead to different results. In this section, we want to analyse if in the constant truncation scheme the obtained results are independent of the choice of $R$. For this, we vary $R$  and investigate how the results change based on the largest number $n$ of moment equations that we consider. In Fig.~\ref{fig: vw_truncation_dependence} we show the dependence of the normalized BAU on the constant $R$ for different wall velocities $v_w$. The different lines show the largest considered number $n$ of moment equations, and the vertical dashed blue line marks the truncation choice $R=-v_w$. \s

We observe that for small wall velocities at low values of $n$ the result strongly depends on the truncation choice $R$. As we increase the number of moments the lines start developing towards constant horizontal lines, i.e.~the computed value of the baryon asymmetry becomes less and less $R$-dependent. Since this calculation is computationally very demanding for large numbers of moment equations, we only go up to $n=50$. To check if the BAU truly converges to a constant $R$-independent value in this framework, larger moments have to be taken into account. However, for large $v_w$, here from $v_w=0.5$ on, we already observe a convergence at $n=50$, making these results more robust than the ones obtained at small $v_w$. We would like to point out that as $R\rightarrow-1$ the numerical results for large $v_w$ can break down. This comes from the fact that the eigenvalues of the exponentially growing eigenmodes are of $\mathcal{O}(100)$ and cannot be stabilized reliably by our algorithm. Although the results suggest that for large $n$ the results become truncation choice independent, we cannot conclude, based on this analysis alone,  that in general it holds true that the BAU will reach a constant value as $n\rightarrow\infty$ for small wall velocities. \s

\begin{figure}[h]
    \centering
    \includegraphics[width=0.9\linewidth]{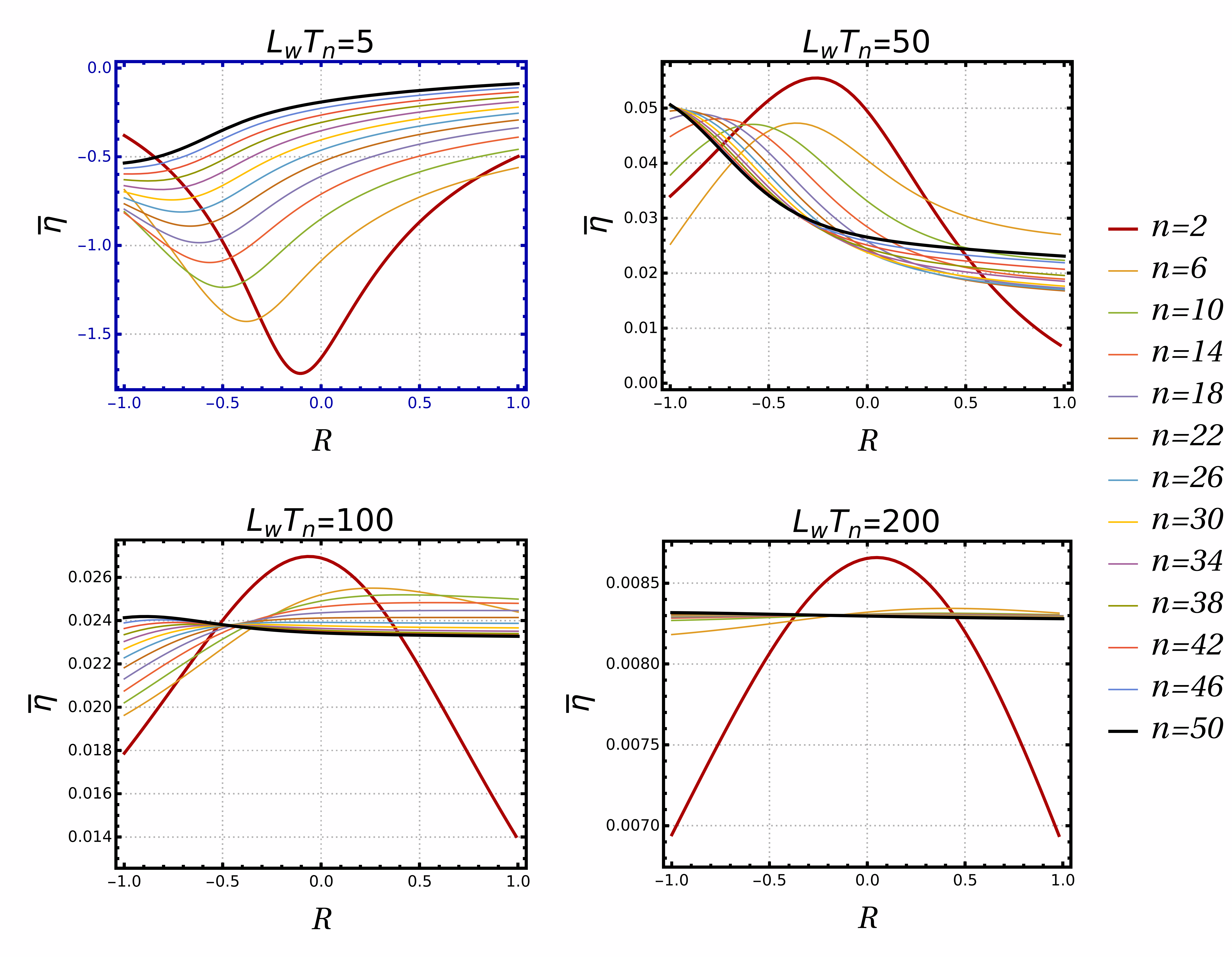}
    \caption{The normalized BAU $\bar{\eta}$ over the $R$-value used in the constant truncation scheme for different values of $L_wT_n$. The lines represent the largest number $n$ of moment equations taken into account. The blue box marks the base scenario of the benchmark model.}
    \label{fig: lw_truncation_dependence}
\end{figure}

We perform the same analysis for different values $L_w T_n$ of the wall thickness $L_w$ times the transition temperature $T_n$, which here is chosen to be the nucleation temperature. The results are displayed in  Fig.~\ref{fig: lw_truncation_dependence}. Also here, the results depend strongly on the truncation choice at low numbers $n$ of the moment equations, while for larger values the curves begin to flatten out and become constant, in particular for large $L_w T_n$. For $L_wT_n=200$, the moment expansion converges to a constant value already at $n=14$. A rapid convergence for large $L_wT_n$ is expected as it is the underlying assumption of the WKB ansatz, i.e. $L_wT_n\gg1$.
Similar to the $v_w$ case, it is not obvious, however, that for small $L_wT_n$, i.e.~here $L_wT_n\lesssim100$, the BAU will reach a constant value as $n\rightarrow\infty$. 
Finally note that the better convergence behavior at large $L_wT_n$ comes at the cost of a decreasing BAU as can be inferred from the scale on the $y$-axis.\s

\begin{figure}[h]
    \centering
    \includegraphics[width=0.9\linewidth]{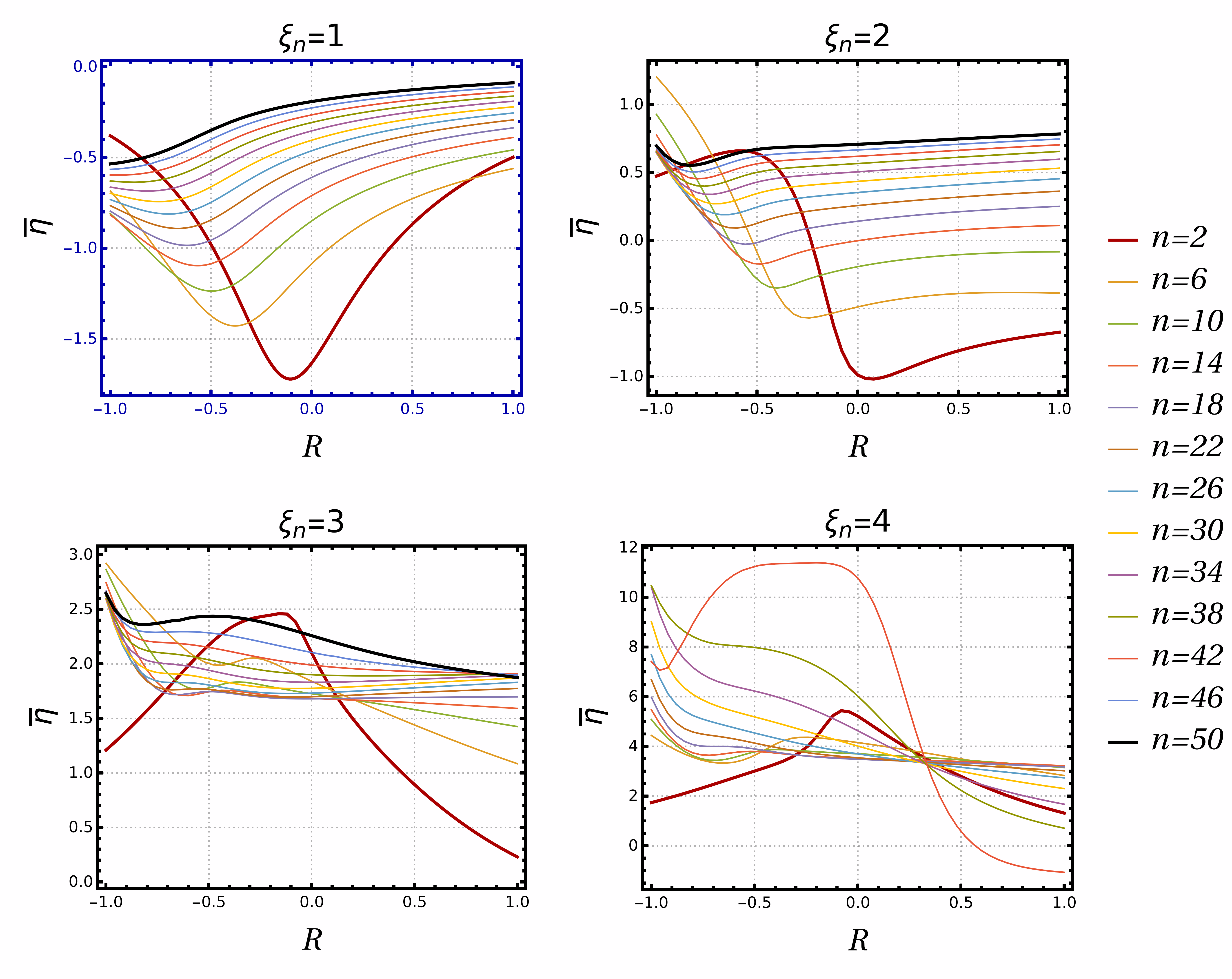}
    \caption{The normalized BAU $\bar{\eta}$ over the $R$-value used in the constant truncation scheme for different values of $\xi_n$. The lines represent the largest number $n$ of moment equations taken into account. The blue box marks the base scenario of the benchmark model.}
    \label{fig: vn_truncation_dependence}
\end{figure}

In Fig.~\ref{fig: vn_truncation_dependence}, we display the corresponding analysis for different phase transition strengths $\xi_n\equiv v_n/T_n$, where $v_n$ denotes the VEV at the nucleation temperature. We see that for the first three values of $\xi_n$, $\xi_n=1,2,3$, the curves begin to flatten out with increasing $n$. However, for $\xi_n=4$ we observe a very erratic behavior and it is not clear that the result will eventually become independent of $R$.

\subsection{Moment Convergence}
\begin{figure}[h]
    \centering
    \includegraphics[width=0.9\linewidth]{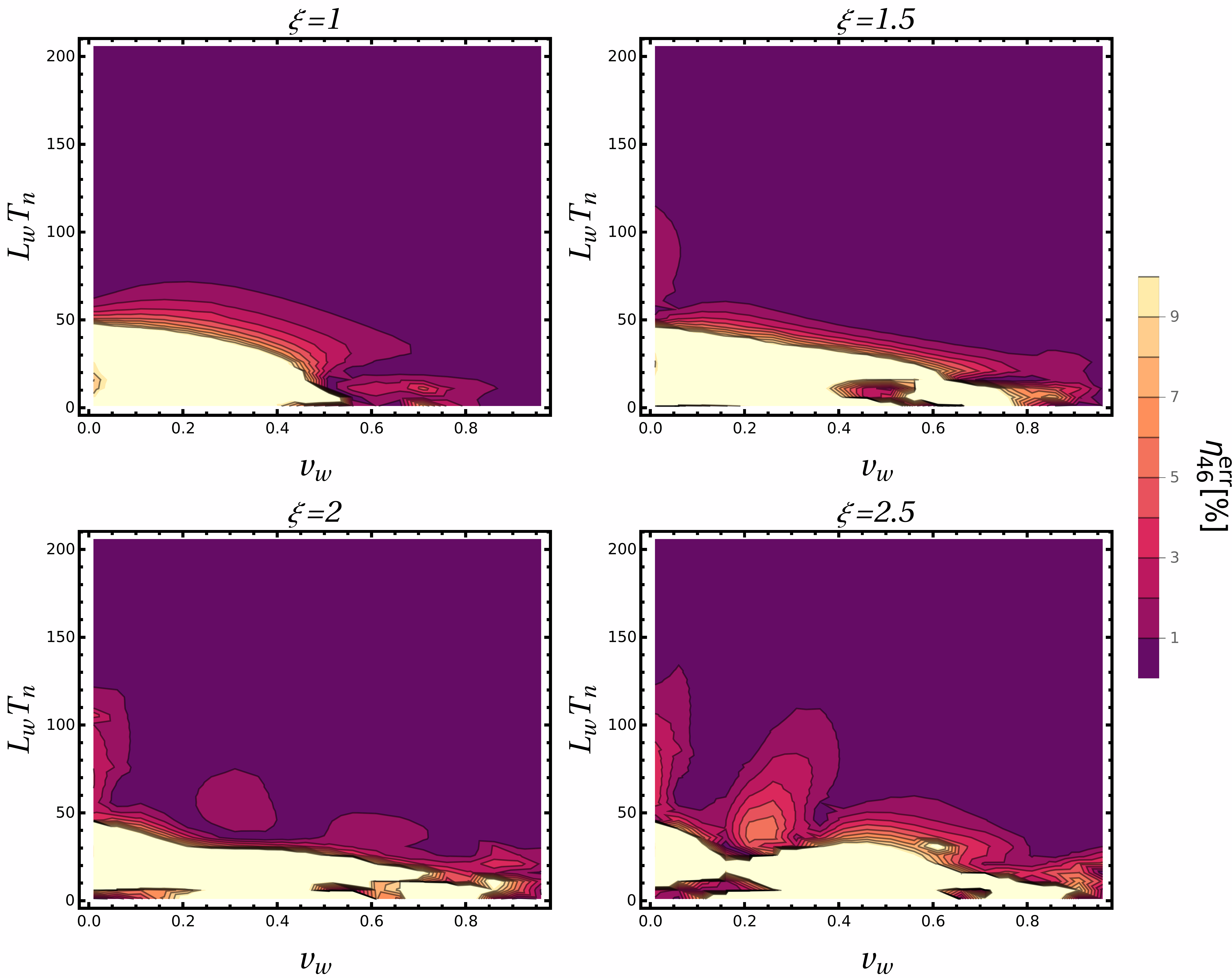}
    \caption{Contour plot of the error quantity $\eta_{46}^\text{err}$ defined in Eq.~(\ref{eq: bau_error}) in terms of $v_w$ and $L_wT_n$ for different transition strengths $\xi_n$. The light yellow regions mark the areas where the error is larger than 10\%. Here, we choose $R=1$ in the constant truncation scheme.}
    \label{fig: error_bench}
\end{figure}

In order to assess more reliably where the computed value of the BAU truly converges to a constant value, we introduce the quantity 
\begin{equation}
\eta^\text{err}_n\equiv\left|\frac{\eta_{n+4}-\eta_n}{\eta_{n}}\right|\;.\label{eq: bau_error}
\end{equation}
If the BAU converges to a constant result, the value of $\eta^\text{err}_n$ should become zero, $\eta_n^\text{err}\rightarrow0$, when we take the limit $n\rightarrow\infty$. In Fig.~\ref{fig: error_bench} we show  contour plots of $\eta_{46}^\text{err}$ as a function of $v_w$ and $L_wT_n$ for different transition strengths $\xi_n$ for $R=1$ in the constant truncation scheme. The light yellow region marks the area where $\eta_{46}^\text{err}>10\%$. We see that, generally, in order to obtain a sufficiently small error throughout the parameter region, we need a value of $L_wT_n\gtrsim50$. Large values of $v_w$ also lead to a better convergence. However, a large $v_w$ by itself is not sufficient to obtain a small error, as can be seen in the case of $\xi_n=2,\, 2.5$, where we need a value of $L_wT_n\gtrsim30$ to obtain a sufficiently small error. Furthermore, we note that within the area where $\eta_{46}^\text{err}>10\%$, there are small islands of stability. They are  outliers that may be due to a convenient choice of input parameters. \s

Note, that these plots are meant to solely give a rough estimate of the parameter regions where we expect the moment expansion to lead to stable results. Thus, we would like to stress that the quantity $\eta_{x}^\text{err}$ is solely a measure of how convergent the current result for $n=x$ is. A value of $\eta_{46}^\text{err}>10\%$ e.g.~means that the result at $n=46$ is off by at least $10\%$ compared to the value at $n=50$. It does not tell us, however, the error compared to the "true" value at $n\rightarrow\infty$. \s

Let us conclude this chapter by stating that these results should not be too surprising, since the WKB method is based on the assumption that $L_wT_n\gg1$. Although for example  Ref.~\cite{Gent:2025csq} suggests that $L_wT_n\gtrsim2$ is sufficient to satisfy this assumption, this is not what we observe when taking into account larger equation moments. In our analysis we obtain a more pessimistic limit of $L_wT_n\gtrsim50$. Additionally, the assumption that the chemical potentials $\mu$ and the velocity disturbance $\delta f$ are sufficiently small might not hold as we increase the strength of the phase transition. We can see this in the increasing error as we increase $\xi_n$ in Fig.~\ref{fig: error_bench} as well as in the erratic behavior seen in Fig.~\ref{fig: vn_truncation_dependence} with increasing $\xi_n$. Unfortunately, the parameter regions which give stable results coincide with the regions that lead to a small BAU. 

\section{Numerical Analysis of the C2HDM}\label{sec:c2hdmresults}

After these general statements and the comparison of our implementation with the literature, we now turn to the investigation of the amount of generated baryogenesis in the C2HDM using our updated computation of the BAU and taking into account all relevant theoretical and experimental constraints on the model. The discussion of the benchmark model showed, however, that there are large uncertainties in the computation of the BAU. Therefore, to assess the validity of the obtained results, we first analyse the impact of the various choices that we have w.r.t.~the application of the transport equations. Thus, we can choose freely
\begin{itemize}
    \item[$\diamond$] the truncation scheme, where we will apply the constant truncation scheme, with $R = -v_w$ and the variance truncation;
    \item[$\diamond$] the number $n$ of moment equations with $n$ chosen as $n = 2+4k,\,k \in \mathbb{N}$;
    \item[$\diamond$] the bubble wall velocity,  $0<v_w<1$; 
    \item[$\diamond$] the VEV profile, where we will choose between the kink profile and the profile deduced from the equations of motion, denoted by field profile in the following;
    \item[$\diamond$] the transition temperature, where we will choose between the critical and the percolation temperature, $T_c$ and $T_p$, respectively.
\end{itemize}
The results will be shown both for the C2HDM Type I and Type II. The benchmark setting for our transport equations is the choice: 
\begin{itemize}
    \item[$\diamond$] percolation temperature $T_p$;
    \item[$\diamond$] constant truncation scheme with  $R = -v_w$;
    \item[$\diamond$] moment truncation at $n = 2$;
    \item[$\diamond$] wall velocity $v_w = 0.5$;
    \item[$\diamond$] field profile.
\end{itemize}
We will study the impact these options by varying them one at a time, while  keeping the others equal to their benchmark values. \s

The parameter points considered in this section were generated using a combination of \texttt{ScannerS}~\cite{Coimbra:2013qq,Muhlleitner:2020wwk} and \texttt{BSMPTv1}~\cite{Basler:2018cwe} applying an Evolutionary Strategy Algorithm~\cite{Hansen2001,hansen:hal-01297037,HBOS} coupled with an anomaly detection method for Novelty Reward~\cite{Romao:2024gjx} to improve the exploration of the parameter space. This machine learning algorithm attempts to fulfill all theoretical and experimental constraints imposed by \texttt{ScannerS} and tries to maximize the phase transition strength at the critical temperature, $\xi_c\equiv\frac{v_c}{T_c}$, calculated using \texttt{BSMPTv1}.
\begin{table}[h!]
\centering
\begin{tabular}{lrrrrrll}\toprule
\multirow{2}{*}{\textbf{Parameter}} & \multicolumn{2}{c}{\textbf{Type I}} & \multicolumn{2}{c}{\textbf{Type II}} & \multirow{2}{*}{\textbf{Scale}} & \multirow{2}{*}{\textbf{Units}}\\
\cmidrule(lr){2-3}\cmidrule(lr){4-5}
 & \textbf{Lower} & \textbf{Upper} & \textbf{Lower} & \textbf{Upper} & & \\
\cmidrule(){1-7}
$m_{1}$   & \multicolumn{2}{c}{----- $125.09$ -----}  & \multicolumn{2}{c}{----- $125.09$ -----}  & linear & $[\text{GeV}]$ \\
$m_{2}$   & $130$   & $1000$ & $130$   & $1000$ & linear & $[\text{GeV}]$ \\
$m_{H^\pm}$ & $80$    & $1000$ & $580$    & $1000$ & linear & $[\text{GeV}]$ \\
\cmidrule(){1-7}
$\tan\beta$ & $0.3$     & $30$   & $0.3$     & $30$   & linear & \\
$\alpha_1$  & $-\pi/2$  & $\pi/2$  & $-\pi/2$  & $\pi/2$  & linear & \\
$\alpha_2$  & $-\pi/2$  & $\pi/2$  & $-\pi/2$  & $\pi/2$  & linear &\\
$\alpha_3$  & $-\pi/2$  & $\pi/2$  & $-\pi/2$  & $\pi/2$  & linear & \\
$m_{12}^2$  & $\pm10^{-1}$ & $\pm10^{7}$ & $\pm10^{-1}$ & $\pm10^{7}$ & log    & $[\text{GeV}]^2$ \\
\bottomrule
\end{tabular}
\caption{C2HDM parameter scan ranges for Type I and Type II. The $m_{12}^2$ parameter is sampled on a logarithmic scale; all others are sampled linearly.}
\label{tab:c2hdm_scan}
\end{table}
For simplicity and computational speed, 
we only considered points with $\xi_c\ge1$, which provides the EW sphaleron suppression required for a successful baryogenesis. Furthermore, we considered the lightest neutral scalar particle to be the SM-like Higgs boson, i.e. $m_{h_1}=m_\text{SM-Higgs}$. The ranges considered in the scan are summarized in Tab.~\ref{tab:c2hdm_scan}. \s

The only further cut applied to our scans was, for the points calculated using the percolation temperature $T_p$, that $\beta/H_*>1$, otherwise it would constitute a gravitational wave wavelength which is larger than
the Hubble horizon and would lead to difficulties with causality
bounds on the amplitude~\cite{Athron:2023xlk,Giblin2014TheDO,Konstandin:2011dr}. A low $\beta/H_*$ would also indicate a slow transition, this would make the transition temperature ill-defined as we would have bubbles appearing at vastly different temperatures. \s

As the $z$ dependence of the top mass value is decisive for the generated BAU, we here display the formula for its calculation in the C2DHM. As it is the second doublet that couples to the up-type quarks in C2DHM Type I and II, we have  
\begin{align}
|m_t(z)| e^{i \theta(t)} = \frac{y_t}{\sqrt{2}} \, (\omega_2 (z) + i \omega_{\text{CP}} (z) ) \;. \label{eq:topvaluecalc}
\end{align}
The CP-violating top-quark phase $\theta$ is hence directly related to $\omega_{\text{CP}}$.

\subsection{Sign of the Baryon Asymmetry}
In the following, we will show scatter plots of the absolute value of the computed baryon asymmetry normalized to the observed value, $\bar{\eta} \equiv |\eta|_{\text{calc}}/\eta_{\text{obs}}$. We take the absolute value of the BAU, because for some parameter points the calculated BAU turned out to be negative.
The sign has a physical meaning, however, in the sense that a negative value would predict a universe with more anti-matter than matter. The C2HDM has a transformation, however, that allows us to consider only the absolute value of the BAU. 
Flipping the sign of the imaginary parts of $m_{12}^2$ and the $\lambda_5$ coupling\footnote{The imaginary parts of these two parameters are related by tree level minimization conditions.} and of $\omega_\text{CP}$ leaves the potential invariant.
Therefore, flipping the sign of $\mathrm{Im}(m_{12}^2)$ and $\mathrm{Im}(\lambda_5)$ is the same as a reflection of the potential at the $\omega_\text{CP}$-axis. Since $\omega_{\text{CP}}$ determines the sign of the source term, its sign flip changes the sign of the BAU. If we find a point with a negative BAU, we can hence turn this into a parameter point that leads to a positive BAU by flipping the sign of $\mathrm{Im}(m_{12}^2)$. This argument has to be taken with caution, however, as it can happen that for some parameter points, when changing $n$, the BAU flips sign again, so that the parameter point becomes invalid. For all scanned points, we  verified that, up to $n =22$, the sign of the BAU remains consistent between all choices for the number $n$ of moment equations.

\subsection{Uncertainties}
\begin{figure}[h!]
	\centering
\includegraphics[width=0.49\textwidth]{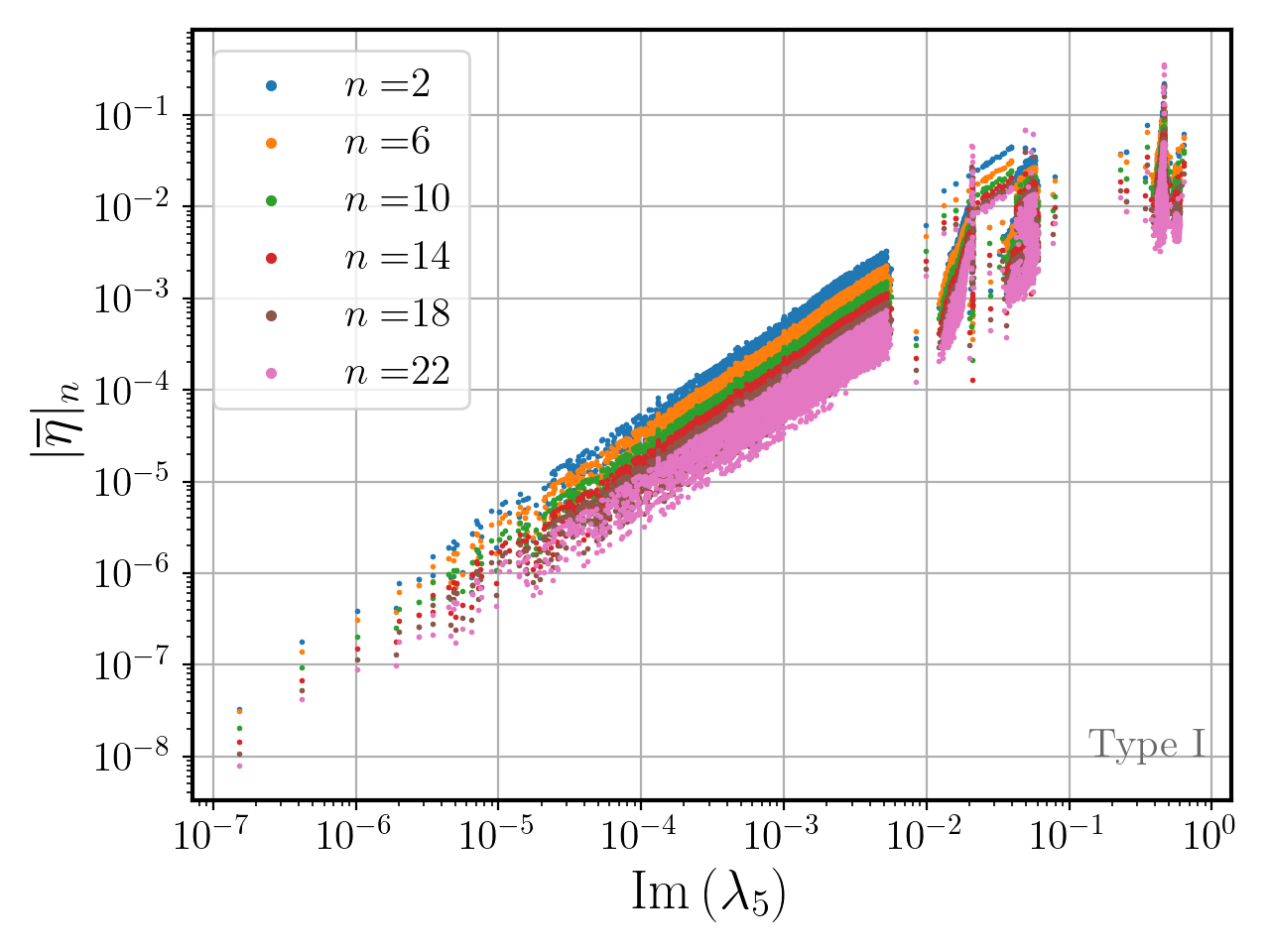} 
\includegraphics[width=0.49\textwidth]{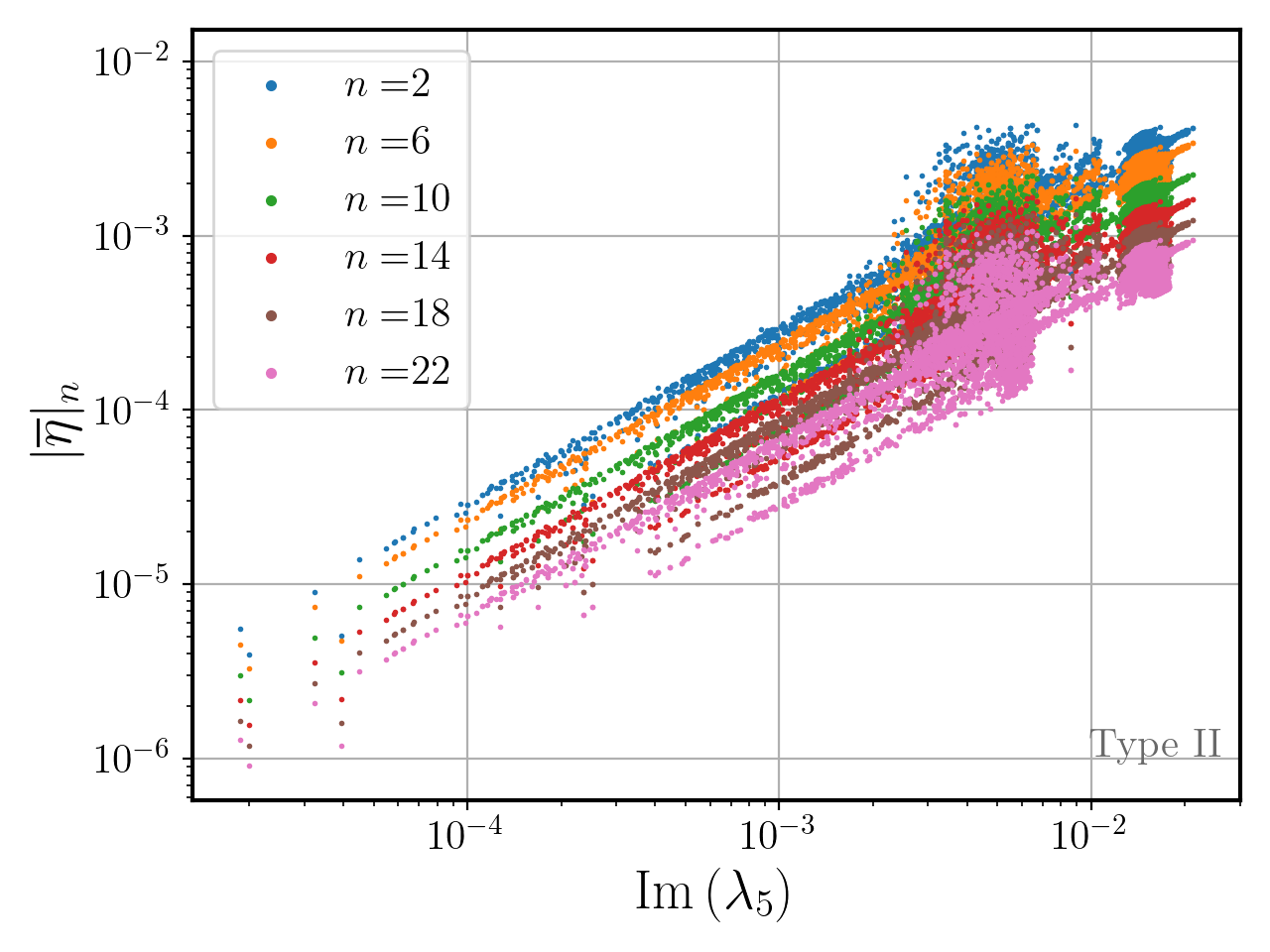} 
	\caption{Scatter plots of the absolute value of the normalized baryon asymmetry for different numbers $n$ of moment equations, as a function of $\mathrm{Im}(\lambda_5)$ for the C2HDM Type I (left) and II (right).}
\label{fig:eta_vs_ell}
\end{figure}

\paragraph{Impact of number $n$ of transport equations}
Due to the large computation times, we will present our results for a maximum of $n=2$ benchmark equations. However, the generated baryon asymmetry depends on the number of $n$ and for large $n$, where we can expect better convergence towards the true result, unfortunately decreases, which we already saw in the benchmark model. This is exemplified in Fig.~\ref{fig:eta_vs_ell}, where we show the impact of $n$ on the absolute value of the BAU normalized to the measured value, $|\bar{\eta}|$, as a function of the imaginary part of $\lambda_5$, for several values of $n$ from 2 to 22 in steps of 4. As can be inferred from the figure, the value of the BAU decreases with increasing $n$.

\paragraph{Truncation scheme dependence}
Another uncertainty arises from the truncation scheme dependence. We saw in the benchmark discussion that this is particularly large for low values of $n$. We confirm this behaviour for the C2HDM, where we find a large dependence on the truncation scheme for $n=2$. 

\begin{figure}[h!]
	\centering
\includegraphics[width=0.49\textwidth]{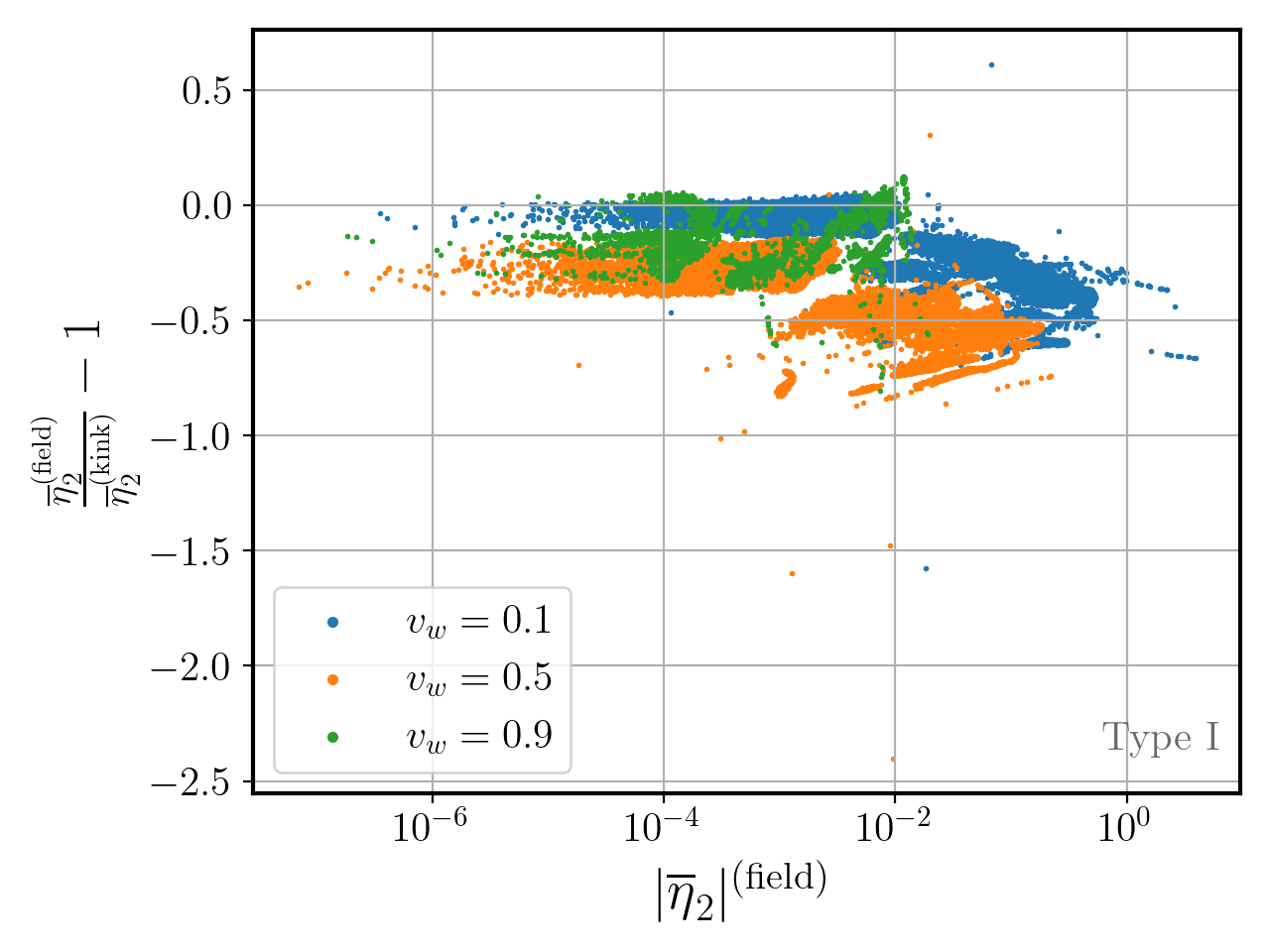} 
\includegraphics[width=0.49\textwidth]{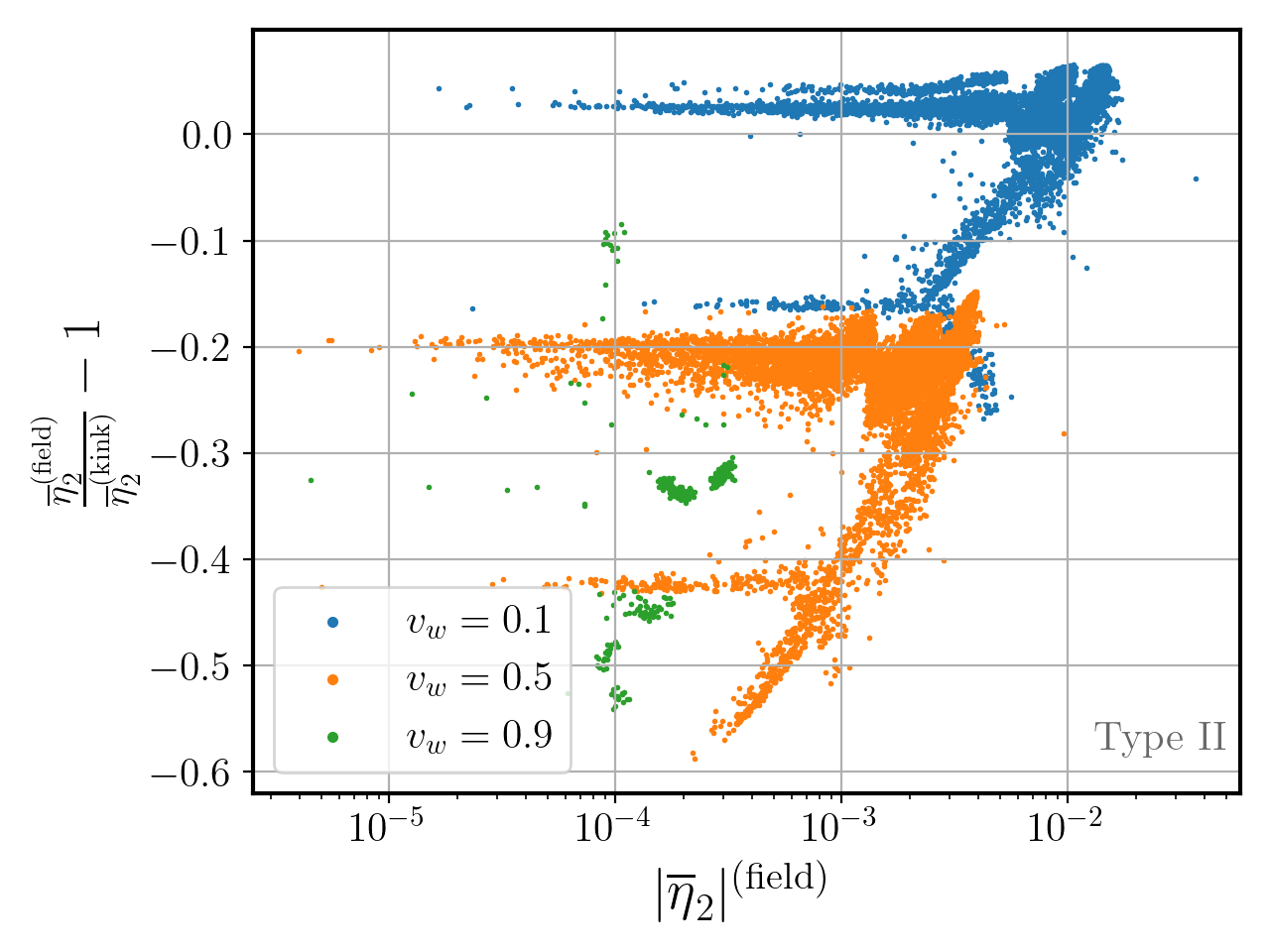} 
\caption{Scatter plot of the relative difference between $|\bar\eta|_{n=2}$ using the kink profile and the solution from the EOMs to model the VEV profile, as a function of $\etab^{\text{(field)}}$, for $v_w = 0.1$ (blue), $0.5$ (orange), and $0.9$ (green). Left (right): C2HDM Type I (II).}
\label{fig:field_vs_kink}
\end{figure}

\paragraph{VEV profile impact}
Another uncertainty is due to the chosen VEV profile. In Fig.~\ref{fig:field_vs_kink}, we explore the impact of using either the field solution or the kink solution for the VEV profile which then models the complex mass of the quarks, in particular of the top quark mass. We show the scatter plots of the relative difference of the normalized BAU when using the solution from the equations of motion ("field") or the kink solution ("kink") to model the VEV profile, for $n=2$ and the C2HDM Type I (left) and Type II (right). For both C2HDM types and all wall velocities the vast majority of points has $|\etab|^\text{(kink)} > |\etab|^\text{(field)}$. This conclusion can only be drawn when $n=2$.  For $n>2$ further investigations would be necessary, cf.~our discussion below of benchmark point \texttt{BP3}. The reason for this hierarchy is that the field solution usually leads to a larger wall width $L_w$ compared to the kink solution. A larger wall width $L_w$, however, decreases $|\bar{\eta}|$, as can be seen in Fig.~\ref{fig: lw_truncation_dependence}.\footnote{Note that we fixed the nucleation temperature to $T_n=100$~GeV.} The obtained BAU values, however, are of the same order of magnitude for the two approaches, so that the related uncertainties are not too large compared to other uncertainties discussed here. 

\begin{figure}[h!]
	\centering
\includegraphics[width=0.49\textwidth]{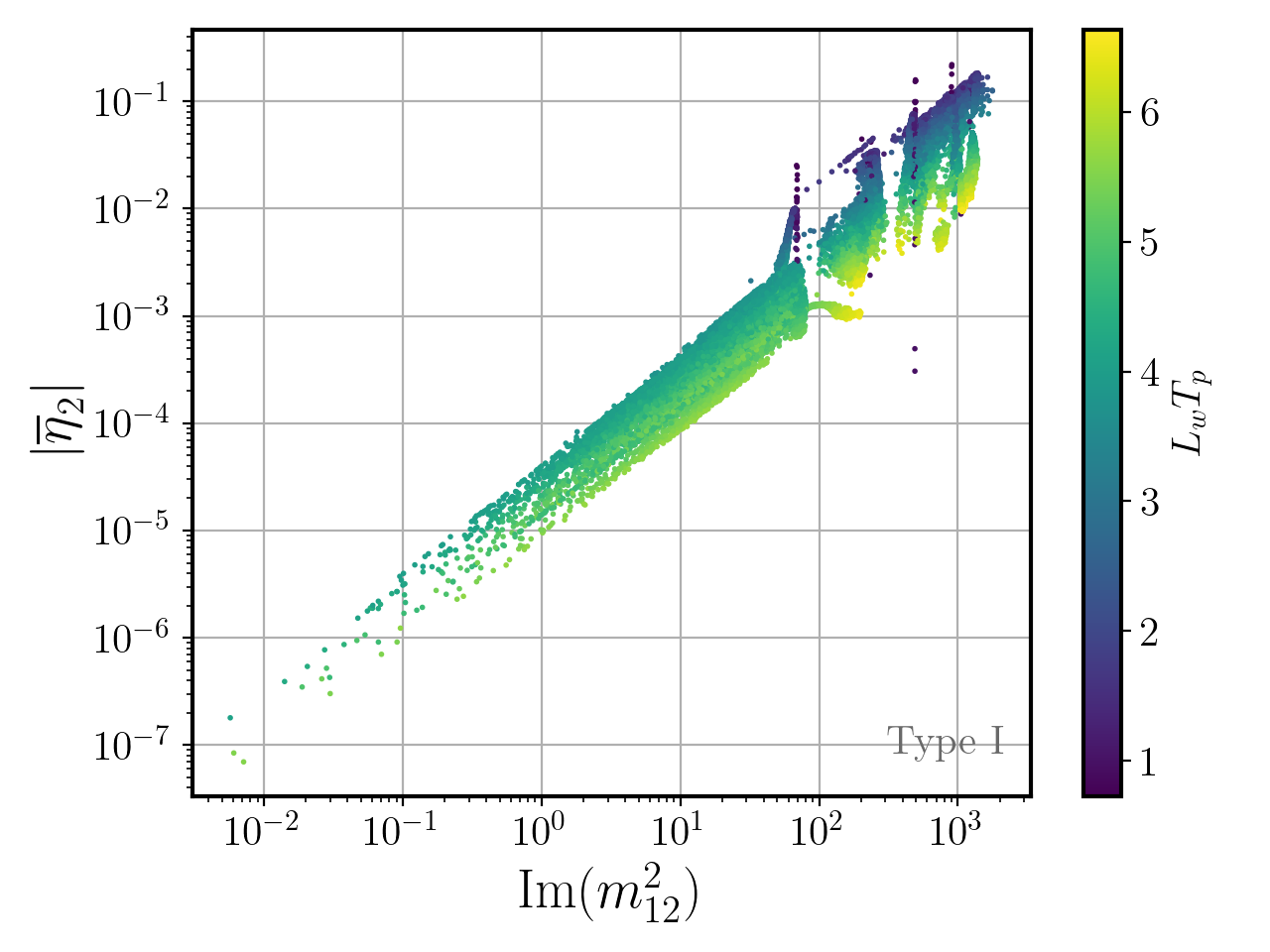} 
\includegraphics[width=0.49\textwidth]{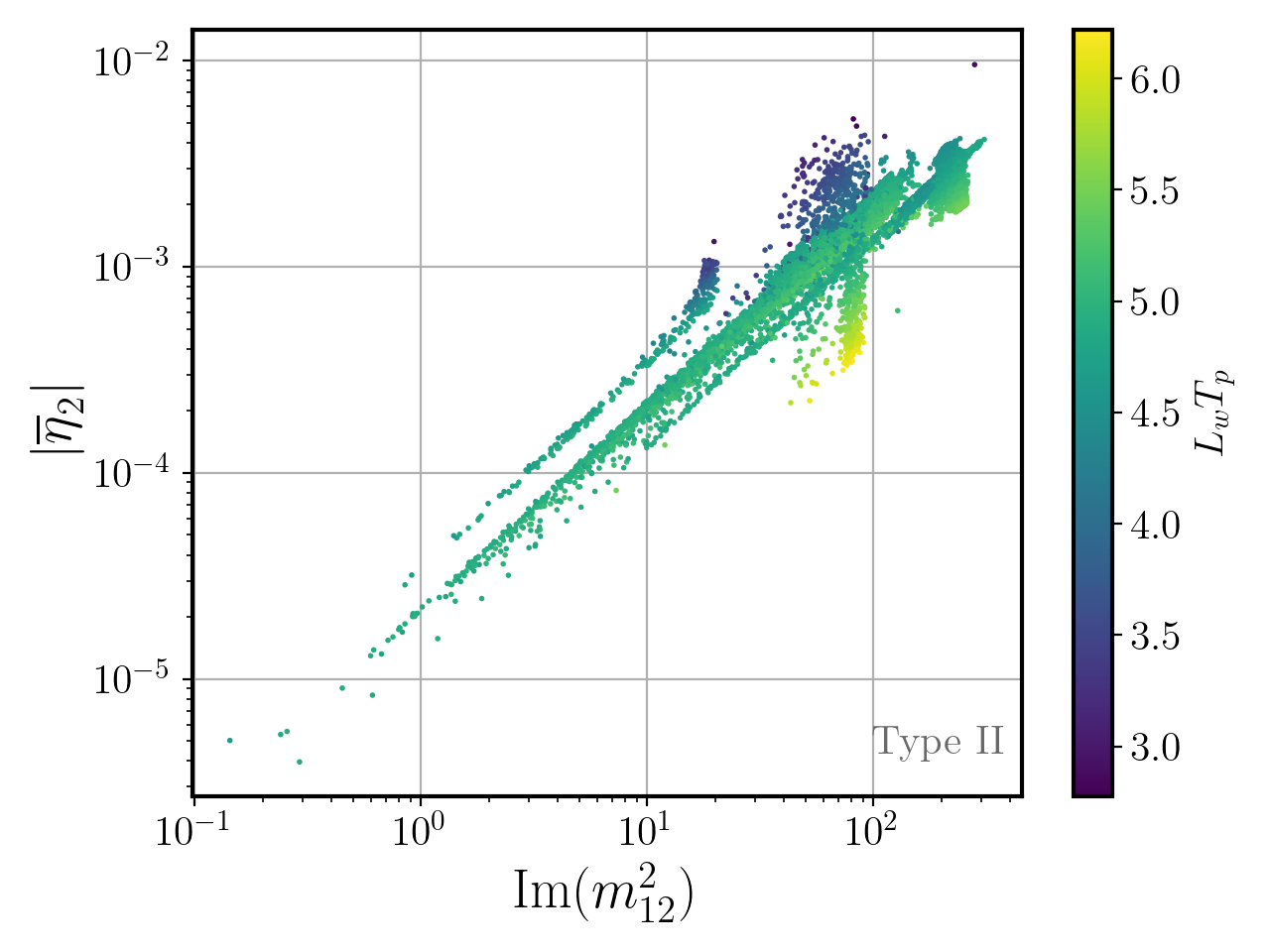} 
\caption{Scatter plot of the
normalized baryon asymmetry for $n=2$, $\bar{\eta}_2$, as a function of $\mathrm{Im}(m_{12}^2)$ for the C2HDM Type I (left) and II (right). The color bar shows $L_w T_p$. The chosen VEV profile is the kink profile.}
\label{fig:lwt}
\end{figure}

\paragraph
{\bf Parameter $L_w T_p$}
Figure~\ref{fig:lwt} shows for the C2HDM Type I (left) and II (right) the normalized baryon asymmetry $\bar{\eta}$ for $n=2$ as a function of $\mathrm{Im}(m_{12}^2)$. The color bar denotes the values of $L_w T_p$, where $T_p$ denotes the percolation temperature, which is our default choice for the transition temperature. For these plots, we applied the kink profile for the VEV, as this allows us to calculate the wall width by applying the formula Eq.~(\ref{eq:lwcalc}). 
 As we can see, in all of our scenarios the wall width times percolation temperature remains below 7 for both C2HDM types. The WKB method, however, is based on the assumption of large values of $L_w T$. We also saw this in our discussion of the benchmark model, where we concluded that we should have $L_w T_n \gtrsim 50$ for stable results. These low values of $L_w T_p$ found here introduce another major source of uncertainties in the results. 

\paragraph{Impact of Transition Temperature and Wall Velocity}

\begin{figure}[h!]
	\centering
\includegraphics[width=0.98\textwidth]{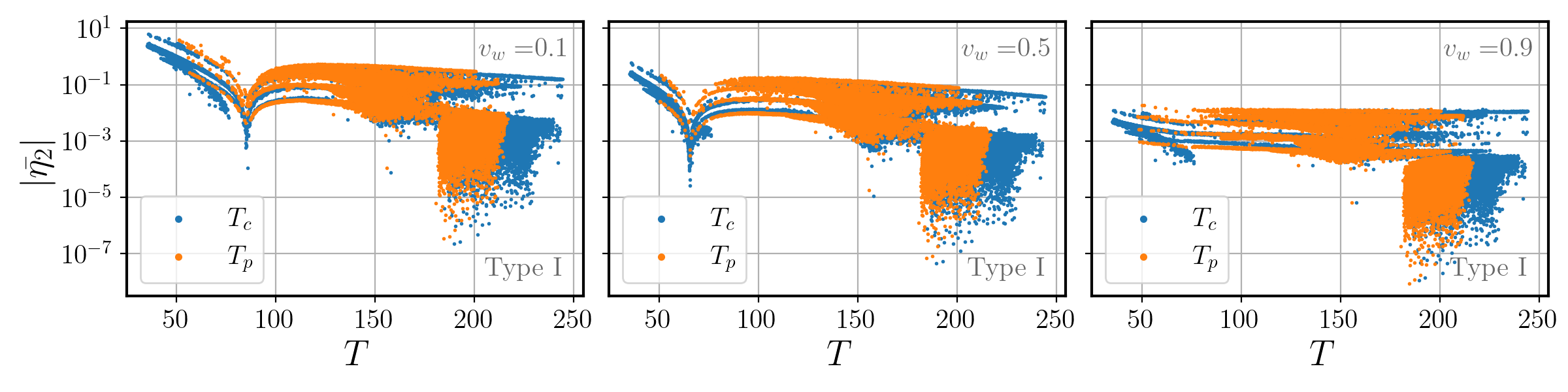}
	\caption{Scatter plot of the BAU $|\etab|$ calculated at the critical (blue points) and the percolation (orange points) temperature  for the C2HDM type I, for the wall velocities $v_w = 0.1$, $v_w = 0.5$ and $v_w = 0.9$ from left to right.}
\label{fig:typeI_eta_vs_TcTp}
\end{figure}

Another source of uncertainty is introduced by  the transition temperature, which we choose to be either the critical or the percolation temperature. In Fig.~\ref{fig:typeI_eta_vs_TcTp}, we show the scatter plots of the calculated BAU, $|\etab|$, at the critical (blue points) and percolation (orange points) temperatures for the C2HDM Type I. 
The left, middle and right panels correspond to the wall velocities $v_w = 0.1$, $v_w = 0.5$ and $v_w = 0.9$, respectively. Since $T_p < T_c$, we expect the overall shape of the distribution of the $T_p$ points to be shifted to the left of the $T_c$ points, and this is indeed what we see except that the points calculated using $T_c$ extend to lower temperature values. Upon further investigation, we found that for these $T_c$ points to the left of the orange region, when calculated using $T_p$ they are either vacuum trapped, i.e. the decay rate is never high enough for the transition to be successful, or are almost vacuum trapped 
and have a $\beta/H_*<1$. 
We also noticed that for $v_w=0.1$ and $v_w=0.5$ the $T_c$ points with maximal BAU are to the left of the $T_p$ points and are therefore not allowed for the same reasons. For this reason, when performing 
scans using $T_c$ instead of $T_p$ for the BAU calculation, we might overestimate the maximum BAU that a specific model can achieve. We can also see that for the same parameter point, when using $v_w=0.1$ and $v_w=0.5$, the calculated BAU using $T_p$ is larger than the BAU calculated using $T_c$ but, for $v_w=0.9$, using $T_p$ yields smaller results. \s

\begin{figure}[h!]
	\centering
\includegraphics[width=0.98\textwidth]{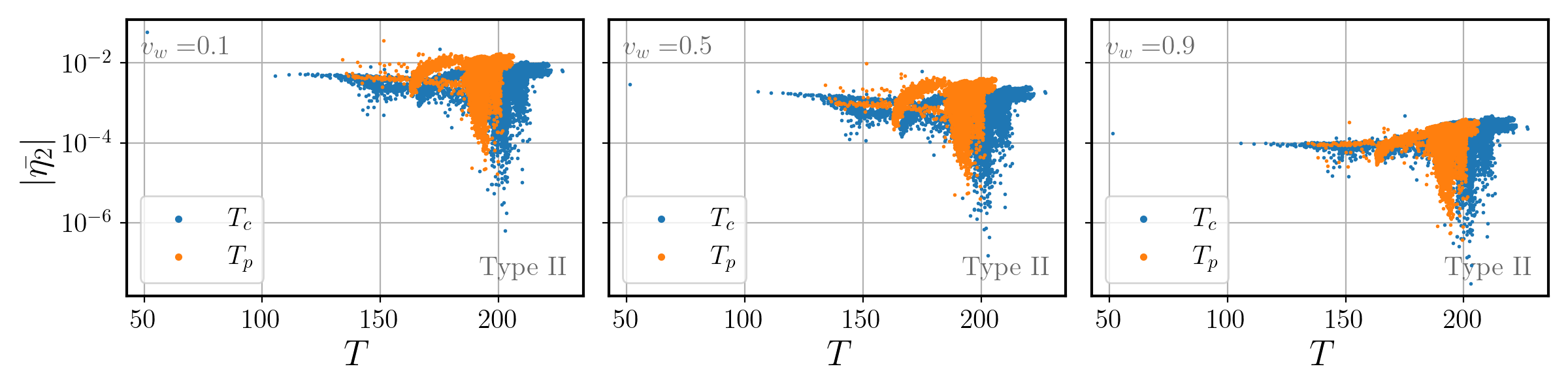}
	\caption{Scatter plot of the BAU $|\etab|$ calculated at the critical (blue points) and the percolation (orange points) temperature  for the C2HDM type II, for the wall velocities $v_w = 0.1$, $v_w = 0.5$ and $v_w = 0.9$ from left to right.}
\label{fig:typeII_eta_vs_TcTp}
\end{figure}

In Fig.~\ref{fig:typeII_eta_vs_TcTp}, we show the corresponding plots for the C2HDM Type II. The results and discussion are identical to those of type I, except that for type II we do not have the $T_{c/p}<50\text{ GeV}$ region that maximizes the BAU for $v_w=0.1$ and $v_w=0.5$. \s

All these plots furthermore show that the computed baryon asymmetry depends also on the wall velocity. We here found, that it overall decreases with increasing wall velocity.  Note, however, that in the present version of \texttt{BSMPT}, the wall velocity is an input parameter. Calculating the wall velocity consistently including the plasma conditions might further change this picture. 

\subsection{Impact of CP Violation}
\begin{figure}[t!]
	\centering
\includegraphics[width=0.49\textwidth]{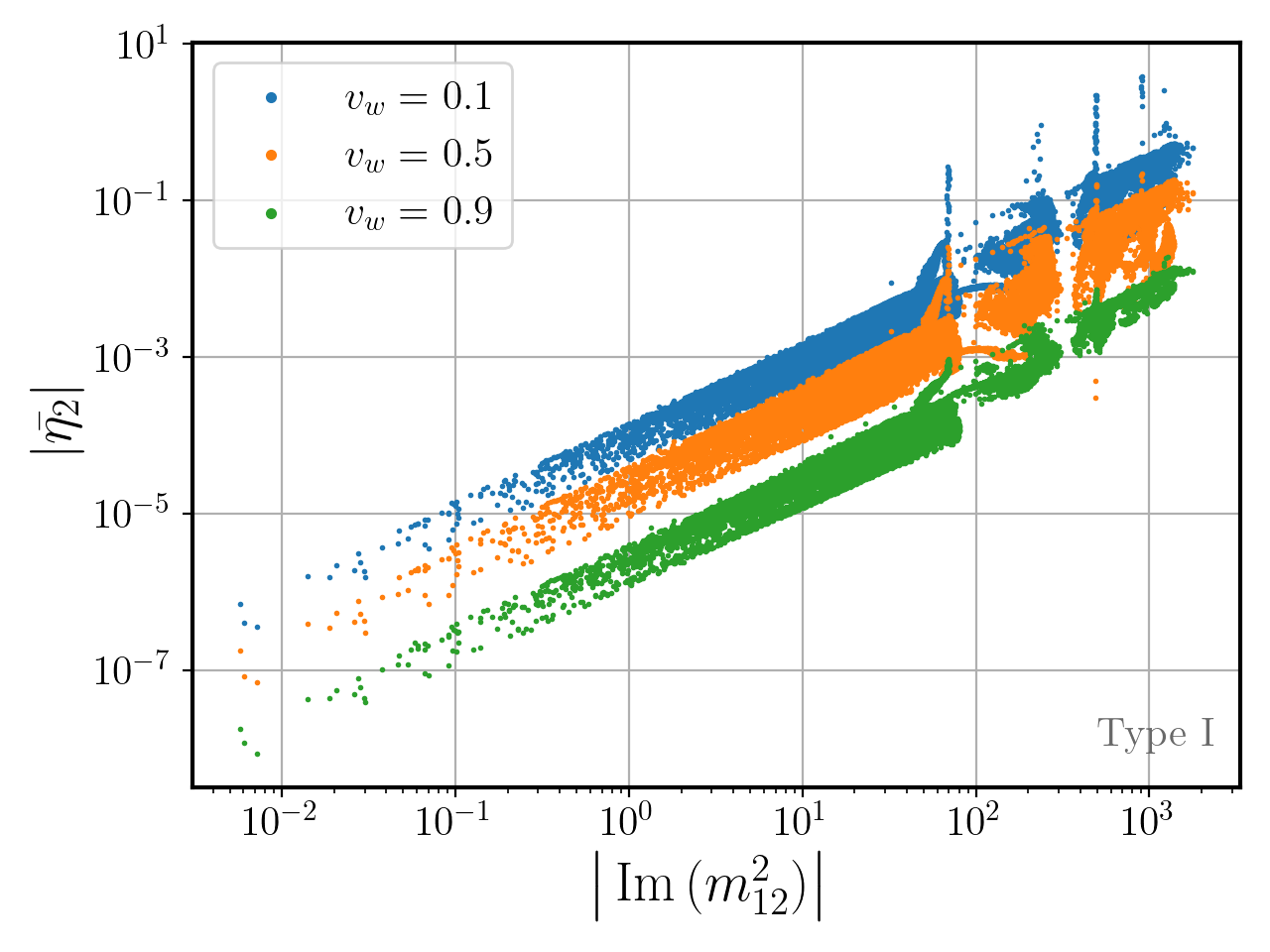} 
\includegraphics[width=0.49\textwidth]{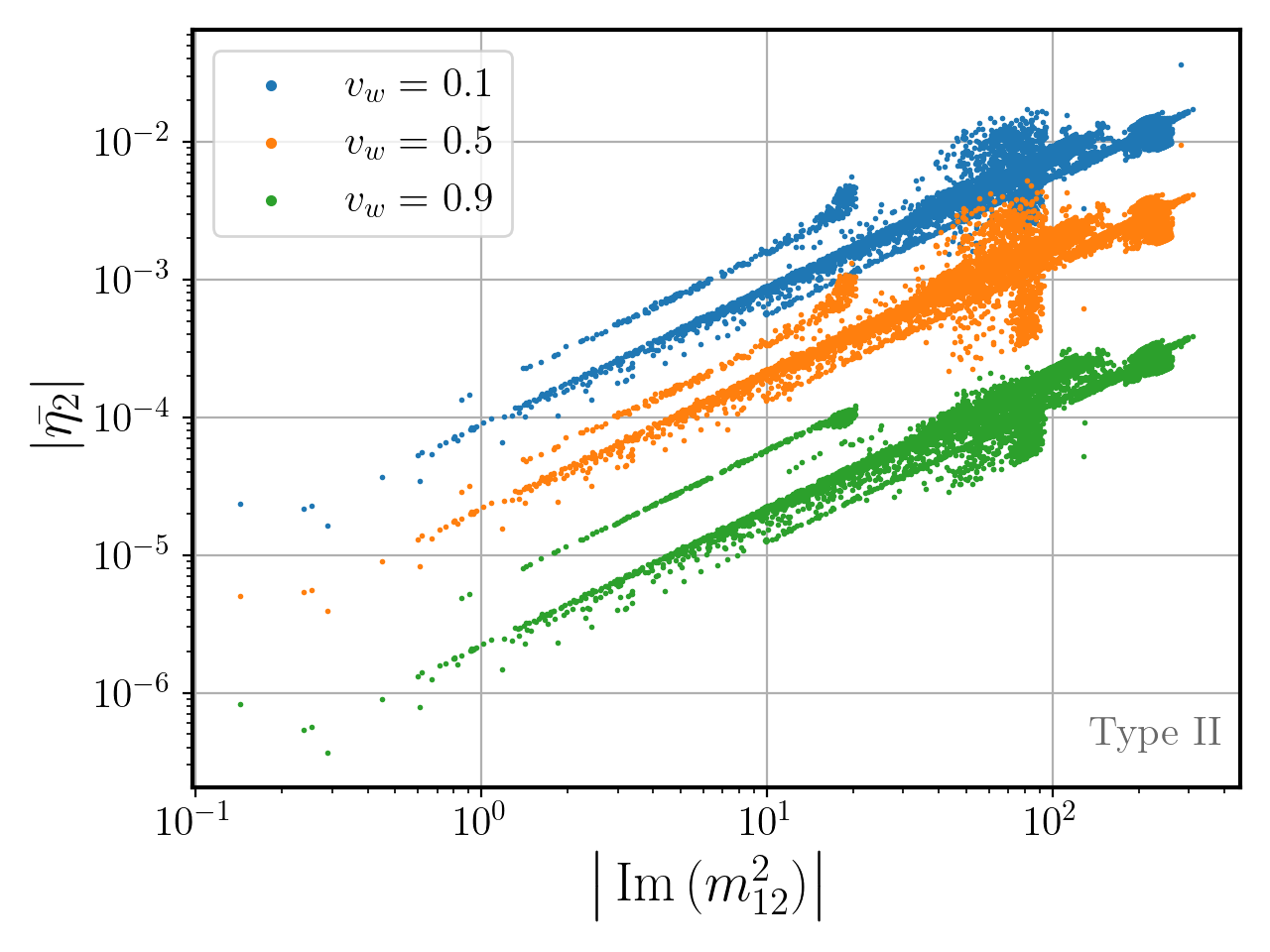} 
	\caption{Scatter plot of the normalized baryon asymmetry $|\etab|$ as a function of $\operatorname{Im}(m_{12}^2)$, for $v_w = 0.1$ (blue), $0.5$ (orange), and $0.9$ (green). Left (right): C2HDM Type I (II). }
\label{fig:imm12}
\end{figure}

With all these caveats in mind regarding the uncertainties related to the obtained values of the baryon asymmetry, we now turn to the phenomenological discussion. We investigate the impact of the amount of CP violation, which is one of the three Sakharov conditions. 
In Fig.~\ref{fig:imm12} we show the dependence of $\bar{\eta}_2$ on the imaginary part of $m_{12}^2$,  $\operatorname{Im}({m_{12}^2})$, which is an input parameter and hence easy to control. In both type I and type II, we can see a very strong positive correlation between the BAU and the explicitly CP-violating parameter $\operatorname{Im}({m_{12}^2})$. This makes it very clear that the amount of CP violation is decisive for the size of the generated BAU, as expected. Also the BAU goes to zero as $\mathrm{Im}(m_{12}^2) \to 0$. The plots suggest that this conclusion is rather robust despite the large uncertainties in the computation of the BAU. Future scans in the C2HDM should hence focus on regions where $\operatorname{Im}({m_{12}^2})$, and consequently $\operatorname{Im}({\lambda_{5}})$, is as large as allowed by the theoretical and experimental constraints. This is also a guideline for future model building, to focus on models with large amounts of CP violation. \s

\begin{figure}[h!]
	\centering
\includegraphics[width=0.49\textwidth]{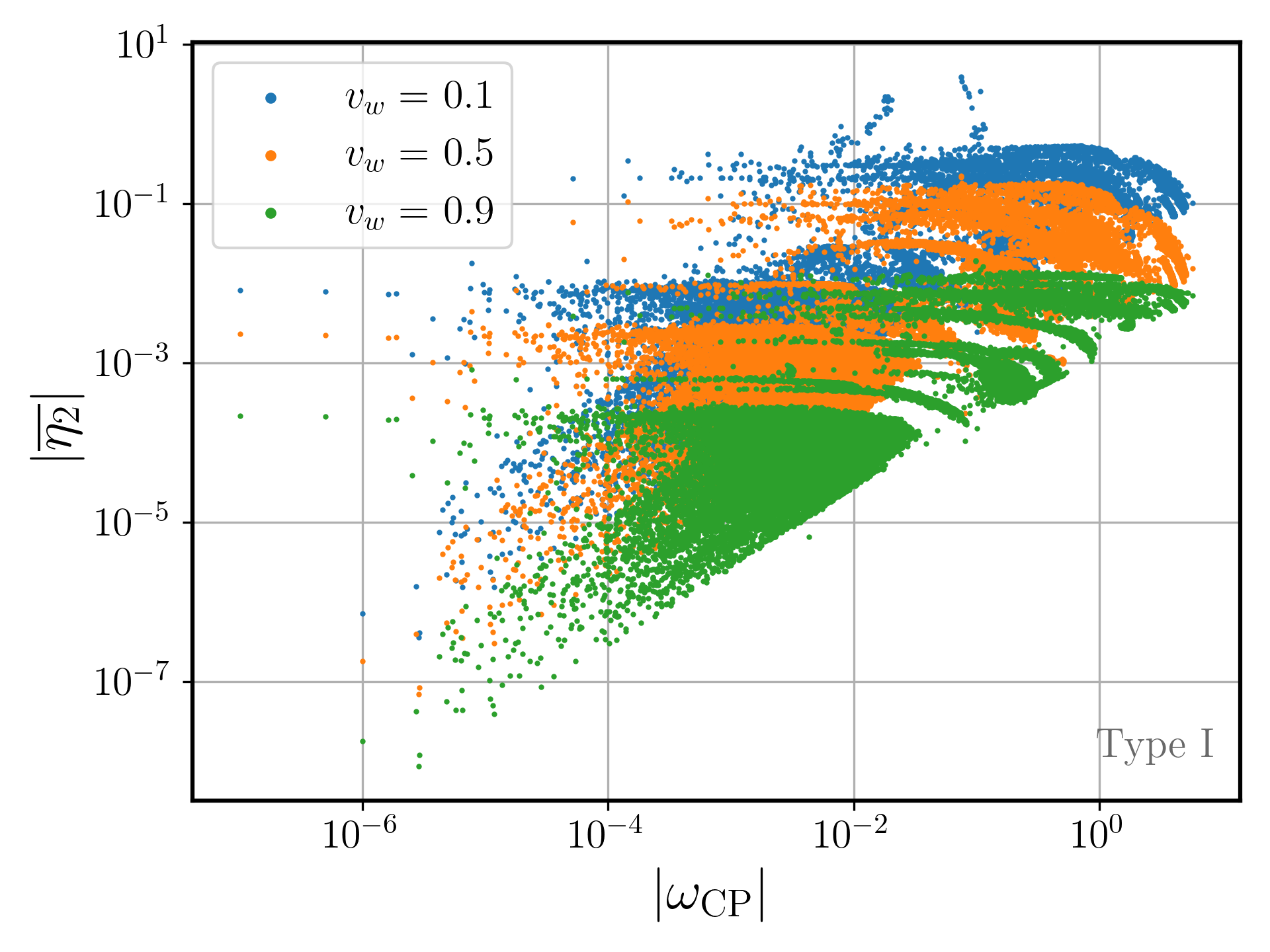} \includegraphics[width=0.49\textwidth]{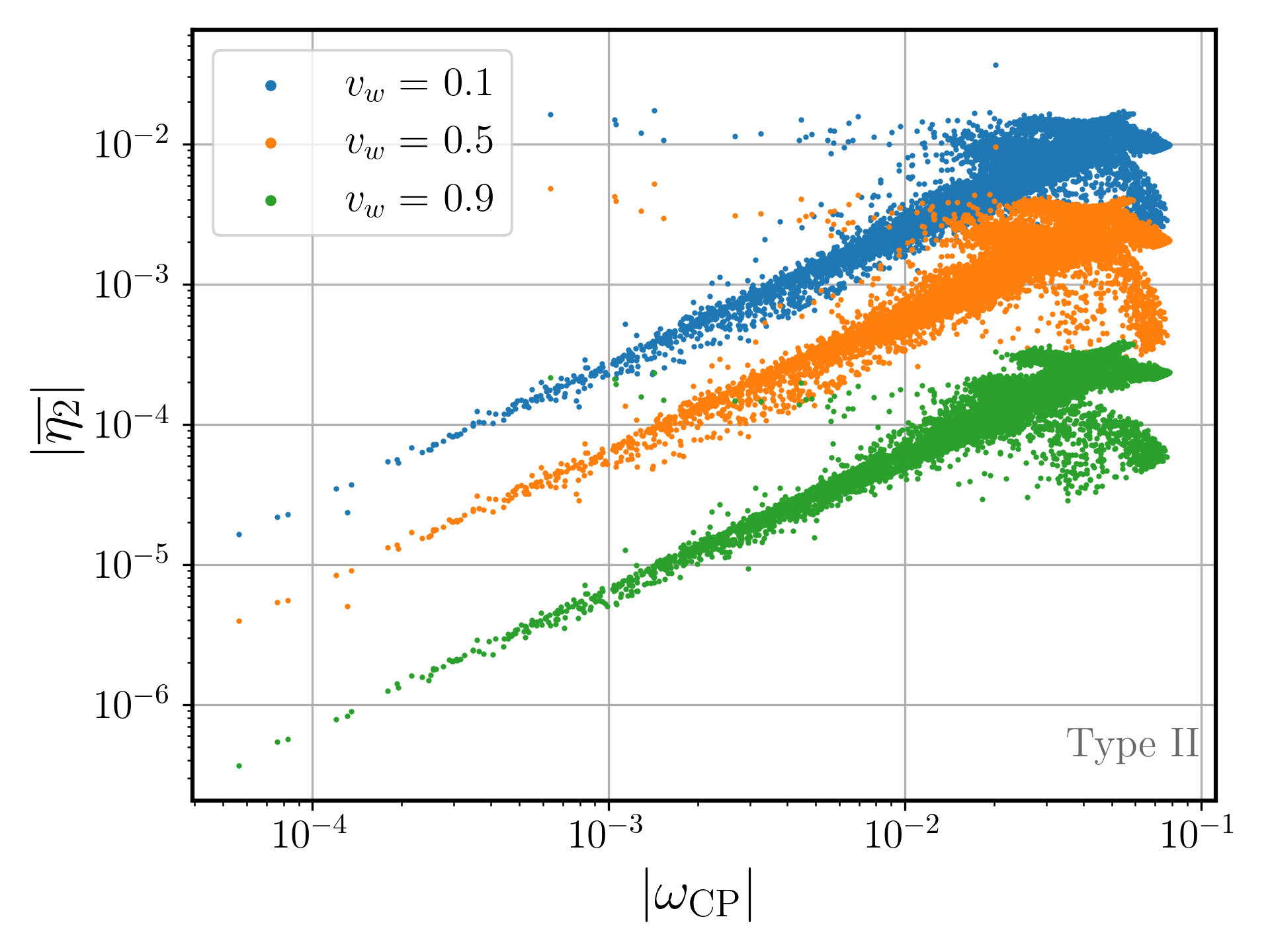} 
	\caption{Scatter plot of the normalized baryon asymmetry $|\etab|$ as a function of $|\omega_{\text{CP}}|$, for $v_w = 0.1$ (blue), $0.5$ (orange), and $0.9$ (green). Left (right): C2HDM Type I (II).}
\label{fig:omega_cp}
\end{figure}

Figure~\ref{fig:omega_cp} displays the scatter plots of $|\bar{\eta}_{2}|$ as a function of the absolute value of the imaginary VEV $|\omega_{\text{CP}}|$ in the true minimum at $T_p$, which spontaneously breaks CP. We found that a non-vanishing CP-violating VEV is only generated in scenarios with non-zero $\mathrm{Im}(m_{12}^2)$ at $T=0$. Consistent with the expectation the generated BAU vanishes as $\omega_{\text{CP}}$ goes to zero. We also note that in the C2HDM Type I the value of CP puts a lower bound on BAU. In the C2HDM Type II the maximum allowed value of $\omega_{\text{CP}}$ is smaller, which is due to the more severe constraints on this C2HDM type at zero temperature. Finally, we remark that we here show the VEV value in the true vacuum. In the wall, however, the VEV value can be different. 

\subsection{Gravitational Waves}
In this section, we briefly investigate the relation between the calculated BAU $|\etab|$ and the signal-to-noise ratio (SNR) of the gravitational wave (GW) spectrum at the Laser Interferometer Space Antenna
(\texttt{LISA}) \cite{Caprini:2019egz,LISA,LISACosmologyWorkingGroup:2022jok}. \s

\begin{figure}[h!]
	\centering
\includegraphics[width=0.49\textwidth]{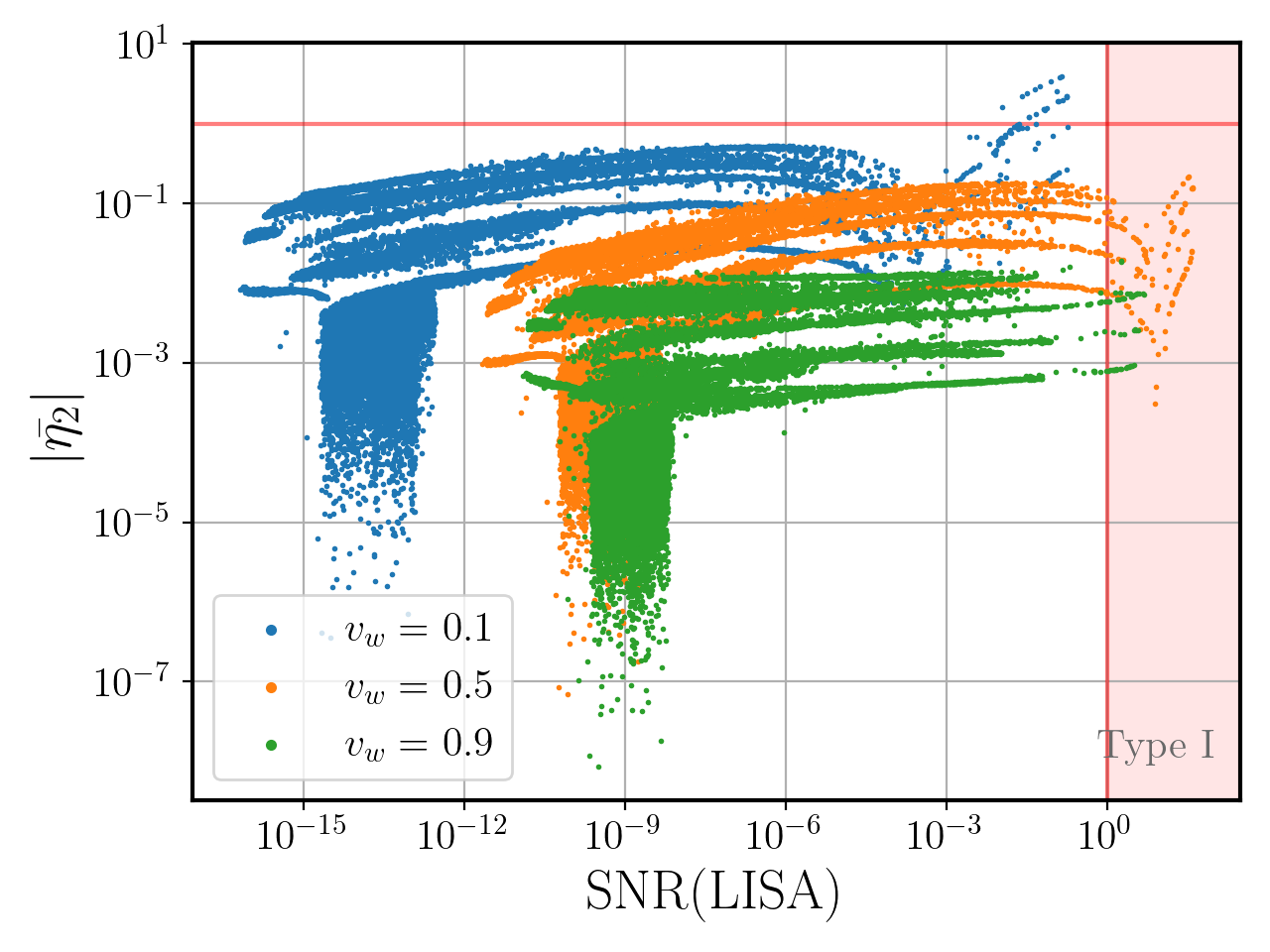} 
\includegraphics[width=0.49\textwidth]{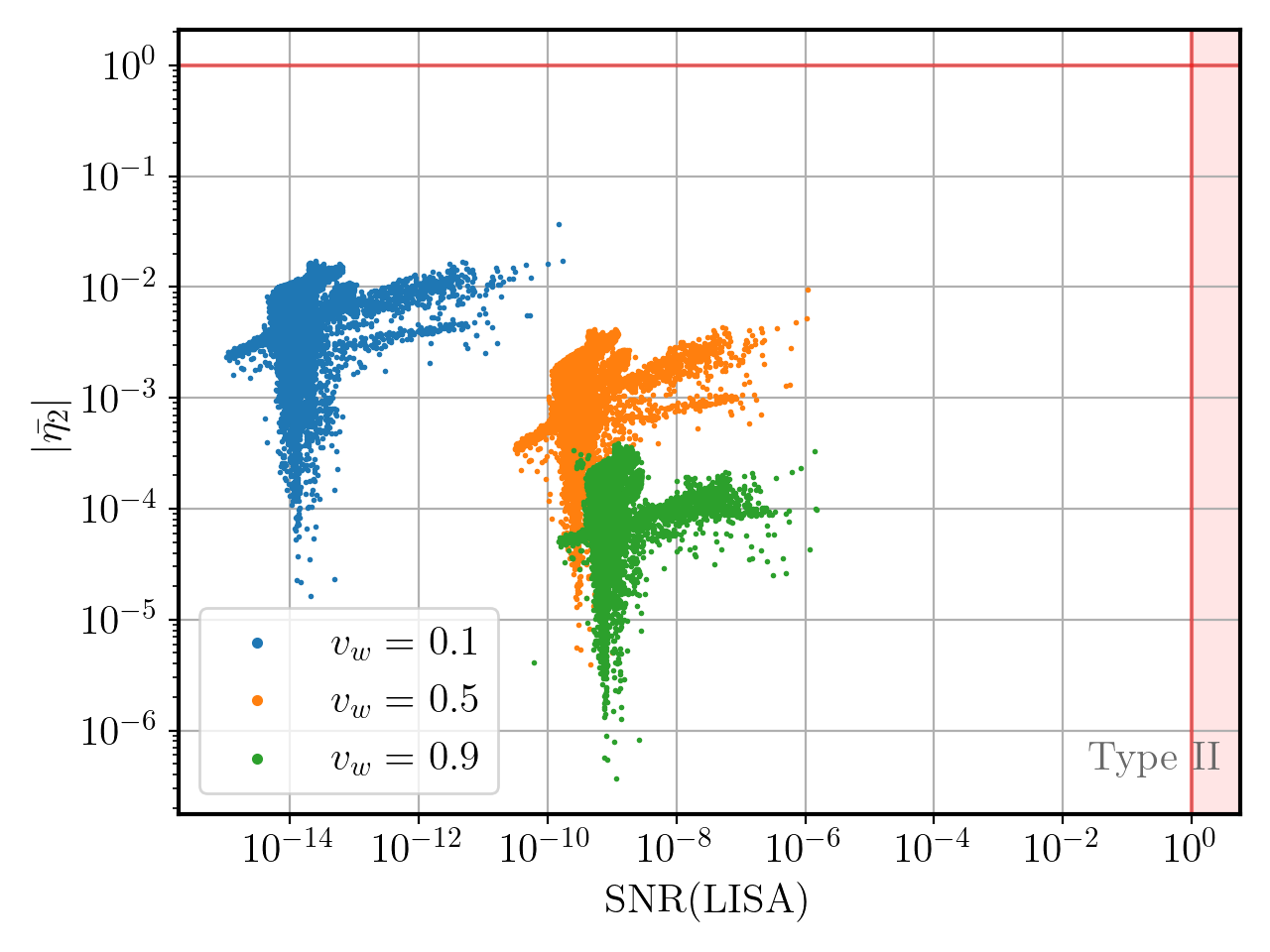} 
	\caption{Scatter plot of the normalized baryon asymmetry $|\etab|$ as a function of the SNR at \texttt{LISA}, for $v_w = 0.1$ (blue), $0.5$ (orange), and $0.9$ (green). Left (right): C2HDM Type I (II). The horizontal red line indicates $|\eta|_2=\eta_\text{obs}$, and the vertical band indicates the region with $\text{SNR}(\text{LISA})\ge1$.}
\label{fig:gw}
\end{figure}

In Fig.~\ref{fig:gw}, we present the scatter plots of $\etab$ as a function of the SNR at \texttt{LISA} with an acquisition time of $4$ years and a duty cycle of $75\%$. More information about the computation of the GW spectra can be found in the \texttt{BSMPT}v3 manual~\cite{Basler:2024aaf}. The horizontal red line corresponds to the observed BAU, $|\etab| = 1$, and the vertical red-colored region has $\text{SNR}>1$, which will, in principle, be detectable by \texttt{LISA}.
As can be inferred from the left plot, for the C2HDM Type I and a wall velocity of $v_w=0.1$, we find parameters that match the measured BAU but have a SNR too low to be detectable by \texttt{LISA}. For the wall velocities of $v_w=0.5$ and $v_w=0.9$, the observed BAU is never matched but in both cases there are parameter combinations that lead to detectable GWs. For the C2HDM Type II, shown in the right plot, the results are much less interesting with a measured BAU of at least two orders of magnitude below the observed value and GW signals four six orders of magnitude below the region with detectable signals. The reason is that the experimental constraints on the C2HDM Type II at $T=0$ severly limit the valid C2HDM parameter space.

\subsection{Benchmark Points}
\label{sec:BP}
In this section, we select some sample benchmark points to perform a dedicated analysis of the impact of the VEV profiles, the $z$ dependence of the top quark mass and the CP-violating phase as well as the applied number $n$ of transport equations. The selected four benchmark points {are given for the C2HDM Type I and distinguish themselves by the following attributes:  
\begin{itemize}
    \item[$\diamond$] \texttt{BP1} - C2HDM example point (shipped with \texttt{BSMPT}).
    \item[$\diamond$] \texttt{BP2} - Largest $L_w T_p$.
    \item[$\diamond$] \texttt{BP3} - Largest $|\overline{\eta}_{2}|$.
    \item[$\diamond$] \texttt{BP4} - Second largest $|\overline{\eta}_{2}|$, with the peculiarity of having a rather low $T_p$.
\end{itemize}
Their input parameters are provided in Tab.~\ref{tab:bp_input} together with some transition parameters of interest. We considered the transition transition temperature to be the percolation temperature, i.e.~$T_* = T_p$, set the wall velocity to $v_w = 0.5$ and applied the constant truncation scheme with $R = -v_w$. For the VEV we applied the kink profile. The charge-breaking vacuum  $\omega_\text{CB}$ was zero for all cases. \s

\begin{table}[h!]
\centering
\begin{tabular}{lrrrrl}\toprule
\textbf{Parameter} & \texttt{BP1} & \texttt{BP2} & \texttt{BP3} & \texttt{BP4} & \textbf{Units}\\
\cmidrule(){1-6}
$\lambda_1$ & $3.29771$ & $0.027295258$ & $1.26844975$ & $0.1842546$ & ---\\
$\lambda_2$ & $0.274365$ & $0.24094447$ & $0.24834156$ & $0.24259527$ & ---\\
$\lambda_3$ & $4.71019$ & $6.8791583$ & $8.52502896$ & $8.9452122$ & ---\\
$\lambda_4$ & $-2.23056$ & $-2.8847011$ & $-3.20054852$ & $-3.2139652$ & ---\\
$\mathrm{Re}(\lambda_5)$ & $-2.43487$ & $0.40229390$ & $-2.28428068$ & $1.7102148$ & ---\\
$\mathrm{Im}(\lambda_5)$ & $0.124948$ & $0.56905137$ & $-0.79304619$ & $0.4602525$ & ---\\
$\mathrm{Re}(m_{12}^2)$ & $2706.86$ & $17318.771$ & $10768.765$ & $19800.912$ & $[\text{GeV}]^2$\\
$\tan\beta$ & $4.64487$ & $16.004550$ & $17.178081$ & $15.37630$ & ---\\
\cmidrule(){1-6}
$\omega_1(-\infty)$ & $52.033$ & $13.910$ & $14.444$ & $15.954$ & $[\text{GeV}]$\\
$\omega_2(-\infty)$ & $208.46$ & $162.21$ & $249.072$ & $247.47$ & $[\text{GeV}]$\\
$\omega_\text{CP}(-\infty)$ & $-0.70255$ & $-4.0745$ & $-0.376$ & $0.074601$ & $[\text{GeV}]$\\
\cmidrule(){1-6}
$T_c$ & $143.71$ & $152.22$ & $167.963$ & $142.74$ & $[\text{GeV}]$\\
$T_p$ & $134.83$ & $150.82$ & $97.923$ & $51.172$ & $[\text{GeV}]$\\
$L_w$(kink) & $0.021589$ & $0.043908$ & $0.014930$ & $0.014135$ & $[\text{GeV}]^{-1}$\\
$\theta_\text{sym (top)}$ & $0.067913$ & $0.055252$ & $-0.038517$ & $0.039267$ & ---\\
$\theta_\text{brk (top)}$ & $0.003370$ & $0.0251129$ & $0.001510$ & $-0.000301$ & ---\\
\bottomrule
\end{tabular}
\caption{\texttt{BSMPT} input parameters for the C2HDM Type I benchmark points \texttt{BP1}, \texttt{BP2}, \texttt{BP3} and \texttt{BP4}. We also display the values of the electroweak VEVs $\omega_{1,2}$ and the CP-violating VEV $\omega_{\text{CP}}$ in the true vacuum at the percolation temperature. We give the critical and the percolation temperatures $T_c$ and $T_p$, respectively, the  bubble wall width $L_w$ obtained from Eq.~(\ref{eq:lwcalc}) and the top mass phase in the false vacuum, $\theta_\text{sym (top)}$,  and in the true vacuum, $\theta_\text{brk (top)}$. We furthermore have $T_* = T_p$, $v_w=0.5$, $R=- v_w$ and use the kink profile for the VEV.}
\label{tab:bp_input}
\end{table}

\begin{figure}[h!]
	\centering
    \vspace*{0.3cm}
\includegraphics[width=0.98\textwidth]
{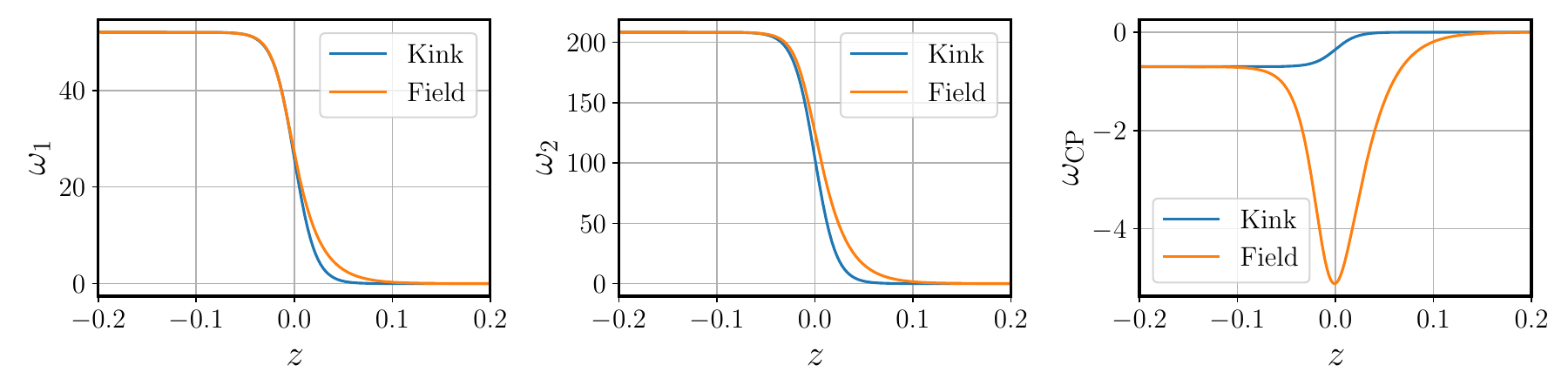}
\vspace*{-0.2cm}
	\caption{\texttt{BP1}: VEV profiles for the kink (blue) and the field (orange) solution as a function of $z$ for $\omega_1(z)$ (left), $\omega_2(z)$ (middle) and $\omega_\text{CP}(z)$ (right).}
\label{fig:bp1:vevprofile}
\end{figure}

In Fig.~\ref{fig:bp1:vevprofile} we display the VEV profiles obtained from both the kink solution (blue) and from the solution of the EOMs (orange) as a function of $z$, the direction perpendicular to the bubble wall. We show the electroweak VEVs $\omega_1$ (left), $\omega_2$ (middle), and $\omega_{\text{CP}}$ (right). As can be inferred from the left and the middle plots, for the electroweak VEVs we  have  rough agreement of the kink and the field solution, although it appears that the field solution has a broader transition region between the false vacuum and the true vacuum, i.e.~a slightly larger effective wall width $L_w$. For $\omega_\text{CP}$ (right plot), the results are clearly different, however. While for the kink solution we have the ``usual" kink-like transition region, for the field solution $\omega_\text{CP}$ reaches almost $\omega_\text{CP}\simeq-5$ inside the bubble wall before arriving at its true vacuum value, $\omega_\text{CP}\simeq-0.7$. It is also evident from the $\omega_\text{CP}$ plot that the effective wall width $L_w$, estimated from the width of the transition region, is now much larger than for the  kink profile where $L_w=0.02$. This suggests that the choice of the same wall width $L_w$ for all VEV profiles does not correctly reproduce the actual situation, namely that each VEV profile may have different length scales. Considering different length scales for each VEV would also remove the need for an independent way of calculating the $\theta$ profiles for the fermion masses, as explained in Sec.~\ref{sec:bubbleprofile}. \s

\begin{figure}[h!]
	\centering
\includegraphics[width=0.49\textwidth]{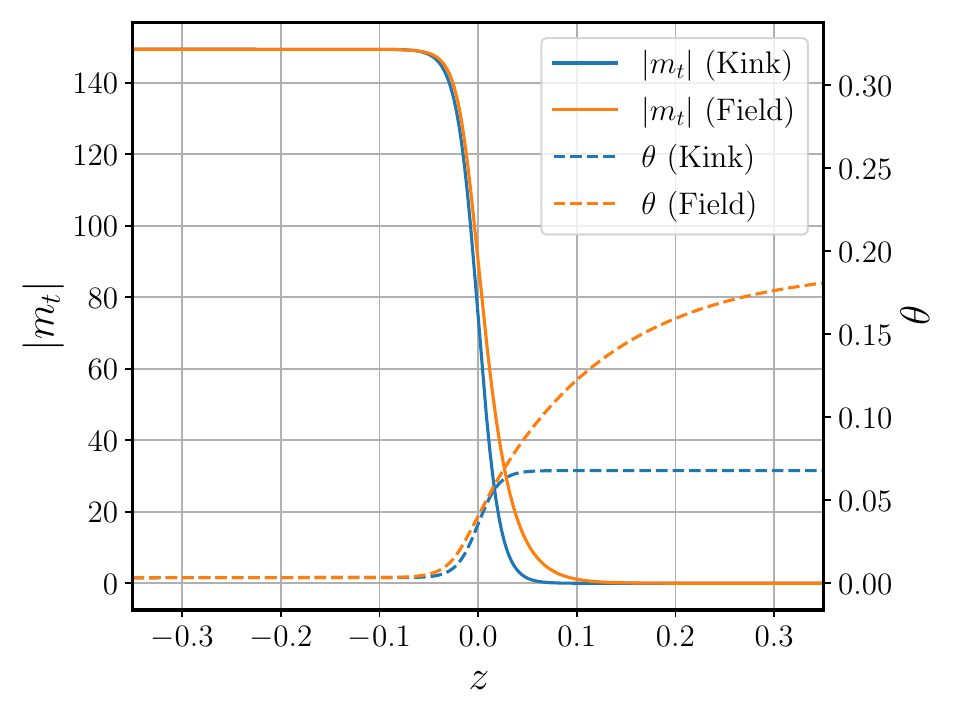} 
\includegraphics[width=0.49\textwidth]{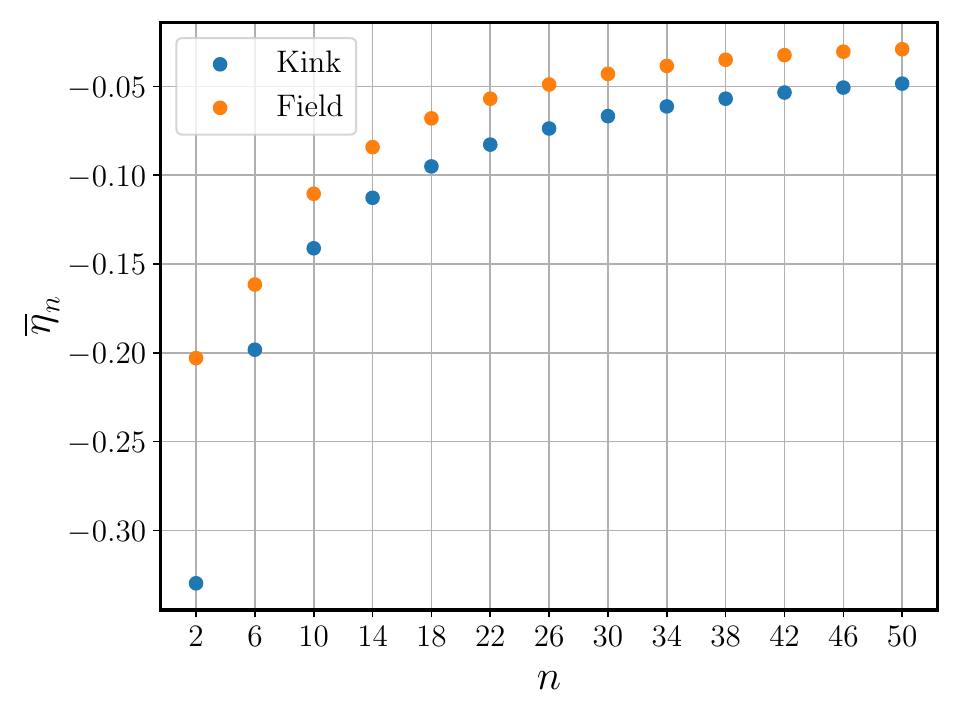} 
	\caption{\texttt{BP1}: Left panel: absolute value of the top mass $|m_t|$ (solid lines) and the top mass phase $\theta$ (dashed lines) as a function of $z$ for the kink solution (blue) and the field solution (orange); right panel: normalized baryon asymmetry $\overline\eta$ as a function of the number $n$ of transport equations for the kink solution (blue) and the field solution (orange).}
\label{fig:bp1:massangle}
\end{figure}

Figure~\ref{fig:bp1:massangle}, shows on the left panel the absolute value of the top mass $|m_t|$ (solid lines) and the top mass phase $\theta$ (dashed lines) at the transition temperature $T_p$ for the kink solution (blue) and the field solution (orange) as a function of $z$ for \texttt{BP1}. We can see that there is an overall agreement between the absolute values of the top mass but a disagreement between the phases $\theta$ in the false vacuum. In the false vacuum the phase $\theta$ is ill-defined, because $\omega_2=\omega_\text{CP}=0$, so its value will depend on the direction from which we approach the false vacuum.  This phase disagreement is not concerning, however, as the source terms for the transport, Eq.~(\ref{eq: dimensionless_source}), vanishes when $(m_t^2)'\to 0$, i.e.~in the region of the disagreement between the phases $\theta$. We note that the peculiar shape of the $\omega_{\text{CP}}$ profile for the field solutions is not visible in the $\theta$ shape, as the top mass (calculated from Eq.~(\ref{eq:topvaluecalc})) is dominated by the $\omega_2$ value, which is much larger than $\omega_{\text{CP}}$. \s

In the right plot we show the normalized baryon asymmetry computed for different numbers $n$ of the transport questions, starting with $n=2$ and increasing up to $n=50$ in steps of four, $n = 2+4k,\,k\in \mathbb{N}$, both for the kink solution (blue) and the field solution (orange) as functions of $z$. We can see in this plot that as we increase $n$ the BAU seems to converge to a fixed value, although higher $n$ are needed before a definite conclusion can be drawn. We can also see that there is a rough agreement in the dependence of the kink solution and the field solution on $n$, with the field solution overall producing a smaller BAU. This is expected as the field solution has a larger effective $L_w$.

\begin{figure}[h!]
	\centering
\includegraphics[width=0.98\textwidth]{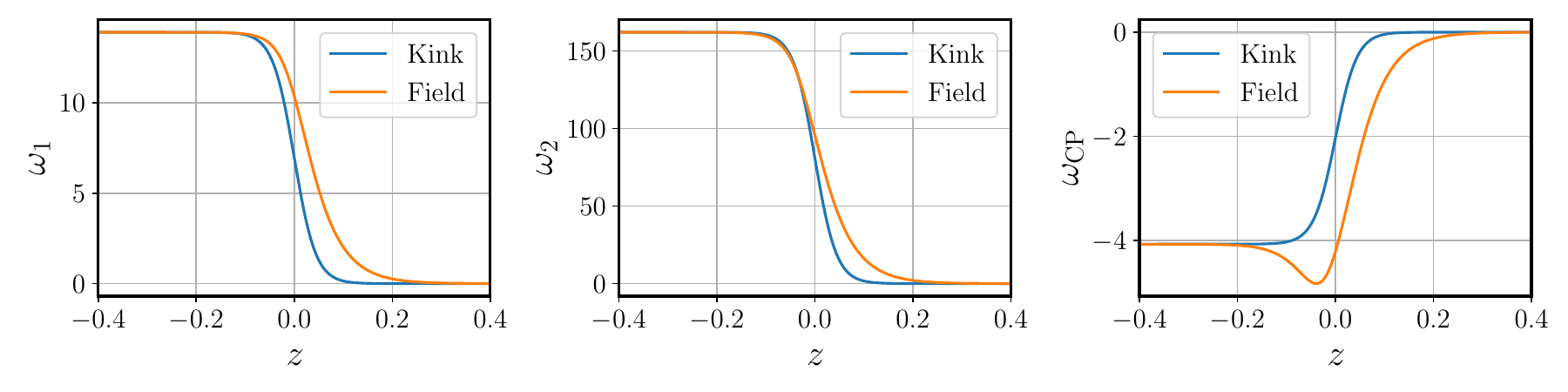}  
\vspace*{-0.3cm}
    \caption{Same as Fig.~\ref{fig:bp1:vevprofile} but for \texttt{BP2}.}
\label{fig:bp2:vevprofile}
\end{figure}

In Fig.~\ref{fig:bp2:vevprofile}, we show the VEV profiles for \texttt{BP2} as a function of $z$ again for the two different VEV profiles. There is a general agreement for the two choices in case of $\omega_1$ and $\omega_2$. 
For $\omega_\text{CP}$ both approaches differ again, although not as extremely as for \texttt{BP1}. \s

\begin{figure}[t!]
	\centering
\includegraphics[width=0.49\textwidth]{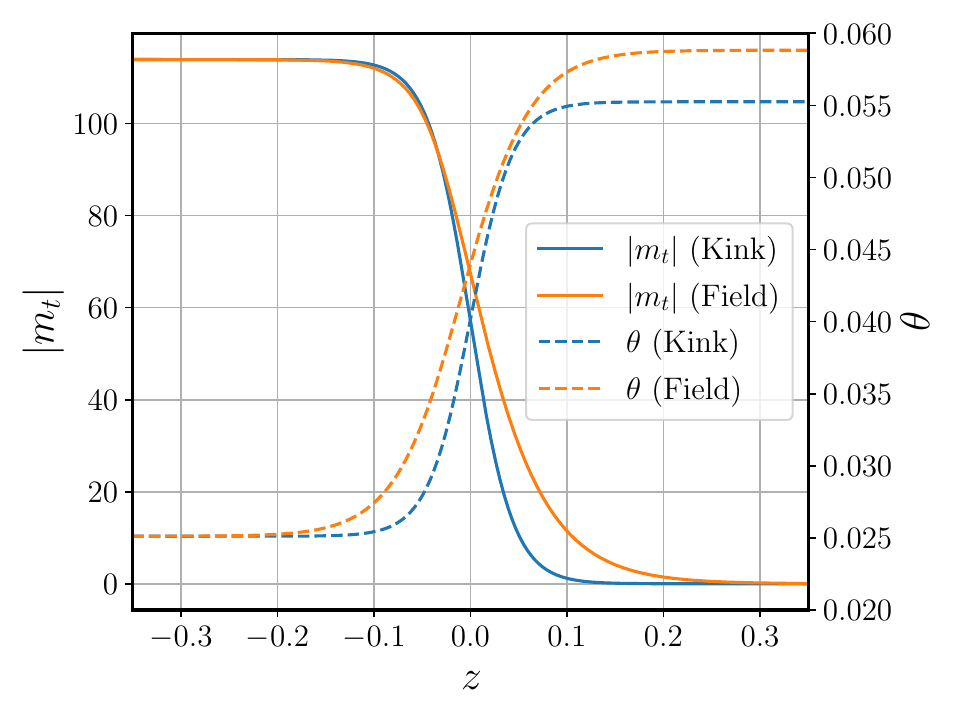}  
\includegraphics[width=0.49\textwidth]{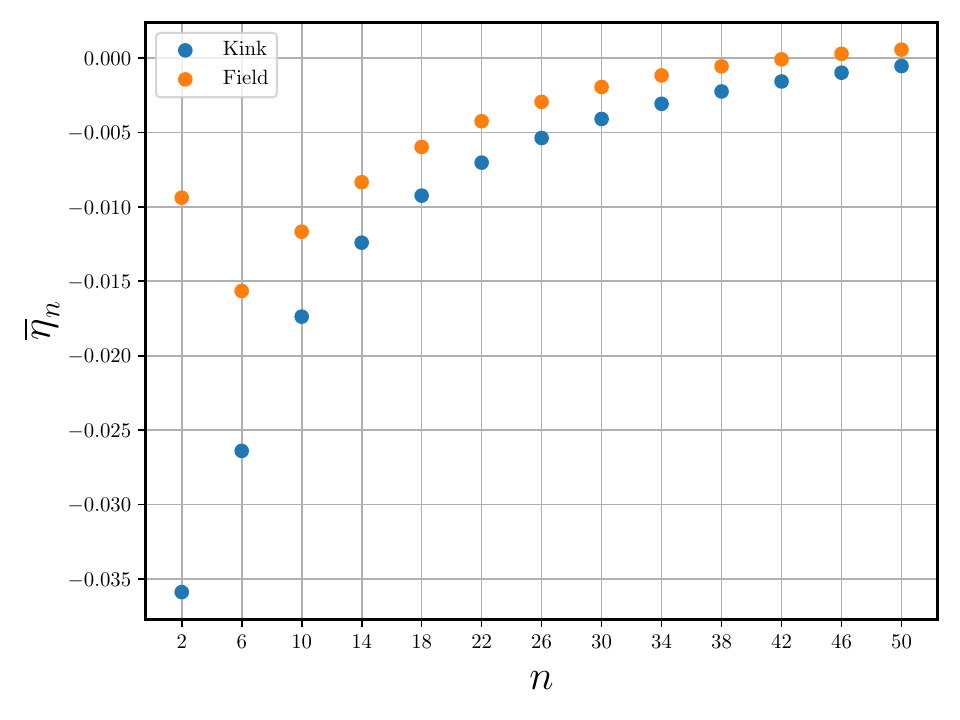} 
\vspace*{-0.3cm}
	\caption{Same as Fig.~\ref{fig:bp1:massangle}, but for \texttt{BP2}.}
\label{fig:bp2:massangle}
\end{figure}

Looking at Fig.~\ref{fig:bp2:massangle} (left), we see that for this benchmark point \texttt{BP2} we have good agreement between the absolute values of the top mass $|m_t|$ in the two VEV approaches. Again the phase $\theta$ shows differences, which now are also more pronounced in the transition region, which is the relevant part for the BAU. In the right panel, we show the $\overline\eta$ as a function of the truncation moment $n$. From Fig.~\ref{fig:bp2:massangle} (right), we can infer that the absolute value of the baryon asymmetry decreases as we increase $n$ (apart from $n=2$ to $n=6$ for the field profile) until, for the field profile, it changes sign to positive values at $n=46$ and $n=50$, a case that we alluded to already above when discussing the sign of the BAU.

\begin{figure}[h!]
	\centering
\includegraphics[width=0.98\textwidth]{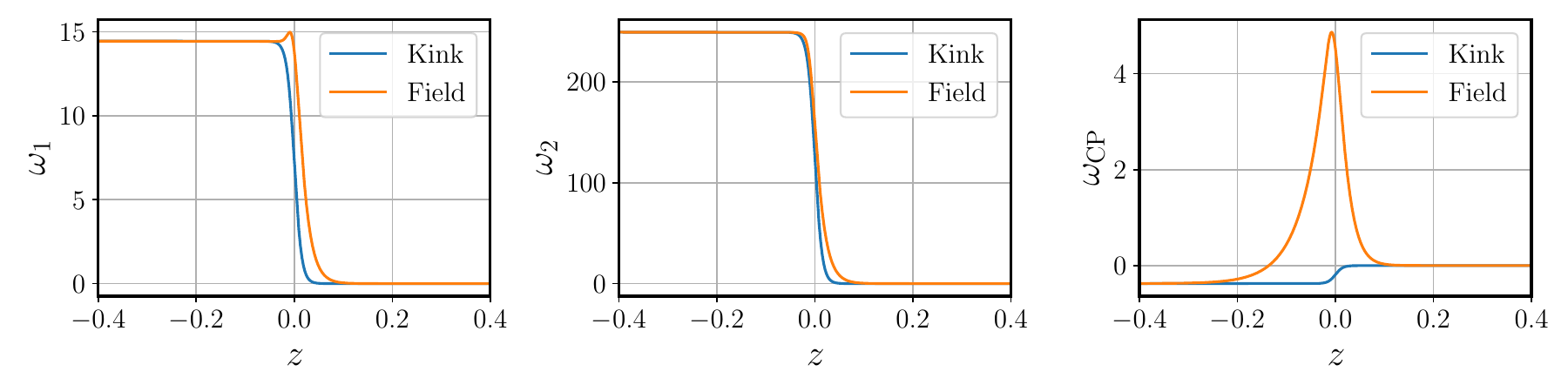}  
\vspace*{-0.3cm}
	\caption{Same as Fig.~\ref{fig:bp1:vevprofile} but for \texttt{BP3}.}
\label{fig:bp3:vevprofile}
\end{figure}

Looking at the VEV profiles for \texttt{BP3} if Fig.~\ref{fig:bp3:vevprofile}, we see that for $\omega_1$ there is a small bump for the field solution; for $\omega_2$ there is an overall agreement between the two VEV solutions although the field profile has a slightly wider profile; for $\omega_\text{CP}$ the differences are much more pronounced with the field solution having a ``bump" several times larger than the difference between the true and false vacuum, $|\omega_\text{CP}(z\to\infty)-\omega_\text{CP}(z\to-\infty)|$.

\begin{figure}[h!]
	\centering
\includegraphics[width=0.49\textwidth]{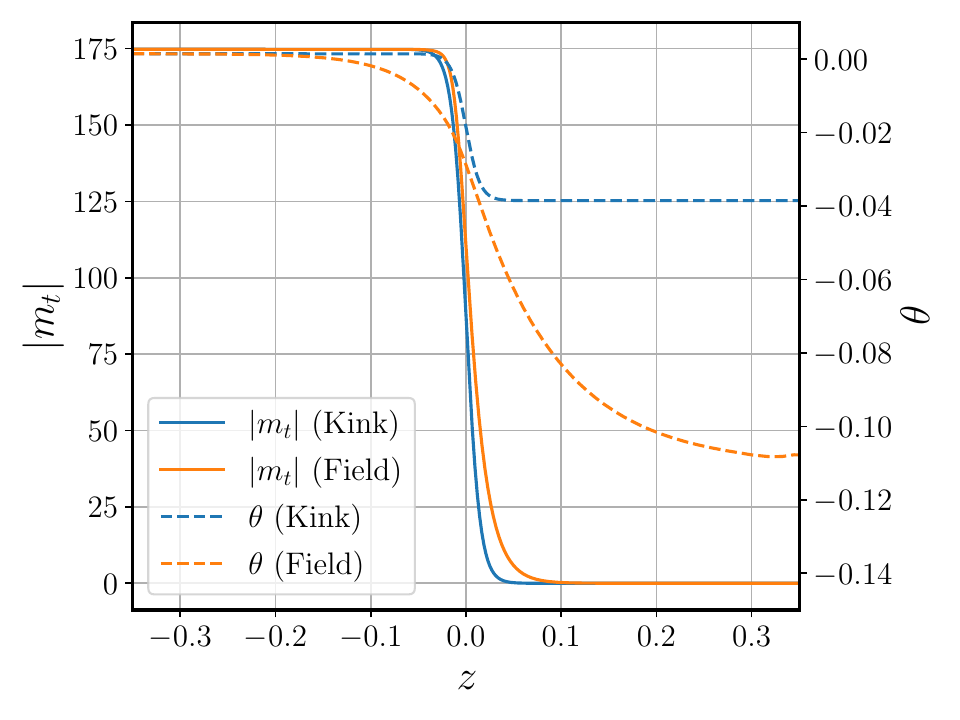}  
\includegraphics[width=0.49\textwidth]{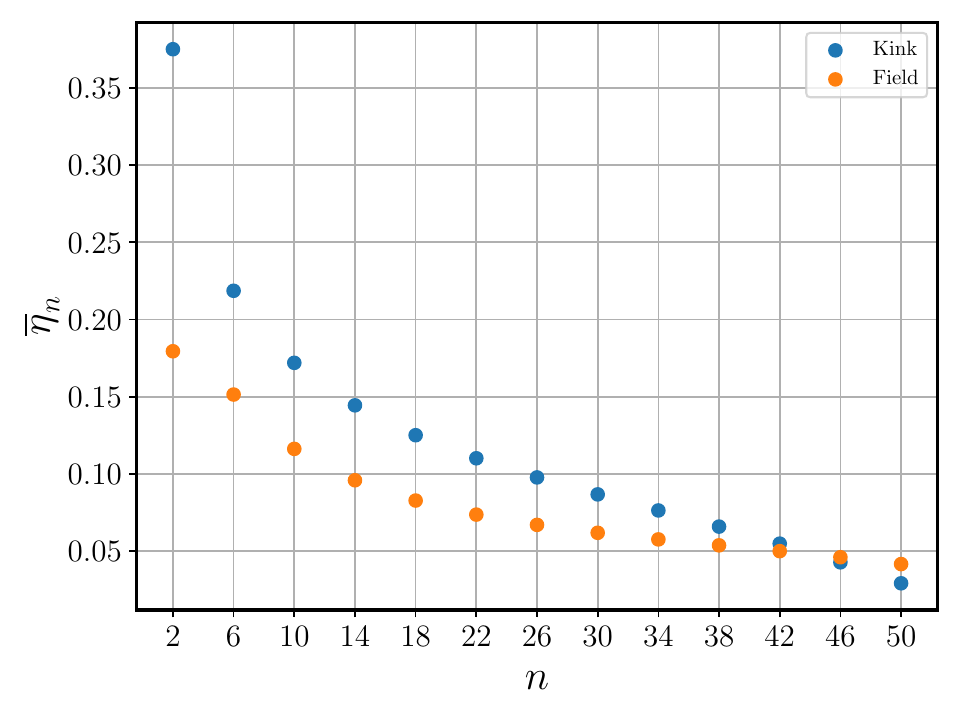} 
\vspace*{-0.3cm}
	\caption{Same as Fig.~\ref{fig:bp1:massangle}, but for \texttt{BP3}.}
\label{fig:bp3:massangle}
\end{figure}

Also for \texttt{BP3} the two top-quark mass profiles are similar for the two approaches, cf.~Fig.~\ref{fig:bp3:massangle} (left). The top phase differs quite significantly, however, in both the amplitude and the effective length scale over which is varies. This difference is not that relevant as $(m_t^2)'\simeq 0$ for a large part of the region where the phases differ and, consequently, the source terms vanishes. Regarding the $n$ dependence of $\bar{\eta}$ shown on the right plot, we can see that for both VEV profiles it decreases with increasing $n$. The solutions furthermore change the hierarchy at $n=46$ and $n=50$, so that now the field profile produces a larger $\overline\eta$. \s

\begin{figure}[h!]
	\centering
    \vspace*{0.4cm}
\includegraphics[width=0.49\textwidth]{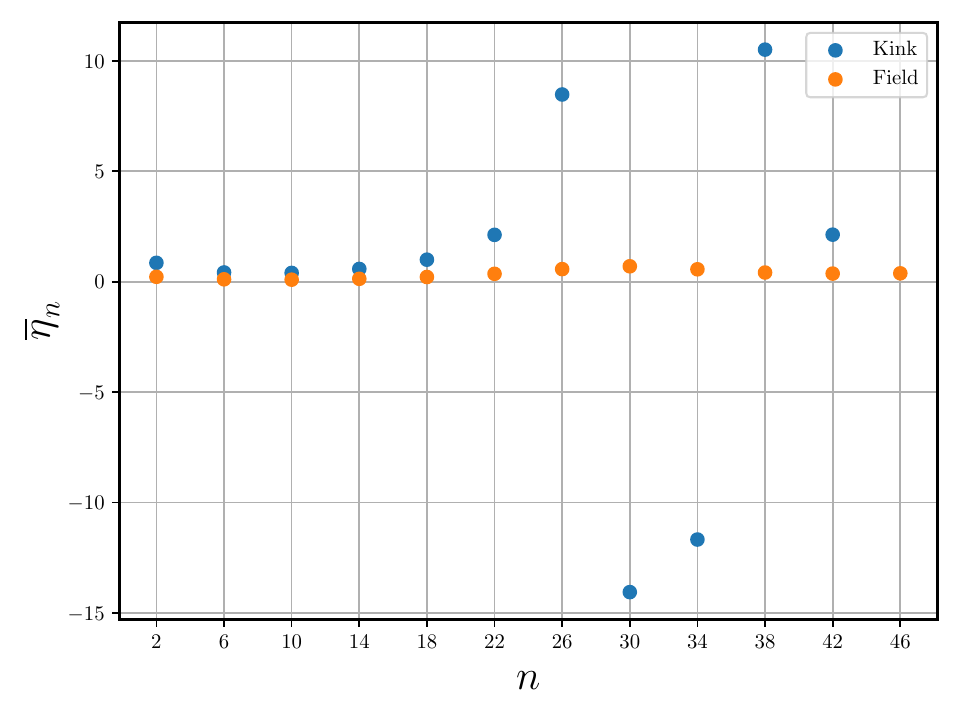} 
\includegraphics[width=0.49\textwidth]{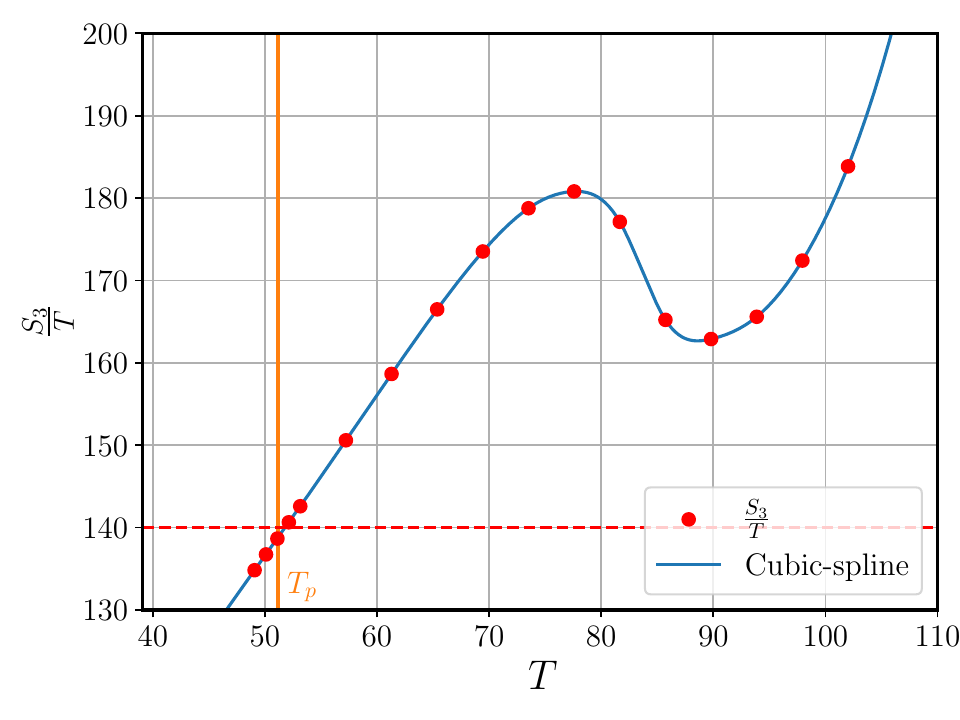}  
	\caption{\texttt{BP4}: Left panel: $\overline\eta$ as a function of the number $n$ of moment equations for the kink solution (blue) and the field solution (orange). right panel: the Euclidian action $\frac{S_3}{T}$ as a function of the temperature, where the red dots are the action calculated by \texttt{BSMPT}v3, the blue line is the cublic spline used to interpolate the action, the horizontal dashed red line is $\frac{S_3}{T}=140$ (which can be used to estimate the nucleation temperature) and the vertical orange line is the percolation temperature.}
\label{fig:BP4}
\end{figure}

In Fig~\ref{fig:BP4}, we show for \texttt{BP4} on the left panel the normalized baryon asymmetry $\overline\eta$ as a function of the number $n$ of moment equations for both the kink and the field solution. We can see that the results obtained from the kink solution $\overline\eta$ are much more erratic compared to those from the field solution. Actually, since the latter has a larger effective $L_w$ we expect more stable results. On the right panel, we plot the Euclidian action $\frac{S_3}{T}$ as a function of the temperature which is used to calculate the tunneling rate. A lower $\frac{S_3}{T}$ produces a larger decay rate with $\frac{S_3}{T}\simeq140$ being a rough way of approximating the nucleation temperature. From the plot, we can infer that the decay rate is increasing from 
the critical temperature down to $T=90 \text{ GeV}$, then decreasing until $T=77.6 \text{ GeV}$ and after that increasing again until we reach the percolation temperature at $T_p = 51.2 \text{ GeV}$. This ``back-and-forth" makes the transition occur at a rather low temperature which boosts the value of $\etab$. This behaviour is much more relevant for lower wall velocities. In Fig.~\ref{fig:typeI_eta_vs_TcTp}, for $v_w=0.1$ and $v_w=0.5$, the left most points which have the lowest percolation temperature, tend to have the largest $\etab$. We verified that also for the other points with low $T_p$ the action showed a ``back-and-forth" behaviour. This discussion also shows the importance of calculating the action from the bounce solution and not applying approximations as it has an impact on the obtained value for the baryon asymmetry. \s

\section{Conclusions \label{sec:concl}}
In this paper, we presented our new implementation of the computation of the baryon asymmetry in the code \texttt{BSMPT}. It is based on the WKB ansatz and generalizes the transport equations systematically to an arbitrary number of moments, as recently developed in Ref.~\cite{Kainulainen:2024qpm}. We properly implemented the boundary conditions and included two different truncation schemes in the solution of the transport equations. The VEV profile, in addition to the kink profile, can now be computed also from the solution of the equations of motion. For the top-Yukawa rate, we used our own new computation at non-zero temperature. \s

We validated the implementation by comparing our results with the literature for a simple benchmark model. We found a reduction in the dependence on the truncation scheme for an increasing number $n$ of moment equations. Furthermore, a larger wall velocity and, as expected, since it is the prerequisite for the validity of the WKB approach, larger values of $L_w T_n$ are beneficial for the convergence of the result. We found that $L_w T_n \gtrsim 50$ is required for meaningful results, in contrast to the literature where $L_w T_n \gtrsim 2$ is quoted. Unfortunately, however, the baryon asymmetry decreases with higher values of both $v_w$ and $L_w T_n$. An increase in the strength of the phase transition turns out to destabilize the results further.
Let us note that in order to clarify whether the results truly converge with larger $n$, one would require going beyond the implemented $n=50$ moment equations. \s

We then moved on to a detailed discussion of the uncertainties in the computation of the baryon asymmetry. We did this within the framework of the C2HDM Type I and II for a representative set of parameter points obtained from a scan in the parameter space of the model taking into account all relevant theoretical and experimental constraints. Barring a change in the collision term, however, the calculation can readily be applied to any extended Higgs sector model with one or more VEV directions. For the C2HDM Type I, we found parameter points that lead to the correct baryon asymmetry, but not for Type II, where the parameter space is more constrained. In accordance with the findings in the benchmark model the uncertainties decrease with increasing number of moment equations and with the parameter $L_w T_p$, which, however, also reduces the baryon asymmetry. We did not find parameter points that lead to $L_w T_p$ values greater than 7 which introduces a large source of uncertainty, as we found that values above 50 are required for sufficiently stable results. Lower transition temperatures and wall velocities on the other hand lead to a larger baryon asymmetry. This underlines the importance of the proper choice of the transition temperature, the consistent computation of the wall velocity, and the derivation of the wall width. The VEV profile has a significant impact on the baryon asymmetry. We found that for most points the kink profile leads to larger asymmetries than the profile deduced from the solution of the equations of motion. However, this also depends on the maximum number of moment equations that is used. \s

Despite these uncertainties, the impact of the amount of CP violation on the baryon asymmetry is clearly visible. A larger value of the input parameter $\mathrm{Im} (m_{12}^2)$ leads to a larger baryon asymmetry. The same is true for an increase in the CP-violating VEV, that can be generated spontaneously at non-zero temperature. Both the baryon asymmetry and the gravitation waves signal from first-order phase transitions are sensitive to the wall velocity with, however, opposite dependencies, such that we found for the C2HDM Type I parameter points at large velocities that lead to detectable GW signals at \texttt{LISA} but to too low baryon asymmetry and vice versa. For the C2HDM Type II, none of the points lead to detectable GW signals. \s

We finally chose four benchmark points to discuss in more detail the role of the VEV profile. We can summarize this investigation by stating that the choice of the VEV profile from the solution of the equations of motion can have a significant impact on the shape of the VEV as function of the wall distance $z$ and in particular shows, that the simple assumption of taking the same bubble width in all VEV directions cannot be maintained. It has a smaller impact on the absolute value of the top mass and a more pronounced one on its CP-violating phase. This has an effect only, however, if the phase changes its shape as function of $z$. The precise details depend on the chosen parameter point. The profile choice impacts the generated amount of baryon asymmetry, where in general the field profile was found to lead to smaller values than the kink profile. We also found the opposite behaviour, however, depending on the number $n$ of transport equations. We furthermore found the field profile to be more stable than the kink profile which is to be expected as it is accompanied by larger effective wall widths. \s

Our new implementation of the baryon asymmetry and detailed investigations of its limitations and dependencies give important insights and outline possible directions for future improvements. The role of the amount of CP violation for the value of the baryon asymmetry reveals itself very clearly, however, despite the uncertainties,  and gives guidance for future model building. Future refinements of the calculation of the baryon asymmetry will hence be crucial for the deduction of the parameters leading to the correct baryon asymmetry and for cornering the underlying model by applying cosmological criteria.

\section*{Acknowlegdements}
MM acknowledges support by the Deutsche Forschungsgemein- schaft (DFG, German Research Foundation) under grant 396021762 - TRR 257. JP is grateful to the Studienstiftung des Deutschen Volkes for financial support. JV and RS acknowledge financial support from the Portuguese Foundation for Science and Technology (FCT)  under contracts UID/PRR2/00618/2025 (https://doi.org/10.54499/UID/PRR2/00618/2025), UID/PRR/00618/2025. JV 
(https://doi.

\noindent
org/10.54499/UID/PRR/00618/2025), and UID/00618/2025 (https://doi.org/10.54499/UID/

\noindent
00618/2025). JV is also supported by FCT under contract PRT/BD/154191/2022 (https://doi.

\noindent
org/10.54499/PRT/BD/154191/2022).
We would like to thank Raphael Boto for providing us with a valid C2HDM parameter sample. We thank K.~Kainulainen and N.~Venkatesan for useful discussions.

\begin{appendix}
\newcommand{\tchannel}[6]{\raisebox{-0.5\height}{
\begin{tikzpicture}
  \begin{feynman}
    \vertex (a) at (0,2)   {$t_L$};
    \vertex (b) at (0,-2)  {$g$};
    \vertex (c) at (3,2)   {$h$};
    \vertex (d) at (3,-2)  {$t_R$};

    \vertex[dot, minimum size=2pt] (ac) at (1.5,1) {};
    \vertex[dot, minimum size=2pt] (bd) at (1.5,-1) {};

    \node[text=black] at (1.3, 1.5) {#1};
    \node[text=black] at (1.7, 1.5) {#2};
    \node[text=black] at (1.7, 0.8) {#3};

    \node[text=black] at (1.7, -0.8) {#4};
    \node[text=black] at (1.3, -1.5) {#5};
    \node[text=black] at (1.7, -1.5) {#6};

    \diagram*{
      (a) -- [fermion] (ac) -- [scalar] (c),
      (b) -- [gluon] (bd) -- [fermion] (d),
      (ac) -- [fermion] (bd),
    };

    \node at ($(ac)!0.5!(bd)$) [draw, circle, fill=gray!30, inner sep=3pt] {$G^R$};
  \end{feynman}
\end{tikzpicture}
}}

\section{Top Yukawa Rate $\Gamma_M$}\label{app: yuk}

\begin{figure}
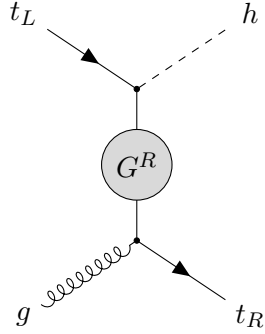

    \centering
    \tchannel{}{}{}{}{}{}
    \caption{Feynman diagram of the process that leads to the top Yukawa rate. $G^R$ denotes the self-energy corrected retarded propagator.}
\label{fig: yukawa_rate_diag}
\end{figure}

In this section, we want to provide an update of the top Yukawa rate $\hat{\Gamma}_y$. The first analytical estimate for this rate was provided in Ref.~\cite{Huet_1996} and since then has been used throughout the literature for nearly two decades. Recently, new estimates were published using the VIA method in Ref.~\cite{Cline_2021} giving $\hat{\Gamma}_y\approx3\cdot10^{-2}T$ and the leading log approximation in Ref.~\cite{Barni:2025ifb} giving $\hat{\Gamma}_y\approx10^{-2}T$. Here, we present our updated calculation where we include the full dispersion relation of the corrected fermion propagator as well as a full numerical integration of the collision operator. The diagram we want to evaluate is shown in Fig.~\ref{fig: yukawa_rate_diag}. Similar to the aforementioned references, we consider only the $t$-channel contribution.\s

To obtain the Yukawa rate, the self-energy of the retarded propagator has to be calculated in finite temperature field theory. We will do so for the massless case as was done in Ref.~\cite{PhysRevD.26.2789}, but without taking the Hard Thermal Loop (HTL) limit. For clarity, we will denote four-vectors of kinematical variables with capital letters and absolute values of the spatial components with small letters in this section. The self-energy $\Sigma$ for a massless fermion at zero temperature has the form 
\begin{equation}
    \Sigma_{T=0}=-a\slashed{K}\;,
\end{equation}
which results in a propagator correction given by $S=\slashed{K}/[(1+a)K^2]$, where the pole remains at $K^2=0$. At finite temperature the situation changes since we have an additional kinematical quantity to be considered, the plasma velocity $U_\mu$ with $U_\mu U^\mu=1$. We will perform our calculation in the plasma rest frame where $U^\mu=(1,0,0,0)$. With this in mind, the self-energy changes to
\begin{equation}
    \Sigma_T=-a\slashed{K}-b\slashed{U}\;,
\end{equation}
where $a$ and $b$ are Lorentz invariant functions, which depend on the energy $k_0$ and the momentum $k$. This changes the propagator structure in a non-trivial way to
\begin{equation}
    S(K)=\frac{(1+a)\slashed{K}+b\slashed{U}}{(1+a)^2K^2+2(1+a)bk_0+b^2}\;.
\end{equation}
The pole of this structure can be determined by solving the equation 
\begin{equation}
    k_0(1+a)+b=k(1+a)\,,
\end{equation}
which in the HTL limit gives $k_0=m_f$ as $k\rightarrow0$, where $m_f$ is the thermal mass given by~\cite{PhysRevD.26.2789}
\begin{equation}
    m_f^2\equiv\frac{g^2C_2}{8}T^2
\end{equation}
and $C_2$ is the quadratic Casimir invariant. The thermal mass was used in the previous results to obtain the top Yukawa rate, where it acted as a regulator mass in the propagator. In the more general case, we obtain the coefficient functions via the projections
\begin{equation}
    t_K\equiv\frac{1}{4}\text{tr}(\slashed{K}i\Sigma_T)=-aK^2-bk_0\quad\text{and}\quad t_U\equiv\frac{1}{4}\text{tr}(\slashed{U}i\Sigma_T)=-ak_0-b\;,\label{eq: trace_projectors}
\end{equation}
which results in
\begin{equation}
    a=\frac{1}{k^2}(t_K-t_Uk_0)\quad\text{and}\quad b=\left(\frac{k_0^2}{k^2}-1\right)t_U-\frac{k_0}{K^2}t_K\;.
\end{equation} 
The self-energy for the retarded fermion propagator with an arbitrary gauge boson in the loop is given by
\begin{align}
i\Sigma^R(K)&=\quad
{\raisebox{-0.28\height}{
\begin{tikzpicture}
  \begin{feynman}
    \vertex (i) at (0,0);
    \vertex (a) at (5,0);
    \vertex (b) at (10,0);
    \vertex (f) at (15,0);
    \diagram*{
      i -- [fermion] a -- [fermion] b -- [fermion] f,
      a -- [gluon, half left, looseness=2]  b,
    };
    \node[text=black] at (0.7,-0.2) {$cl$};
    \node[text=black] at (1.2,-0.2) {$cl$};
    \node[text=black] at (0.7,0.2) {$q$};
    \node[text=black] at (1.8,-0.2) {$cl$};
    \node[text=black] at (2.3,0.2) {$cl$};
    \node[text=black] at (2.3,-0.2) {$q$};
  \end{feynman}
\end{tikzpicture}}}
\quad
+
\quad
{\raisebox{-0.28\height}{
\begin{tikzpicture}
  \begin{feynman}
    \vertex (i) at (0,0);
    \vertex (a) at (5,0);
    \vertex (b) at (10,0);
    \vertex (f) at (15,0);
    \diagram*{
      i -- [fermion] a -- [fermion] b -- [fermion] f,
      a -- [gluon, half left, looseness=2]  b,
    };
    \node[text=black] at (0.7,-0.2) {$cl$};
    \node[text=black] at (1.2,-0.2) {$q$};
    \node[text=black] at (0.7,0.2) {$cl$};
    \node[text=black] at (1.8,-0.2) {$cl$};
    \node[text=black] at (2.3,0.2) {$cl$};
    \node[text=black] at (2.3,-0.2) {$q$};
  \end{feynman}
\end{tikzpicture}}}
\nonumber\\
&=-\frac{1}{2}ig^ 2C_2\int\frac{\text{d}^4P}{(2\pi)^4}G_{\mu\nu}^R(P)\gamma^\mu S^K(P-K)\gamma^\nu+G_{\mu\nu}^K(P)\gamma^\mu S^R(P-K)\gamma^\nu\;,
\end{align}
with loop momentum $P$. Inserting the expressions for the propagators and extracting the temperature-dependent contributions we obtain
\begin{align}
    i\Sigma^R_T(K)=&-g^2C_2\int\frac{\text{d}^4P}{(2\pi)^3}(\slashed{P}-\slashed{K})\bigg[\frac{-2n_F(p_0-k_0)}{P^2+i\text{sgn}(p_0)\epsilon}\delta\left((P-K)^2\right)\nonumber\\
    &+\frac{2n_B(p_0)}{(P-K)^2+i\text{sgn}(p_0-k_0)\epsilon}\delta(P^2)\bigg]\;,
\end{align}
where $n_{F/B}(x)\equiv f_\text{FD/BE}(|x|)$ (see Eq.~(\ref{eq: fermi/bose})). After performing the transformation $P\rightarrow P+K$ on the first term and $P\rightarrow-P$ on the second term the expression simplifies to
\begin{align}
    i\Sigma^R_T(K)&=2g^2C_2\int\frac{\text{d}^4P}{(2\pi)^3}\frac{\delta(P^2)}{(P+K)^2+i\text{sgn}(p_0+k_0)\epsilon}\left[(\slashed{P}+\slashed{K})n_B(p_0)+\slashed{P}n_F(p_0)\right]\nonumber\\
    &=g^2C_2\int\frac{\text{d}p\text{d}x}{(2\pi)^2}\frac{p}{(P+K)^2+i\text{sgn}(p_0+k_0)}\left[(\slashed{P}+\slashed{K})n_B(p_0)+\slashed{P}n_F(p_0)\right]\bigg|_{p_0=\pm p}\;,\label{eq: retarded_fermion_correction}
\end{align}
where $x\equiv\text{cos}\theta$ and $\theta$ is the angle between the three-momentum vectors $\mathbf{p}$ and $\mathbf{k}$. Next, we apply the projectors defined in Eq.~(\ref{eq: trace_projectors}) on Eq.~(\ref{eq: retarded_fermion_correction}) as well as the identity 
\begin{equation}
\lim_{\epsilon \to 0^+}\frac{1}{x\pm i\epsilon}=\mp i\pi\delta(x)+\mathcal{P}\left(\frac{1}{x}\right),\label{eq: cauchy_princip}
\end{equation}
where $\mathcal{P}$ is the Cauchy principal value to obtain
\begin{align}
    \text{Re}(t_K)&=\frac{g^2C_2}{8\pi^2}\int\text{d}p\left[\left(4p+\frac{K^2}{2k}L_1(P)\right)n_B+\left(4p-\frac{K^2}{2k}L_1(P)\right)n_F\right]\;,\nonumber\\
    \text{Im}(t_K)&=\frac{g^2C_2K^2}{16\pi k}\int_\frac{|k_0-k|}{2}^\frac{k_0+k}{2}\text{d}p(n_F-n_B)\;,\nonumber\\
    \text{Re}(t_U)&=\frac{g^2C_2}{8\pi^2k}\int\text{d}p\left[(pL_2(P)+k_0L_1(P))n_B+pL_2(P)n_F\right]\;,\nonumber\\
    \text{Im}(t_U)&=\frac{g^2C_2}{8\pi k}\begin{cases}
        \int_{\frac{k_0-k}{2}}^\frac{k_0+k}{2}\text{d}p\left[(p-k_0)n_B+pn_F\right] & k_0>k\\
        \int_{\frac{k_0+k}{2}}^\infty\text{d}p\left[(p-k_0)n_B+pn_F\right]+\int_{\frac{k-k_0}{2}}^\infty\text{d}p\left[(p+k_0)n_B+pn_F\right] & k_0<k
    \end{cases}\;
\end{align}
with $L_1(P)$ and $L_1(P)$ given by 
\begin{align}
    L_1(P)&\equiv\text{log}\left|\frac{(k-2p)^2-k_0^2}{(k+2p)^2-k_0^2}\right|\quad,\quad L_2(P)\equiv\text{log}\left|\frac{(k_0+k)^2((k_0-k)^2-4p^2)}{(k_0-k)^2((k_0+k)^2-4p^2)}\right|\;.
\end{align}
The real parts of the above expressions agree with the results obtained in Ref.~\cite{Peshier_1998}, in which the imaginary time formalism was used. However, for the Im$(t_U)$ our result differs in the $k_0<k$ regime by an overall minus sign.
\begin{figure}
    \centering
    \includegraphics[width=0.49\linewidth]{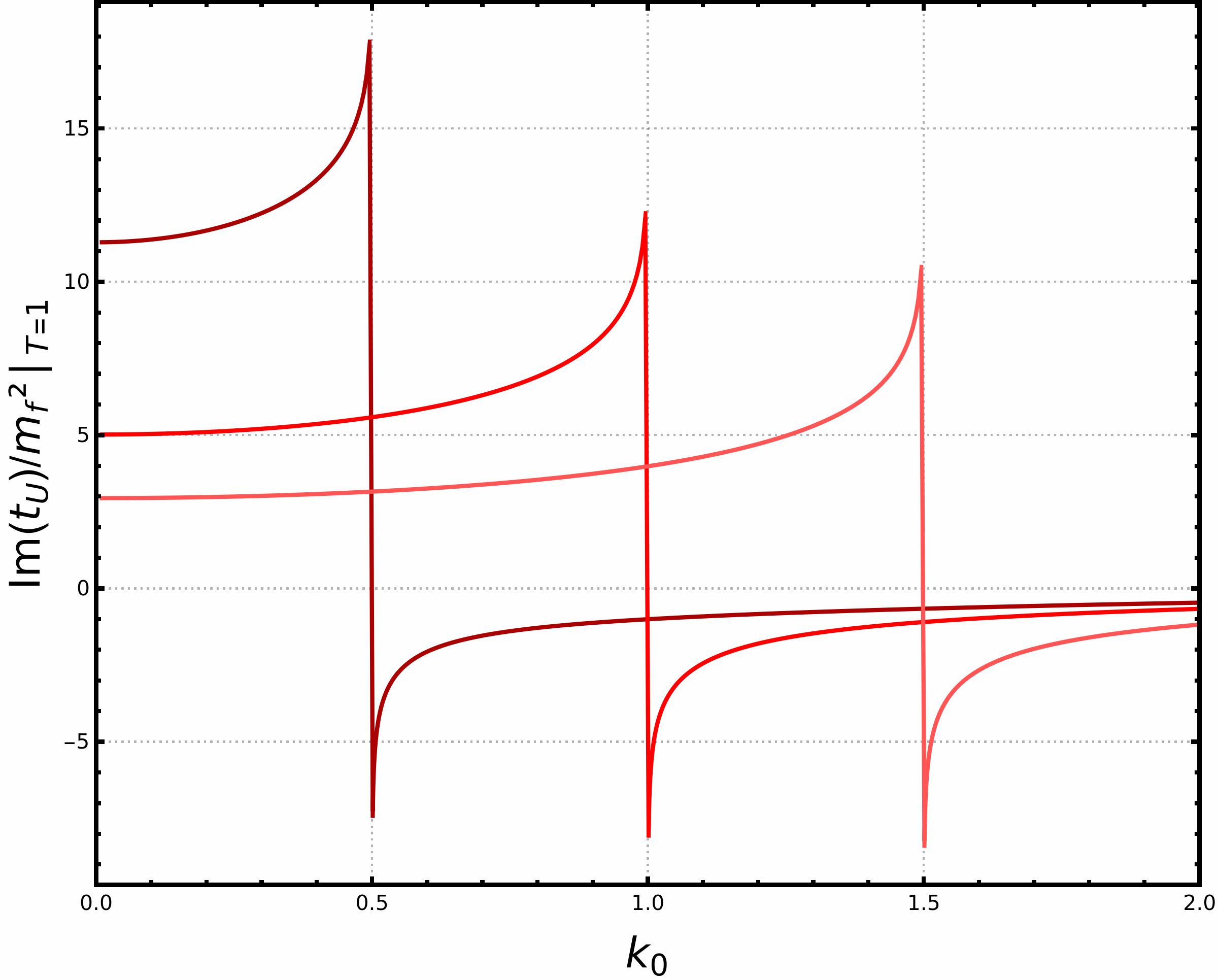}
    \includegraphics[width=0.49\linewidth]{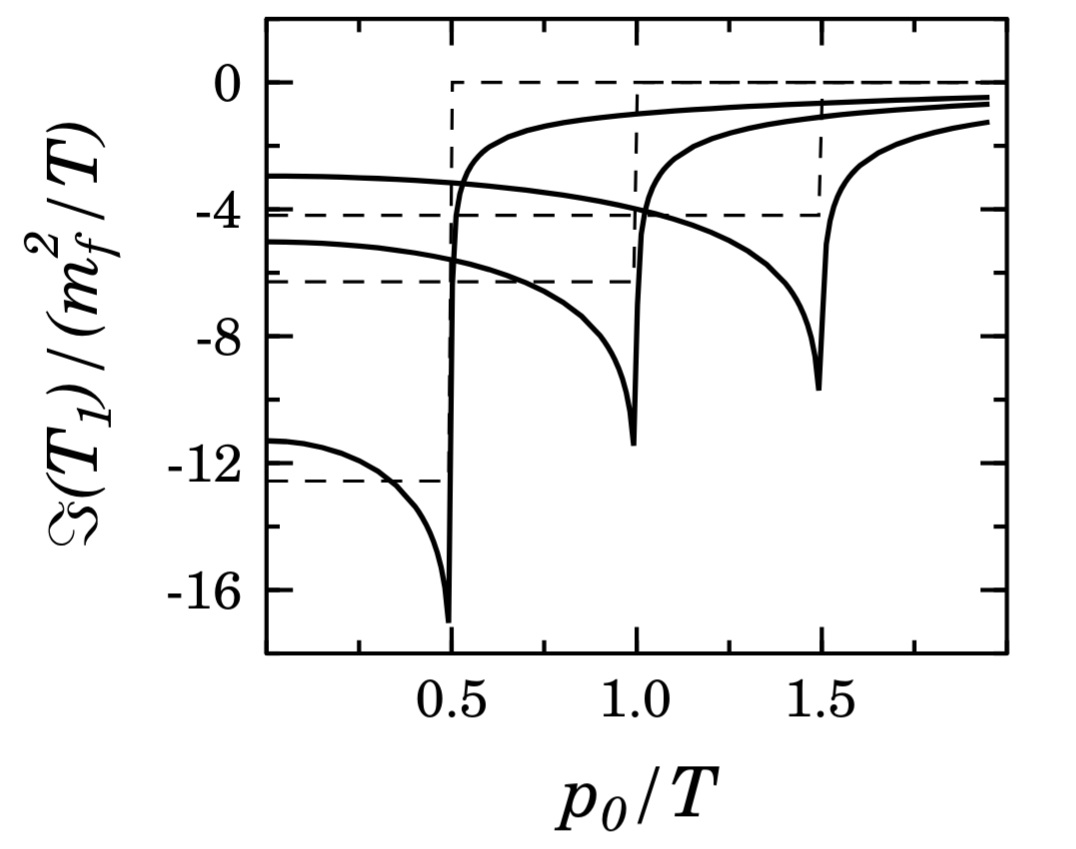}
    \caption{Imaginary part of $t_U$ divided by $m_f^2$ at $T=1$ over $k_0$. On the left our result and on the right the result obtained in Ref.~\protect\cite{Peshier_1998} is displayed. In both plots, the different lines shown from left to right denote the respective result for a fixed value of $k=0.5,1,1.5$.}
    \label{fig: Im_diff}
\end{figure}
This difference can be seen in Fig.~\ref{fig: Im_diff}, where we show our computed imaginary part of $t_U$ normalized by $m_f^2$ as a function of $k_0$ for different values of $k$ in the left plot and the same quantity computed in Ref.~\cite{Peshier_1998} on the right. We checked that this has only a negligible effect on the final result.
\s

We want to consider the collision operator defined for a $2\rightarrow2$ process averaged via Eq.~(\ref{eq: average}). For a general $2\rightarrow2$ process this gives
\begin{equation}
    \langle\mathcal{C}[f]\rangle=\frac{1}{N_1}\int\text{d}\Pi_{\{1..4\}}(2\pi)^4\delta^4\left(P_1+P_2-P_3-P_4\right)|\mathcal{M}|^2_{12\rightarrow 34}\mathcal{P}[f]\;,
\end{equation}
with 
\begin{equation}
    \mathcal{P}[f]=f_1f_2(1\pm f_3)(1\pm f_4)-f_3f_4(1\pm f_1)(1\pm f_2)\;.
\end{equation}
Since the collision operator is Lorentz invariant, we will calculate it in the plasma frame. With the identity
\begin{equation}
    (1\pm f_i)=f_ie^{\beta E_i-\beta\mu_i}\;,
\end{equation}
we can write $\mathcal{P}[f]$ as
\begin{align}
    \mathcal{P}[f]&=f_1f_2f_3f_4\left(e^{\beta(E_3+E_4-\mu_3-\mu_4)}-e^{\beta(E_1+E_2-\mu_1-\mu_2)}\right)\nonumber\\
    &\approx f_1^0f_2^0(1\pm f_3^0)(1\pm f_4^0)\left(\frac{\mu_1+\mu_2-\mu_3-\mu_4}{T}\right)\;,
\end{align}
where we expanded around $\mu\approx0$ and used energy conservation in the second line. For our purposes we only consider a scenario where all particles are massless and will therefore perform the derivation with this in mind. We extract the temperature dependence of the collision operator by rescaling the momenta via $P_i\rightarrow P_iT$ such that
\begin{equation}
    \langle\mathcal{C}[f]\rangle=\frac{T^4}{N_1}\int\text{d}\Pi_{\{1..4\}}(2\pi)^4\delta^4\left(P_1+P_2-P_3-P_4\right)|\mathcal{M}|^2_{12\rightarrow 34}\mathcal{P}[f]\;.
\end{equation}
We can immediately integrate out the fourth particle, such that
\begin{equation}
    \langle\mathcal{C}[f]\rangle=\frac{T^4}{N_1}\int\text{d}\Pi_{\{1..3\}}2\pi\delta\left((P_1+P_2-P_3)^2\right)|\mathcal{M}|^2_{12\rightarrow 34}\mathcal{P}[f]\;.
\end{equation}
With the remaining $\delta$ distribution we can integrate out the momentum of the third particle using
\begin{equation}
    \delta\left((P_1+P_2-P_3)^2\right)=\frac{\delta\left(p_3-\frac{p_1p_2-\mathbf{p}_1\mathbf{p}_2}{F}\right)}{2|F|}\;,
\end{equation}
where 
\begin{equation}
    F\equiv p_1+p_2+\hat{\mathbf{p}}_3(\mathbf{p_1}+\mathbf{p}_2)\quad\text{with}\quad\hat{\mathbf{p}}_3=\frac{\mathbf{p}_3}{p_3}\;.
\end{equation}
This leaves us with 
\begin{align}
    \langle\mathcal{C}[f]\rangle&=\frac{T^4}{N_1}\int\text{d}\Pi_{\{1,2\}}\int\text{d}\Omega\frac{p_3}{16\pi^2}|\mathcal{M}|^2_{12\rightarrow 34}\mathcal{P}[f]\bigg|_{p_3=\frac{p_1p_2-\mathbf{p}_1\mathbf{p}_2}{2|F|}}\;\nonumber\\
    &=\hat{\Gamma}_{12\rightarrow34}(\mu_1+\mu_2-\mu_3-\mu_4)\;,
\end{align}
where
\begin{equation}
    \hat{\Gamma}_{12\rightarrow34}\equiv\frac{T^3}{N_1}\int\text{d}\Pi_{\{1,2\}}\int\text{d}\Omega\frac{p_3}{16\pi^2}|\mathcal{M}|^2_{12\rightarrow 34}f_1^0f_2^0(1\pm f_3^0)(1\pm f_4^0)\;.
\end{equation}
All that is left to do is to plug in the amplitude for the processes $t_Lh\rightarrow t_Rg$, c.f. Fig.~\ref{fig: yukawa_rate_diag}, which in terms of the coefficients $a$ and $b$ as well as the Nandelstam variables $s=(P_1+P_2)^2$ and $t=(P_1-P_3)^2$ is given by
\begin{align} |\mathcal{M}|_{t_Lh\rightarrow t_Rg}=&-4g_s^2y_t^2\big[|1+a|^2st+|b|^2(s+t-4p _2p_3)+2\text{Re}(ab^*+b)(p_1(s+t)+p_2t-p_3s)\nonumber\\
    &-4\text{Im}(a^*b+b)U^\mu P_1^\nu P_2^\rho P_3^\sigma\epsilon_{\mu\nu\rho\sigma}\big]/|(1+a)^2t+2(1+a)b(p_1-p_3)+b^2|^2\;,
\end{align}
After plugging in the numerical values $g_s=\sqrt{4\pi\alpha_s}$, $\alpha_s=0.12$, $y_t=1$, $C_2=4/3$ and integrating the collision operator using the Cuba library~\cite{Hahn_2005} we obtain a value of 
\begin{equation}
    \hat{\Gamma}_y=(5967 \pm 6) 10^{-6} \, T\;.
\end{equation}
Our result differs by 41\% compared to the most recent result using the leading log approximation calculated in Ref.~\cite{Barni:2025ifb}.
\section{Universal Transport Functions}\label{app: universal_funcs}
In this appendix we want to derive explicit expressions for the universal functions defined in Eq.~(\ref{eq: universal_kernels}). We start by considering the generic integral form
\begin{align}
    \mathcal{K}(V,n,m,k)&\equiv\left\langle\frac{p_z^n}{E^m}V(E,p_z)\mathcal{F}^k_{0w}\right\rangle\;\nonumber\\
    &=\frac{1}{N_1}\int\text{d}^3p\frac{p_z^n}{E^m}V\mathcal{F}^k_{0w}[\gamma_w(E+v_wp_z)] \;, \label{eq: universal_kernel_def}
\end{align}
where $V(E,p_z)$ is an arbitrary function of $E$ and $p_z$, $N_1$ is a normalization constant given by Eq.~(\ref{eq: average}) and 
\begin{equation}
    \mathcal{F}^k_{0w}\equiv\frac{\partial^kf_{0w}}{\partial(\gamma_w(E+v_wp_z))^k}\;,
\end{equation}
where $f_{0w}$ is defined in Eq.~(\ref{eq: equilibrium_distribution}). First, we perform a Lorentz transformation 
\begin{align}
    E=\gamma_w(E'-v_wp_z')\quad,\quad p_z=\gamma_w(p_z'-v_wE')\quad,\quad p_x=p_x'\quad,\quad p_y=p_y'\;,
\end{align}
where our new variables $E'$ and $p_z'$ are the energy and momentum in the plasma frame resulting in
\begin{equation}
    \mathcal{K}(V,n,m,k)=\frac{1}{N_1}\int\frac{\text{d}^3p'}{E'}\frac{(\gamma_w(p_z'-v_wE'))^n}{(\gamma_w(E'-v_wp_z'))^{m-1}}V\mathcal{F}^k_{0}[E']\;.\label{eq: kernel_boosted}
\end{equation}
Next, we perform the coordinate transformation $(p_x',p_y',p_z')\rightarrow(E',p_z',\theta)$ via
\begin{equation}
    E'=\sqrt{p_x^{\prime2}+p_y^{\prime2}+p_z^{\prime2}+m^2}\quad,\quad\text{tan}(\theta)=\frac{p_y'}{p_x'}\quad,\quad p_z'=p_z'\;.
\end{equation}
This leads to the Jacobian $\text{det}(J)=E'$ and turns Eq.~(\ref{eq: kernel_boosted}) into
\begin{align}
    \mathcal{K}(V,n,m,k)&=\frac{1}{N_1}\int_m^\infty\text{d}E'\int_{-p_{E'}}^{p_{E'}}\text{d}p_z'\int_{-\pi}^\pi\text{d}\theta\frac{(\gamma_w(p_z'-v_wE'))^n}{(\gamma_w(E'-v_wp_z'))^{m-1}}V\mathcal{F}^k_{0}\nonumber\\
    &=\frac{2\pi}{N_1}\int_m^\infty\text{d}E\int_{-p_E}^{p_E}\text{d}p_z\frac{(\gamma_w(p_z'-v_wE'))^n}{(\gamma_w(E'-v_wp_z'))^{m-1}}V\mathcal{F}^k_{0}\;,
\end{align}
where $p_{E'}\equiv\sqrt{E^{\prime2}-m^2}$. 
Next, we make the substitutions $p_z'=yp_{E'}$ and $E'=\omega T$ to arrive at the final expression
\begin{equation}
    \mathcal{K}(V,n,m,k)=-\frac{3}{\pi^2\gamma_w}T^{n-m-k+1}\int_x^\infty\text{d}\omega\int_{-1}^1\text{d}y\frac{\tilde{p}_\omega\tilde{p}_z^n}{\tilde{E}^{m-1}}V\mathcal{F}^k_0[\omega]\;,
\end{equation}
with 
\begin{equation}
    x=m/T\quad,\quad\tilde{p}_\omega=\sqrt{\omega^2-x^2}\quad,\quad \tilde{p}_z=\gamma_w(y\tilde{p}_\omega-v_w\omega)\quad,\quad\tilde{E}=\gamma_w(\omega-v_wy\tilde{p}_\omega)\;.
\end{equation}
We can now define the universal functions in terms of $\mathcal{K}$ as follows
\begin{align}
    K_\ell&=T^{-1}\mathcal{K}(1,\ell,\ell,0)\;,\\
    D_\ell&=\mathcal{K}(1,\ell,\ell,1)\;,\\
    Q_\ell&=\frac{T^2}{2}\mathcal{K}(1,\ell-1,\ell,2)\;,\\
    Q^{8o}_\ell&=\frac{T^2}{2}\mathcal{K}(V_{s/h},\ell-2,\ell,1)\;,\\
    Q^{9o}_\ell&=\frac{T^4}{4}\left[\mathcal{K}(V_{s/h},\ell-2,\ell+2,1)-\gamma_w\mathcal{K}(V_{s/h},\ell-2,\ell+1,2)\right]\;,
\end{align}
where $V_{s/h}$ are functions that depend on  whether we use the spin or helicity basis and are given by~\cite{Barni:2025ifb}
\begin{equation}
    V_s=\frac{|\tilde{p}_z|}{\sqrt{\tilde{p}_z^2+x^2}}\quad,\quad V_h=V_s^2\left(1-\frac{x^2}{\omega^2}\right)^{-1/2}\;.
\end{equation}
Note that the $T$ dependence of the universal functions cancels, making them dimensionless and solely dependent on $x$ and $v_w$. 

\end{appendix}
\vspace*{0.5cm}
\bibliographystyle{h-physrev}
\addcontentsline{toc}{section}{References}
\bibliography{main.bib}

@article{Basler:2021kgq,
    author = {Basler, Philipp and Biermann, Lisa and M{\"u}hlleitner, Margarete and M{\"u}ller, Jonas},
    title = "{Electroweak baryogenesis in the CP-violating two-Higgs doublet model}",
    eprint = "2108.03580",
    archivePrefix = "arXiv",
    primaryClass = "hep-ph",
    reportNumber = "KA-TP-14-2021",
    doi = "10.1140/epjc/s10052-023-11192-9",
    journal = "Eur. Phys. J. C",
    volume = "83",
    number = "1",
    pages = "57",
    year = "2023"
}

@article{Kuzmin:1985mm,
      author         = "Kuzmin, V. A. and Rubakov, V. A. and Shaposhnikov, M. E.",
      title          = "{On the Anomalous Electroweak Baryon Number
                        Nonconservation in the Early Universe}",
      journal        = "Phys. Lett.",
      volume         = "155B",
      year           = "1985",
      pages          = "36",
      doi            = "10.1016/0370-2693(85)91028-7",
      reportNumber   = "IC/85/8",
      SLACcitation   = "%%CITATION = PHLTA,155B,36;%%"
}

@article{Cohen:1990it,
      author         = "Cohen, Andrew G. and Kaplan, David B. and Nelson, Ann E.",
      title          = "{Baryogenesis at the weak phase transition}",
      journal        = "Nucl. Phys.",
      volume         = "B349",
      year           = "1991",
      pages          = "727-742",
      doi            = "10.1016/0550-3213(91)90395-E",
      reportNumber   = "NSF-ITP-90-85, UCSD-PTH-90-09, BUHEP-90-15",
      SLACcitation   = "%%CITATION = NUPHA,B349,727;%%"
}

@article{Cohen:1993nk,
      author         = "Cohen, Andrew G. and Kaplan, D. B. and Nelson, A. E.",
      title          = "{Progress in electroweak baryogenesis}",
      journal        = "Ann. Rev. Nucl. Part. Sci.",
      volume         = "43",
      year           = "1993",
      pages          = "27-70",
      doi            = "10.1146/annurev.ns.43.120193.000331",
      eprint         = "hep-ph/9302210",
      archivePrefix  = "arXiv",
      primaryClass   = "hep-ph",
      reportNumber   = "UCSD-PTH-93-02, BUHEP-93-4",
      SLACcitation   = "%%CITATION = HEP-PH/9302210;%%"
}

@article{Quiros:1994dr,
      author         = "Quiros, M.",
      title          = "{Field theory at finite temperature and phase
                        transitions}",
      journal        = "Helv. Phys. Acta",
      volume         = "67",
      year           = "1994",
      pages          = "451-583",
      SLACcitation   = "%%CITATION = HPACA,67,451;%%"
}

@article{Rubakov:1996vz,
      author         = "Rubakov, V. A. and Shaposhnikov, M. E.",
      title          = "{Electroweak baryon number nonconservation in the early
                        universe and in high-energy collisions}",
      journal        = "Usp. Fiz. Nauk",
      volume         = "166",
      year           = "1996",
      pages          = "493-537",
      doi            = "10.1070/PU1996v039n05ABEH000145",
      note           = "[Phys. Usp.39,461(1996)]",
      eprint         = "hep-ph/9603208",
      archivePrefix  = "arXiv",
      primaryClass   = "hep-ph",
      reportNumber   = "CERN-TH-96-13, INR-0913-96",
      SLACcitation   = "%%CITATION = HEP-PH/9603208;%%"
}

@article{Funakubo:1996dw,
      author         = "Funakubo, Koichi",
      title          = "{CP violation and baryogenesis at the electroweak phase
                        transition}",
      journal        = "Prog. Theor. Phys.",
      volume         = "96",
      year           = "1996",
      pages          = "475-520",
      doi            = "10.1143/PTP.96.475",
      eprint         = "hep-ph/9608358",
      archivePrefix  = "arXiv",
      primaryClass   = "hep-ph",
      reportNumber   = "SAGA-HE-108",
      SLACcitation   = "%%CITATION = HEP-PH/9608358;%%"
}

@article{Trodden:1998ym,
      author         = "Trodden, Mark",
      title          = "{Electroweak baryogenesis}",
      journal        = "Rev. Mod. Phys.",
      volume         = "71",
      year           = "1999",
      pages          = "1463-1500",
      doi            = "10.1103/RevModPhys.71.1463",
      eprint         = "hep-ph/9803479",
      archivePrefix  = "arXiv",
      primaryClass   = "hep-ph",
      reportNumber   = "CWRU-P6-98",
      SLACcitation   = "%%CITATION = HEP-PH/9803479;%%"
}

@article{Bernreuther:2002uj,
      author         = "Bernreuther, Werner",
      title          = "{CP violation and baryogenesis}",
      booktitle      = "{Workshop of the Graduate College of Elementary Particle
                        Physics Berlin, Germany, April 2-5, 2001}",
      journal        = "Lect. Notes Phys.",
      volume         = "591",
      year           = "2002",
      pages          = "237-293",
      note           = "[,237(2002)]",
      eprint         = "hep-ph/0205279",
      archivePrefix  = "arXiv",
      primaryClass   = "hep-ph",
      reportNumber   = "PITHA-02-08",
      SLACcitation   = "%%CITATION = HEP-PH/0205279;%%"
}

@article{Morrissey:2012db,
      author         = "Morrissey, David E. and Ramsey-Musolf, Michael J.",
      title          = "{Electroweak baryogenesis}",
      journal        = "New J. Phys.",
      volume         = "14",
      year           = "2012",
      pages          = "125003",
      doi            = "10.1088/1367-2630/14/12/125003",
      eprint         = "1206.2942",
      archivePrefix  = "arXiv",
      primaryClass   = "hep-ph",
      reportNumber   = "NPAC-12-08",
      SLACcitation   = "%%CITATION = ARXIV:1206.2942;%%"
}

@article{Coimbra:2013qq,
    author = "Coimbra, Rita and Sampaio, Marco O. P. and Santos, Rui",
    title = "{ScannerS: Constraining the phase diagram of a complex scalar singlet at the LHC}",
    eprint = "1301.2599",
    archivePrefix = "arXiv",
    primaryClass = "hep-ph",
    doi = "10.1140/epjc/s10052-013-2428-4",
    journal = "Eur. Phys. J. C",
    volume = "73",
    pages = "2428",
    year = "2013"
}

@article{Ferreira:2014dya,
    author = "Ferreira, P. M. and Guedes, Renato and Sampaio, Marco O. P. and Santos, Rui",
    title = "{Wrong sign and symmetric limits and non-decoupling in 2HDMs}",
    eprint = "1409.6723",
    archivePrefix = "arXiv",
    primaryClass = "hep-ph",
    doi = "10.1007/JHEP12(2014)067",
    journal = "JHEP",
    volume = "12",
    pages = "067",
    year = "2014"
}

@article{Costa:2015llh,
    author = {Costa, Raul and M\"uhlleitner, Margarete and Sampaio, Marco O. P. and Santos, Rui},
    title = "{Singlet Extensions of the Standard Model at LHC Run 2: Benchmarks and Comparison with the NMSSM}",
    eprint = "1512.05355",
    archivePrefix = "arXiv",
    primaryClass = "hep-ph",
    doi = "10.1007/JHEP06(2016)034",
    journal = "JHEP",
    volume = "06",
    pages = "034",
    year = "2016"
}

@article{Muhlleitner:2016mzt,
    author = "Muhlleitner, Margarete and Sampaio, Marco O. P. and Santos, Rui and Wittbrodt, Jonas",
    title = "{The N2HDM under Theoretical and Experimental Scrutiny}",
    eprint = "1612.01309",
    archivePrefix = "arXiv",
    primaryClass = "hep-ph",
    doi = "10.1007/JHEP03(2017)094",
    journal = "JHEP",
    volume = "03",
    pages = "094",
    year = "2017"
}

@article{Muhlleitner:2020wwk,
    author = {M\"uhlleitner, Margarete and Sampaio, Marco O. P. and Santos, Rui and Wittbrodt, Jonas},
    title = "{ScannerS: parameter scans in extended scalar sectors}",
    eprint = "2007.02985",
    archivePrefix = "arXiv",
    primaryClass = "hep-ph",
    reportNumber = "KA-TP-05-2020, LU TP 20-38",
    doi = "10.1140/epjc/s10052-022-10139-w",
    journal = "Eur. Phys. J. C",
    volume = "82",
    number = "3",
    pages = "198",
    year = "2022"
}

@article{Planck:2018vyg,
    author = "Aghanim, N. and others",
    collaboration = "Planck",
    title = "{Planck 2018 results. VI. Cosmological parameters}",
    eprint = "1807.06209",
    archivePrefix = "arXiv",
    primaryClass = "astro-ph.CO",
    doi = "10.1051/0004-6361/201833910",
    journal = "Astron. Astrophys.",
    volume = "641",
    pages = "A6",
    year = "2020",
    note = "[Erratum: Astron.Astrophys. 652, C4 (2021)]"
}

@article{Cline:1997vk,
    author = "Cline, James M. and Joyce, Michael and Kainulainen, Kimmo",
    title = "{Supersymmetric electroweak baryogenesis in the WKB approximation}",
    eprint = "hep-ph/9708393",
    archivePrefix = "arXiv",
    reportNumber = "MCGILL-97-26, HIP-1997-44-TH",
    doi = "10.1016/S0370-2693(97)01361-0",
    journal = "Phys. Lett. B",
    volume = "417",
    pages = "79--86",
    year = "1998",
    note = "[Erratum: Phys.Lett.B 448, 321--321 (1999)]"
}

@article{Sakharov:1967dj,
    author = "Sakharov, A. D.",
    title = "{Violation of CP Invariance, C asymmetry, and baryon asymmetry of the universe}",
    doi = "10.1070/PU1991v034n05ABEH002497",
    journal = "Pisma Zh. Eksp. Teor. Fiz.",
    volume = "5",
    pages = "32--35",
    year = "1967"
}

@article{Branco:1985aq,
    author = "Branco, G. C. and Rebelo, M. N.",
    title = "{The Higgs Mass in a Model With Two Scalar Doublets and Spontaneous {CP} Violation}",
    reportNumber = "IFM-7/85",
    doi = "10.1016/0370-2693(85)91476-5",
    journal = "Phys. Lett. B",
    volume = "160",
    pages = "117--120",
    year = "1985"
}

@article{Ginzburg:2002wt,
    author = "Ginzburg, Ilya F. and Krawczyk, Maria and Osland, Per",
    title = "{Two Higgs doublet models with CP violation}",
    eprint = "hep-ph/0211371",
    archivePrefix = "arXiv",
    reportNumber = "CERN-TH-2002-330, IFT-40-2002",
    pages = "703--706",
    month = "11",
    year = "2002"
}

@article{Khater:2003wq,
    author = "Khater, Wafaa and Osland, Per",
    title = "{CP violation in top quark production at the LHC and two Higgs doublet models}",
    eprint = "hep-ph/0302004",
    archivePrefix = "arXiv",
    doi = "10.1016/S0550-3213(03)00300-6",
    journal = "Nucl. Phys. B",
    volume = "661",
    pages = "209--234",
    year = "2003"
}

@article{Kainulainen:2024qpm,
    author = "Kainulainen, Kimmo and Venkatesan, Niyati",
    title = "{Systematic moment expansion for electroweak baryogenesis}",
    eprint = "2407.13639",
    archivePrefix = "arXiv",
    primaryClass = "hep-ph",
    doi = "10.1088/1475-7516/2024/08/058",
    journal = "JCAP",
    volume = "08",
    pages = "058",
    year = "2024"
}

@article{Liu:1992tn,
    author = "Liu, Bao-Hua and McLerran, Larry D. and Turok, Neil",
    title = "{Bubble nucleation and growth at a baryon number producing electroweak phase transition}",
    reportNumber = "TPI-MINN-92-18-T",
    doi = "10.1103/PhysRevD.46.2668",
    journal = "Phys. Rev. D",
    volume = "46",
    pages = "2668--2688",
    year = "1992"
}

@article{Moore:1995si,
    author = "Moore, Guy D. and Prokopec, Tomislav",
    title = "{How fast can the wall move? A Study of the electroweak phase transition dynamics}",
    eprint = "hep-ph/9506475",
    archivePrefix = "arXiv",
    reportNumber = "PUPT-1544, PUP-TH-1544, LANCS-TH-9517",
    doi = "10.1103/PhysRevD.52.7182",
    journal = "Phys. Rev. D",
    volume = "52",
    pages = "7182--7204",
    year = "1995"
}

@article{Basler:2020nrq,
    author = {Basler, Philipp and M{\"u}hlleitner, Margarete and M{\"u}ller, Jonas},
    title = "{BSMPT v2 a tool for the electroweak phase transition and the baryon asymmetry of the universe in extended Higgs Sectors}",
    eprint = "2007.01725",
    archivePrefix = "arXiv",
    primaryClass = "hep-ph",
    doi = "10.1016/j.cpc.2021.108124",
    journal = "Comput. Phys. Commun.",
    volume = "269",
    pages = "108124",
    year = "2021"
}

@article{Cline_2000,
   title={Supersymmetric electroweak baryogenesis},
   volume={2000},
   ISSN={1029-8479},
   url={http://dx.doi.org/10.1088/1126-6708/2000/07/018},
   DOI={10.1088/1126-6708/2000/07/018},
   number={07},
   journal={Journal of High Energy Physics},
   publisher={Springer Science and Business Media LLC},
   author={Cline, James M and Joyce, Michael and Kainulainen, Kimmo},
   year={2000},
   month=jul, pages={018–018} 
}

@article{Kainulainen_2001,
   title={First principle derivation of semiclassical force for electroweak baryogenesis},
   volume={2001},
   ISSN={1029-8479},
   url={http://dx.doi.org/10.1088/1126-6708/2001/06/031},
   DOI={10.1088/1126-6708/2001/06/031},
   number={06},
   journal={Journal of High Energy Physics},
   publisher={Springer Science and Business Media LLC},
   author={Kainulainen, Kimmo and Prokopec, Tomislav and Schmidt, Michael G and Weinstock, Steffen},
   year={2001},
   month=jun, pages={031–031} 
}

@article{Kainulainen_2002,
   title={Semiclassical force for electroweak baryogenesis: Three-dimensional derivation},
   volume={66},
   ISSN={1089-4918},
   url={http://dx.doi.org/10.1103/PhysRevD.66.043502},
   DOI={10.1103/physrevd.66.043502},
   number={4},
   journal={Physical Review D},
   publisher={American Physical Society (APS)},
   author={Kainulainen, Kimmo and Prokopec, Tomislav and Schmidt, Michael G. and Weinstock, Steffen},
   year={2002},
   month=aug 
}

@article{Fromme_2007,
   title={Top transport in electroweak baryogenesis},
   volume={2007},
   ISSN={1029-8479},
   url={http://dx.doi.org/10.1088/1126-6708/2007/03/049},
   DOI={10.1088/1126-6708/2007/03/049},
   number={03},
   journal={Journal of High Energy Physics},
   publisher={Springer Science and Business Media LLC},
   author={Fromme, Lars and Huber, Stephan J},
   year={2007},
   month=mar, pages={049–049} 
}

@article{PhysRevD.101.063525,
  title = {Electroweak baryogenesis at high bubble wall velocities},
  author = {Cline, James M. and Kainulainen, Kimmo},
  journal = {Phys. Rev. D},
  volume = {101},
  issue = {6},
  pages = {063525},
  numpages = {16},
  year = {2020},
  month = {Mar},
  publisher = {American Physical Society},
  doi = {10.1103/PhysRevD.101.063525},
  url = {https://link.aps.org/doi/10.1103/PhysRevD.101.063525}
}

@article{Barni:2025ifb,
    author = "Barni, Giulio",
    title = "{Electroweak Baryogenesis with BARYONET: a self-contained review of the WKB approach}",
    eprint = "2510.21915",
    archivePrefix = "arXiv",
    primaryClass = "hep-ph",
    month = "10",
    year = "2025"
}

@article{Gunion:2002zf,
    author = "Gunion, John F. and Haber, Howard E.",
    title = "{The CP conserving two Higgs doublet model: The Approach to the decoupling limit}",
    eprint = "hep-ph/0207010",
    archivePrefix = "arXiv",
    reportNumber = "SCIPP-02-10",
    doi = "10.1103/PhysRevD.67.075019",
    journal = "Phys. Rev. D",
    volume = "67",
    pages = "075019",
    year = "2003"
}

@article{Barroso:2012wz,
    author = "Barroso, A. and Ferreira, P. M. and Santos, Rui and Silva, Joao P.",
    title = "{Probing the scalar-pseudoscalar mixing in the 125 GeV Higgs particle with current data}",
    eprint = "1205.4247",
    archivePrefix = "arXiv",
    primaryClass = "hep-ph",
    doi = "10.1103/PhysRevD.86.015022",
    journal = "Phys. Rev. D",
    volume = "86",
    pages = "015022",
    year = "2012"
}

@article{Ginzburg:2004vp,
    author = "Ginzburg, Ilya F. and Krawczyk, Maria",
    title = "{Symmetries of two Higgs doublet model and CP violation}",
    eprint = "hep-ph/0408011",
    archivePrefix = "arXiv",
    doi = "10.1103/PhysRevD.72.115013",
    journal = "Phys. Rev. D",
    volume = "72",
    pages = "115013",
    year = "2005"
}

@article{ElKaffas:2006gdt,
    author = "El Kaffas, Abdul Wahab and Khater, Wafaa and Ogreid, Odd Magne and Osland, Per",
    title = "{Consistency of the two Higgs doublet model and CP violation in top production at the LHC}",
    eprint = "hep-ph/0605142",
    archivePrefix = "arXiv",
    doi = "10.1016/j.nuclphysb.2007.03.041",
    journal = "Nucl. Phys. B",
    volume = "775",
    pages = "45--77",
    year = "2007"
}

@article{Arhrib:2010ju,
    author = "Arhrib, A. and Christova, E. and Eberl, H. and Ginina, E.",
    title = "{CP violation in charged Higgs production and decays in the Complex Two Higgs Doublet Model}",
    eprint = "1011.6560",
    archivePrefix = "arXiv",
    primaryClass = "hep-ph",
    reportNumber = "HEPHY-PUB-896-10",
    doi = "10.1007/JHEP04(2011)089",
    journal = "JHEP",
    volume = "04",
    pages = "089",
    year = "2011"
}

@article{Inoue:2014nva,
    author = "Inoue, Satoru and Ramsey-Musolf, Michael J. and Zhang, Yue",
    title = "{CP-violating phenomenology of flavor conserving two Higgs doublet models}",
    eprint = "1403.4257",
    archivePrefix = "arXiv",
    primaryClass = "hep-ph",
    doi = "10.1103/PhysRevD.89.115023",
    journal = "Phys. Rev. D",
    volume = "89",
    number = "11",
    pages = "115023",
    year = "2014"
}

@article{Fontes:2014xva,
    author = "Fontes, Duarte and Rom{\~a}o, J. C. and Silva, Jo{\~a}o P.",
    title = "{$h \rightarrow Z \gamma$ in the complex two Higgs doublet model}",
    eprint = "1408.2534",
    archivePrefix = "arXiv",
    primaryClass = "hep-ph",
    doi = "10.1007/JHEP12(2014)043",
    journal = "JHEP",
    volume = "12",
    pages = "043",
    year = "2014"
}

@article{Grzadkowski:2014ada,
    author = "Grzadkowski, B. and Ogreid, O. M. and Osland, P.",
    title = "{Measuring CP violation in Two-Higgs-Doublet models in light of the LHC Higgs data}",
    eprint = "1409.7265",
    archivePrefix = "arXiv",
    primaryClass = "hep-ph",
    doi = "10.1007/JHEP11(2014)084",
    journal = "JHEP",
    volume = "11",
    pages = "084",
    year = "2014"
}

@article{Fontes:2017zfn,
    author = {Fontes, Duarte and M{\"u}hlleitner, Margarete and Rom{\~a}o, Jorge C. and Santos, Rui and Silva, Jo{\~a}o P. and Wittbrodt, Jonas},
    title = "{The C2HDM revisited}",
    eprint = "1711.09419",
    archivePrefix = "arXiv",
    primaryClass = "hep-ph",
    reportNumber = "CFTP-17-008, DESY-17-207, KA-TP-40-2017",
    doi = "10.1007/JHEP02(2018)073",
    journal = "JHEP",
    volume = "02",
    pages = "073",
    year = "2018"
}

@article{Georgi:1978ri,
    author = "Georgi, Howard and Nanopoulos, Dimitri V.",
    title = "{Suppression of Flavor Changing Effects From Neutral Spinless Meson Exchange in Gauge Theories}",
    reportNumber = "HUTP-78/A055",
    doi = "10.1016/0370-2693(79)90433-7",
    journal = "Phys. Lett. B",
    volume = "82",
    pages = "95--96",
    year = "1979"
}

@article{Donoghue:1978cj,
    author = "Donoghue, John F. and Li, Ling Fong",
    title = "{Properties of Charged Higgs Bosons}",
    reportNumber = "COO-3066-113",
    doi = "10.1103/PhysRevD.19.945",
    journal = "Phys. Rev. D",
    volume = "19",
    pages = "945",
    year = "1979"
}

@article{Lavoura:1994fv,
    author = "Lavoura, L. and Silva, Joao P.",
    title = "{Fundamental CP violating quantities in a SU(2) x U(1) model with many Higgs doublets}",
    eprint = "hep-ph/9404276",
    archivePrefix = "arXiv",
    doi = "10.1103/PhysRevD.50.4619",
    journal = "Phys. Rev. D",
    volume = "50",
    pages = "4619--4624",
    year = "1994"
}

@article{Botella:1994cs,
    author = "Botella, F. J. and Silva, Joao P.",
    title = "{Jarlskog - like invariants for theories with scalars and fermions}",
    eprint = "hep-ph/9411288",
    archivePrefix = "arXiv",
    reportNumber = "FTUV-94-68, IFIC-94-65",
    doi = "10.1103/PhysRevD.51.3870",
    journal = "Phys. Rev. D",
    volume = "51",
    pages = "3870--3875",
    year = "1995"
}

@article{ElKaffas:2007rq,
    author = "El Kaffas, Abdul Wahab and Osland, Per and Ogreid, Odd Magne",
    title = "{CP violation, stability and unitarity of the two Higgs doublet model}",
    eprint = "hep-ph/0702097",
    archivePrefix = "arXiv",
    journal = "Nonlin. Phenom. Complex Syst.",
    volume = "10",
    pages = "347--357",
    year = "2007"
}

@article{deSouza:2025uxb,
    author = "de Souza, Fernando Abreu and Castro, Nuno Filipe and Crispim Rom{\~a}o, Miguel and Porod, Werner",
    title = "{Exploring scotogenic parameter spaces and mapping uncharted dark matter phenomenology with multi-objective search algorithms}",
    eprint = "2505.08862",
    archivePrefix = "arXiv",
    primaryClass = "hep-ph",
    reportNumber = "IPPP/25/29",
    doi = "10.1007/JHEP10(2025)116",
    journal = "JHEP",
    volume = "10",
    pages = "116",
    year = "2025"
}

@article{deSouza:2025bpl,
    author = "de Souza, Fernando Abreu and Boto, Rafael and Crispim Rom{\~a}o, Miguel and Figueiredo, Pedro N. and Rom{\~a}o, Jorge C. and Silva, Jo{\~a}o P.",
    title = "{Unearthing large pseudoscalar Yukawa couplings with machine learning}",
    eprint = "2505.10625",
    archivePrefix = "arXiv",
    primaryClass = "hep-ph",
    reportNumber = "IPPP/25/28",
    doi = "10.1007/JHEP07(2025)268",
    journal = "JHEP",
    volume = "07",
    pages = "268",
    year = "2025"
}

@article{Boto:2025ovp,
    author = "Boto, Rafael and Matos, Jo{\~a}o A. C. and Rom{\~a}o, Jorge C. and Silva, Jo{\~a}o P.",
    title = "{Surveying the complex three Higgs doublet model with Machine Learning}",
    eprint = "2510.02445",
    archivePrefix = "arXiv",
    primaryClass = "hep-ph",
    month = "10",
    year = "2025"
}

@article{Glashow:1976nt,
    author = "Glashow, Sheldon L. and Weinberg, Steven",
    title = "{Natural Conservation Laws for Neutral Currents}",
    reportNumber = "HUTP-76-A158",
    doi = "10.1103/PhysRevD.15.1958",
    journal = "Phys. Rev. D",
    volume = "15",
    pages = "1958",
    year = "1977"
}

@article{Paschos:1976ay,
    author = "Paschos, E. A.",
    title = "{Diagonal Neutral Currents}",
    reportNumber = "BNL-21870",
    doi = "10.1103/PhysRevD.15.1966",
    journal = "Phys. Rev. D",
    volume = "15",
    pages = "1966",
    year = "1977"
}

@article{Basler:2024aaf,
    author = {Basler, Philipp and Biermann, Lisa and M{\"u}hlleitner, Margarete and M{\"u}ller, Jonas and Santos, Rui and Viana, Jo{\~a}o},
    title = "{BSMPT v3 a tool for phase transitions and primordial gravitational waves in extended Higgs sectors}",
    eprint = "2404.19037",
    archivePrefix = "arXiv",
    primaryClass = "hep-ph",
    reportNumber = "KA-TP-08-2024",
    doi = "10.1016/j.cpc.2025.109766",
    journal = "Comput. Phys. Commun.",
    volume = "316",
    pages = "109766",
    year = "2025"
}

@article{LISA,
      title={Laser Interferometer Space Antenna}, 
      author={Pau Amaro-Seoane and Heather Audley and Stanislav Babak and John Baker and Enrico Barausse and Peter Bender and Emanuele Berti and Pierre Binetruy and Michael Born and Daniele Bortoluzzi and Jordan Camp and Chiara Caprini and Vitor Cardoso and Monica Colpi and John Conklin and Neil Cornish and Curt Cutler and Karsten Danzmann and Rita Dolesi and Luigi Ferraioli and Valerio Ferroni and Ewan Fitzsimons and Jonathan Gair and Lluis Gesa Bote and Domenico Giardini and Ferran Gibert and Catia Grimani and Hubert Halloin and Gerhard Heinzel and Thomas Hertog and Martin Hewitson and Kelly Holley-Bockelmann and Daniel Hollington and Mauro Hueller and Henri Inchauspe and Philippe Jetzer and Nikos Karnesis and Christian Killow and Antoine Klein and Bill Klipstein and Natalia Korsakova and Shane L Larson and Jeffrey Livas and Ivan Lloro and Nary Man and Davor Mance and Joseph Martino and Ignacio Mateos and Kirk McKenzie and Sean T McWilliams and Cole Miller and Guido Mueller and Germano Nardini and Gijs Nelemans and Miquel Nofrarias and Antoine Petiteau and Paolo Pivato and Eric Plagnol and Ed Porter and Jens Reiche and David Robertson and Norna Robertson and Elena Rossi and Giuliana Russano and Bernard Schutz and Alberto Sesana and David Shoemaker and Jacob Slutsky and Carlos F. Sopuerta and Tim Sumner and Nicola Tamanini and Ira Thorpe and Michael Troebs and Michele Vallisneri and Alberto Vecchio and Daniele Vetrugno and Stefano Vitale and Marta Volonteri and Gudrun Wanner and Harry Ward and Peter Wass and William Weber and John Ziemer and Peter Zweifel},
      year={2017},
      eprint={1702.00786},
      archivePrefix={arXiv},
      primaryClass={astro-ph.IM}
}

@article{Caprini:2019egz,
      author         = "Caprini, Chiara and others",
      title          = "{Detecting gravitational waves from cosmological phase
                        transitions with LISA: an update}",
      journal        = "JCAP",
      volume         = "2003",
      year           = "2020",
      number         = "03",
      pages          = "024",
      doi            = "10.1088/1475-7516/2020/03/024",
      eprint         = "1910.13125",
      archivePrefix  = "arXiv",
      primaryClass   = "astro-ph.CO",
      reportNumber   = "DESY-19-159, IPPP/19/27, HIP-2019-14/TH, MITP/19-066,
                        IFT-UAM/CSIC-19-139",
      SLACcitation   = "%%CITATION = ARXIV:1910.13125;%%"
}

@article{LISACosmologyWorkingGroup:2022jok,
    author = "Auclair, Pierre and others",
    collaboration = "LISA Cosmology Working Group",
    title = "{Cosmology with the Laser Interferometer Space Antenna}",
    eprint = "2204.05434",
    archivePrefix = "arXiv",
    primaryClass = "astro-ph.CO",
    reportNumber = "LISA CosWG-22-03, FERMILAB-PUB-22-349-SCD",
    doi = "10.1007/s41114-023-00045-2",
    journal = "Living Rev. Rel.",
    volume = "26",
    number = "1",
    pages = "5",
    year = "2023"
}

@article{Athron:2023xlk,
    author = "Athron, Peter and Bal\'azs, Csaba and Fowlie, Andrew and Morris, Lachlan and Wu, Lei",
    title = "{Cosmological phase transitions: From perturbative particle physics to gravitational waves}",
    eprint = "2305.02357",
    archivePrefix = "arXiv",
    primaryClass = "hep-ph",
    doi = "10.1016/j.ppnp.2023.104094",
    journal = "Prog. Part. Nucl. Phys.",
    volume = "135",
    pages = "104094",
    year = "2024"
}

@article{Giblin2014TheDO,
  title={The detectability of cosmological gravitational-wave backgrounds: a rule of thumb},
  author={John T. Giblin and Eric Thrane},
  year={2014},
  url={https://api.semanticscholar.org/CorpusID:118954150}
}

@article{Konstandin:2011dr,
    author = "Konstandin, Thomas and Servant, Geraldine",
    title = "{Cosmological Consequences of Nearly Conformal Dynamics at the TeV scale}",
    eprint = "1104.4791",
    archivePrefix = "arXiv",
    primaryClass = "hep-ph",
    doi = "10.1088/1475-7516/2011/12/009",
    journal = "JCAP",
    volume = "12",
    pages = "009",
    year = "2011"
}

@article{Gent:2025csq,
    author = "Gent, T. and Huber, S. and Mimasu, K. and No, J. M.",
    title = "{Towards precise baryogenesis in the 2HDM$+a$}",
    eprint = "2512.22081",
    archivePrefix = "arXiv",
    primaryClass = "hep-ph",
    reportNumber = "IFT-UAM/CSIC-25-167",
    month = "12",
    year = "2025"
}

@article{Huet_1996,
   title={Electroweak baryogenesis in supersymmetric models},
   volume={53},
   ISSN={1089-4918},
   url={http://dx.doi.org/10.1103/PhysRevD.53.4578},
   DOI={10.1103/physrevd.53.4578},
   number={8},
   journal={Physical Review D},
   publisher={American Physical Society (APS)},
   author={Huet, Patrick and Nelson, Ann E.},
   year={1996},
   month=apr, pages={4578–4597} }

@article{Cline_2021,
   title={Electroweak baryogenesis from light fermion sources: A critical study},
   volume={104},
   ISSN={2470-0029},
   url={http://dx.doi.org/10.1103/PhysRevD.104.083507},
   DOI={10.1103/physrevd.104.083507},
   number={8},
   journal={Physical Review D},
   publisher={American Physical Society (APS)},
   author={Cline, James M. and Laurent, Benoit},
   year={2021},
   month=oct }

@article{PhysRevD.26.2789,
  title = {Effective fermion masses of order $\mathrm{gT}$ in high-temperature gauge theories with exact chiral invariance},
  author = {Weldon, H. Arthur},
  journal = {Phys. Rev. D},
  volume = {26},
  issue = {10},
  pages = {2789--2796},
  numpages = {0},
  year = {1982},
  month = {Nov},
  publisher = {American Physical Society},
  doi = {10.1103/PhysRevD.26.2789},
  url = {https://link.aps.org/doi/10.1103/PhysRevD.26.2789}
}

@article{Peshier_1998,
   title={One-Loop Self Energies at Finite Temperature},
   volume={266},
   ISSN={0003-4916},
   url={http://dx.doi.org/10.1006/aphy.1997.5781},
   DOI={10.1006/aphy.1997.5781},
   number={1},
   journal={Annals of Physics},
   publisher={Elsevier BV},
   author={Peshier, André and Schertler, Klaus and Thoma, Markus H.},
   year={1998},
   month=jun, pages={162–177} }

@article{Hahn_2005,
   title={Cuba—a library for multidimensional numerical integration},
   volume={168},
   ISSN={0010-4655},
   url={http://dx.doi.org/10.1016/j.cpc.2005.01.010},
   DOI={10.1016/j.cpc.2005.01.010},
   number={2},
   journal={Computer Physics Communications},
   publisher={Elsevier BV},
   author={Hahn, T.},
   year={2005},
   month=jun, pages={78–95} }

@article{Basler:2018cwe,
    author = {Basler, Philipp and M{\"u}hlleitner, Margarete},
    title = "{BSMPT (Beyond the Standard Model Phase Transitions): A tool for the electroweak phase transition in extended Higgs sectors}",
    eprint = "1803.02846",
    archivePrefix = "arXiv",
    primaryClass = "hep-ph",
    doi = "10.1016/j.cpc.2018.11.006",
    journal = "Comput. Phys. Commun.",
    volume = "237",
    pages = "62--85",
    year = "2019"
}

@article{Hansen2001,
  author={Hansen, Nikolaus and Ostermeier, Andreas},
  journal={Evolutionary Computation}, 
  title={Completely Derandomized Self-Adaptation in Evolution Strategies}, 
  year={2001},
  volume={9},
  number={2},
  pages={159-195},
  doi={10.1162/106365601750190398}}

@unpublished{hansen:hal-01297037,
  TITLE = {{The CMA Evolution Strategy: A Tutorial}},
  AUTHOR = {Hansen, Nikolaus},
  URL = {https://inria.hal.science/hal-01297037},
  NOTE = {ArXiv e-prints, arXiv:1604.00772, 2016, pp.1-39},
  YEAR = {2005},
  HAL_ID = {hal-01297037},
  HAL_VERSION = {v2},
}

@inproceedings{HBOS,
author = {Goldstein, Markus and Dengel, Andreas},
year = {2012},
month = {09},
pages = {},
title = {Histogram-based Outlier Score (HBOS): A fast Unsupervised Anomaly Detection Algorithm}
}

@article{Romao:2024gjx,
 archiveprefix = {arXiv},
 author = {Romão, Jorge Crispim and Crispim Rom{\~a}õ, Miguel},
 doi = {10.1103/PhysRevD.109.095040},
 eprint = {2402.07661},
 journal = {Phys. Rev. D},
 number = {9},
 pages = {095040},
 primaryclass = {hep-ph},
 reportnumber = {IPPP/24/04, CFTP/24-002},
 title = {Combining evolutionary strategies and novelty detection to go beyond the alignment limit of the Z3 3HDM},
 volume = {109},
 year = {2024}
}

@article{Biekotter:2025fjx,
    author = {Biek{\"o}tter, Thomas and Olea-Romacho, Mar{\'\i}a Olalla},
    title = "{Benchmarking a fading window: electroweak baryogenesis in the C2HDM, LHC constraints after Run 2 and prospects for LISA}",
    eprint = "2505.09670",
    archivePrefix = "arXiv",
    primaryClass = "hep-ph",
    reportNumber = "IFT-UAM/CSIC-25-42",
    doi = "10.1007/JHEP12(2025)040",
    journal = "JHEP",
    volume = "12",
    pages = "040",
    year = "2025"
}

@article{Aiko:2025tbk,
    author = "Aiko, Masashi and Endo, Motoi and Kanemura, Shinya and Mura, Yushi",
    title = "{Electroweak baryogenesis in 2HDM without EDM cancellation}",
    eprint = "2504.07705",
    archivePrefix = "arXiv",
    primaryClass = "hep-ph",
    reportNumber = "KEK-TH-2695, OU-HET-1268",
    doi = "10.1007/JHEP07(2025)236",
    journal = "JHEP",
    volume = "07",
    pages = "236",
    year = "2025"
}

@article{Goncalves:2023svb,
    author = "Gon{\c{c}}alves, Dorival and Kaladharan, Ajay and Wu, Yongcheng",
    title = "{Gravitational waves, bubble profile, and baryon asymmetry in the complex 2HDM}",
    eprint = "2307.03224",
    archivePrefix = "arXiv",
    primaryClass = "hep-ph",
    doi = "10.1103/PhysRevD.108.075010",
    journal = "Phys. Rev. D",
    volume = "108",
    number = "7",
    pages = "075010",
    year = "2023"
}

@article{Cline:2011mm,
    author = "Cline, James M. and Kainulainen, Kimmo and Trott, Michael",
    title = "{Electroweak Baryogenesis in Two Higgs Doublet Models and B meson anomalies}",
    eprint = "1107.3559",
    archivePrefix = "arXiv",
    primaryClass = "hep-ph",
    doi = "10.1007/JHEP11(2011)089",
    journal = "JHEP",
    volume = "11",
    pages = "089",
    year = "2011"
}

@article{Kanemura:2023juv,
    author = "Kanemura, Shinya and Mura, Yushi",
    title = "{Electroweak baryogenesis via top-charm mixing}",
    eprint = "2303.11252",
    archivePrefix = "arXiv",
    primaryClass = "hep-ph",
    reportNumber = "OU-HET-1174",
    doi = "10.1007/JHEP09(2023)153",
    journal = "JHEP",
    volume = "09",
    pages = "153",
    year = "2023"
}

@article{Fromme:2006cm,
    author = "Fromme, Lars and Huber, Stephan J. and Seniuch, Michael",
    title = "{Baryogenesis in the two-Higgs doublet model}",
    eprint = "hep-ph/0605242",
    archivePrefix = "arXiv",
    reportNumber = "CERN-PH-TH-2006-094, BI-TP-2006-18",
    doi = "10.1088/1126-6708/2006/11/038",
    journal = "JHEP",
    volume = "11",
    pages = "038",
    year = "2006"
}

\end{document}